\documentclass[a4paper,11pt]{article}
\pdfoutput=1
\usepackage{jheppub}
\usepackage{natbib}
\setcitestyle{mcite,numbers,sort&compress}

\usepackage{graphicx}  % needed for figures
\usepackage{float}
\usepackage{dcolumn}   % needed for some tables
\usepackage{bm}        % for math
\usepackage{amssymb}   % for math
\usepackage{slashed}   % for Dirac Slash
\usepackage{amsmath}   % for mutiline eqn
\usepackage{simplewick} % for contraction
\usepackage{verbatim}  % for multi-line comment
\usepackage{color} % for textcolor
\usepackage{xcolor} % for textcolor
\usepackage{appendix} % for appendix
\usepackage{subfig} % for sub fig
\usepackage{epstopdf}

\definecolor{Blue}{rgb}{0, 0.1, 0.5}

\usepackage{hyperref}

\newcounter{YJC}

\begin{document}
%\widetext

\title{Study of the Roberge-Weiss phase caused by external uniform classical electric field using lattice QCD approach}
%\affiliation{Department of Physics, Liaoning Normal University, Dalian 116029, China}

\author[a,b]{Ji-Chong Yang}
%\email{yangjichong@fudan.edu.cn}
\emailAdd{yangjichong@lnnu.edu.cn}
\affiliation[a]{Department of Physics, Liaoning Normal University, No. 850 Huanghe Road, Dalian 116029, P.R. China.}
\affiliation[b]{Center for Theoretical and Experimental High Energy Physics, Liaoning Normal University, No. 850 Huanghe Road, Dalian 116029, China}

\author[a,b]{Xiao-Ting Chang}
\emailAdd{cxtinglns@163.com}
%\affiliation{Department of Physics, Liaoning Normal University, Dalian 116029, China}

\author[a,b]{Jian-Xing Chen}
\emailAdd{13614090213@163.com}
%\affiliation{Department of Physics, Liaoning Normal University, Dalian 116029, China}

%\begin{abstract}
\abstract{
The effect of an external electric field on the quark matter is an important question due to the presence of strong electric fields in heavy ion collisions.
In the lattice QCD approach, the case of a real electric field suffers from the `sign problem', and a classical electric field is often used similar as the case of chemical potential.
Interestingly, in axial gauge a uniform classical electric field actually can correspond to an inhomogeneous imaginary chemical potential that varies with coordinate.
On the other hand, with imaginary chemical potential, Roberge-Weiss~(R-W) phase transition occurs.
In this work, the case of a uniform classical electric field is studied by using lattice QCD approach, with the emphasis on the properties of the R-W phase.
Novel phenomena show up at high temperatures.
It is found that, the chiral condensation oscillates with $z$ at high temperatures, and so is the absolute value of the Polyakov loop.
It is verified that the charge density also oscillates with $z$ at high temperatures.
The Polyakov loop can be described by an ansatz $A_p+\sum _{q=u,d} C_q\exp\left(L_{\tau} Q_q iazeE_z\right)$, where $A_p$ is a complex number and $C_d>0,C_u\geq 0$ are real numbers that are fitted for different temperatures and electric field strengths.
As a consequence, the behavior of the phase of Polyakov loop is different depending on whether the Polyakov loop encloses the origin, which implies a possible phase transition.
}
%\end{abstract}

\maketitle

\section{\label{sec1}Introduction}

The study of Quantum Chromodynamics~(QCD) matter is essential for a deeper understanding of the nature of strong interactions.
In recent years, the effects of electromagnetic fields have also become a hot topic due to the strong electromagnetic fields that can be generated in heavy ion collision experiments~\cite{magneticCatalysis1,magneticCatalysis2,magneticCatalysis3,magneticreview1,magneticreview2,electricmagneticreview}.
The effect of electromagnetic field on chiral condensation has been investigated by using various low-energy effective models ~\cite{Babansky:1997zh,Goyal:1999ye,Klevansky:1992qe,Ebert:1999ht,Klimenko:2003ci}, which shows that the magnetic field may induces magnetic catalysis.
However, inverse catalysis is observed around pseudo critical temperature~\cite{inversecatalysis} in the lattice calculation, and then the presence of an external magnetic field is investigated intensively in both lattice approach~\cite{latticemag9,Bali:2013esa,latticemag1,latticemag2,latticemag3,latticemag4,latticemag5,magneticreview2,latticemag8} and using effective models~\cite{Mao:2016lsr,Chernodub:2010xce,Ferreira:2014kpa,Chao:2013qpa,Bazavov:2012vg,MagneticInhibition,polyakovloop,polyakovloop2,Chiralityimbalance,Liu:2014uwa,Liu:2018zag}.
By now the community has developed a good understanding of inverse magnetic catalysis, but there is still no theory that can explain all the phenomena observed in lattice simulations at one time, for example the phenomena with respect to diamagnetism, paramagnetism, meson mass, meson condensation, etc.

Not only magnetic fields but also strong electric fields are generated in non-central high-energy heavy ion collisions~\cite{largeElectricFieldInHIC1,largeElectricFieldInHIC2,largeElectricFieldInHIC3}, which can reach as large as about $10m_{\pi}^2$, where $m_{\pi}$ is the pion mass.
In the electric field case, it has been shown that, the external electric field restores the chiral symmetry~\cite{electricAndMagnetic1,electricAndMagnetic2,electricPNJL,Babansky:1997zh,Cao:2015dya,Ruggieri:2016xww,Ruggieri:2016jrt}.
In the lattice QCD approach, the case of an external real~(Minkowski) electric field suffers from the notorious `sign problem', except for the case of isospin electric charges~\cite{isospincharge}.
Similar as the case of chemical potential~\cite{imaginarychemical1,imaginarychemical2}, the analytical extension is often used to study the external electric field, which is also known as Euclidean electric field, or classical electric field~\cite{u1phaseisimaginary1,u1phaseisimaginary2,u1phaseisimaginary3}.
In the previous studies on hadron electromagnetic polarizabilities~\cite{electricpolarizability1,electricpolarizability2,electricpolarizability3}, lattice QCD with an external electric field has been shown to be a reliable tool.
The electric susceptibility is also studied with the presence an external classical electric field~\cite{chargedistrib2}, and it is found that a non-constant charge distribution is required to maintain equilibrium~\cite{chargedistrib1}.

Another interesting phenomena connecting the case of external classical electric field and the case of imaginary chemical potential is the presence of Roberge-Weiss~(R-W) transition~\cite{RWtransition}, which has been investigated by using lattice QCD approach~\cite{RWlattice1}.
In fact, except for the boundary, a uniform~(homogeneous) static external classical electric field in axial gauge is equivalent to an inhomogeneous imaginary chemical potential.
Since the case of a homogeneous imaginary chemical potential and the corresponding R-W transition has been verified in the lattice QCD approach,
and a uniform external classical electric field corresponds to the case of imaginary chemical potential which varies according to coordinates linearly, the presence of the R-W transition is also expected in the case of external classical electric field.
A phase transition similar to the R-W transition in the presence of fermionic fields coupled to magnetic backgrounds is studied~\cite{pzcomplexplane}.

In this work, the effect of a strong external uniform classical electric field is studied by using the lattice QCD approach, with the emphasis on the R-W phase caused by the external electric field.
The case of $N_f=1+1$ Kogut-Susskind staggered fermions with the same bare mass and different electric charges are investigated.
The chiral condensation and charge distribution are also investigated.

The remained of this paper is organized as follows.
In section~\ref{sec2}, the model with a uniform external electric field is presented, and the connection between the electric field and imaginary chemical potential is discussed.
The numerical results are established in section~\ref{sec3}.
Section~\ref{sec4} is a summary.

\section{\label{sec2}External electric field}

Considering an external electric field at ${\bf z}$ direction, ${\bf E}=(0,0,E_z)$, in the axial gauge $A^{\rm EM}_z=0$, the gauge field can be written as $A^{\rm EM}_{\mu}=(-E_z z, 0, 0, 0)$ such that ${\bf E}=(F^{\rm EM}_{tx},F^{\rm EM}_{ty},F^{\rm EM}_{tz})$ where $F^{\rm EM}_{\mu\nu}=\partial _{\mu}A^{\rm EM}_{\nu}-\partial _{\nu}A^{\rm EM}_{\mu}$, the superscript `EM' is added to distinguish with the QCD gauge field.
The Lagrangian with one massless fermion is
\begin{equation}
\begin{split}
&\mathcal{L}_{q}=\bar{\psi} _q\slashed{\partial} _{\mu}\psi _q+ \bar{\psi} _q i \slashed{A}\psi _q -i Q_q e E_z z \bar{\psi}_q \gamma _0 \psi _q,
\end{split}
\label{eq.2.1}
\end{equation}
where $A_{\mu}=g\sum _a T^a A^a_{\mu}$ is the QCD gauge field, $Q_q$ is the electric charge of the fermion.

A Wick rotation is performed to put the Lagrangian into a Euclidean space which applies a substation that $t\to -i \tau$, $\partial _t \to i\partial _{\tau}$, $A_0 \to iA_4$ and $A^{\rm EM}_0 \to iA^{\rm EM}_4$, so that
\begin{equation}
\begin{split}
&\int d^4 x \mathcal{L}_{q}\to S_q=\int d^4x^E \left(\bar{\psi} \sum _{j=1}^4 \gamma _j^E \partial _j \psi + \sum _{j=1}^4 \bar{\psi} i g \gamma _j^E A_j\psi -i Q_q e E_z z \bar{\psi} \gamma _4^E \psi\right),
\end{split}
\label{eq.2.2}
\end{equation}
where $\gamma _{j=1,2,3}=i \gamma _j^E$ and $\gamma _4^E = \gamma _0$.
The tangent space Wick rotation~\cite{tangentspacewr} also yields the same result.

Note that, the substitution $A^{\rm EM}_0 \to iA^{\rm EM}_4$ corresponds to an imaginary electric field, or Euclidean electric field, which can also be viewed as an analytical extension.

\begin{figure}
\begin{center}
\includegraphics[width=0.7\textwidth]{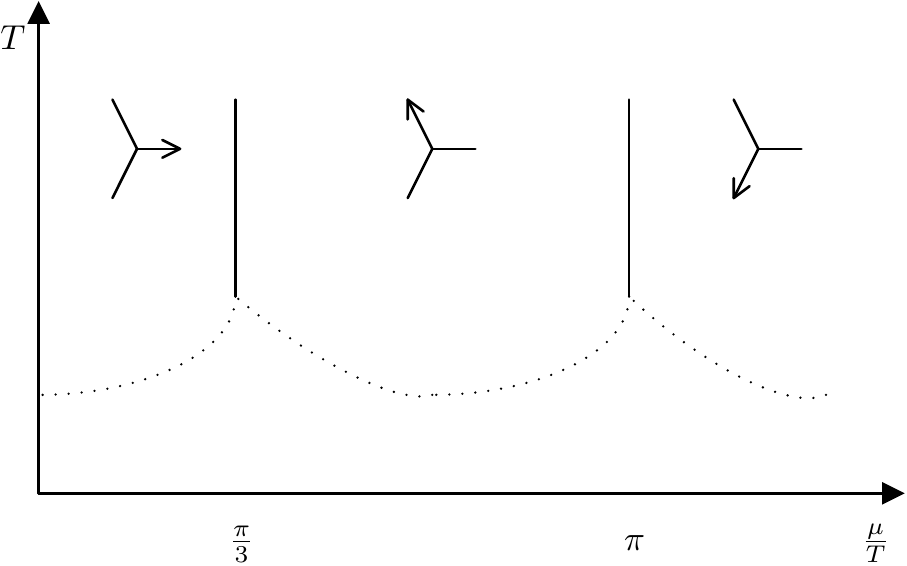}
\caption{\label{fig:rw}A Sketch of the phase diagram of R-W transition.}
\end{center}
\end{figure}
On the other hand, the action with imaginary chemical potential can be written as
\begin{equation}
\begin{split}
&\mathcal{L}_{q}=\bar{\psi}_q \slashed{\partial} _{\mu}\psi _q + \bar{\psi} _q i \slashed{A}\psi _q -i\mu \bar{\psi} _q \gamma _0 \psi _q.
\end{split}
\label{eq.2.3}
\end{equation}
A simple observation is that the case of the presence of an external electric field can be viewed as a stacking of volumes with different imaginary chemical potentials $\mu =Q_q e E_z z$ extending the ${\bf z}$-axis.
The Lagrangian in Eq.~(\ref{eq.2.3}) has been studied and an R-W transition is predicted and verified by lattice simulations.
The definition feature of the R-W transition is the presence of imaginary part of the Polyakov loop.
The sketch of the phase diagram of R-W transition is shown in Fig.~\ref{fig:rw}.
At high temperatures, there is a first order phase transition that the phase of the Polyakov loop is $2n\pi/3$, when $(2n-1)\pi/3 < \mu/T < (2n+1)\pi/3$, where $n$ are integers.
For the case of Euclidean electric field, some questions arise.
Is there also R-W transition induced by an external uniform electric field?
Is it appropriate to study the case of external electric field as a stacking of volumes with different imaginary chemical potentials?
To answer those questions, the R-W phase induced by the external electric field is studied using the lattice approach.

By using the staggered fermion~\cite{ks,book2010}, the action can be discretized as
\begin{equation}
\begin{split}
&S_G=\frac{\beta}{N_c}\sum _n \sum _{\mu>\nu}{\rm Retr}\left[1-U_{\mu\nu}(n)\right],\\
&S_q=\sum _n \left(\sum _{\mu}\sum _{\delta = \pm \mu}\bar{\chi}(n) U_{\delta}(n)V_{\delta}(n)\eta _{\delta}(n)\chi (n+\delta) +2am\bar{\chi}\chi \right),
\end{split}
\label{eq.2.4}
\end{equation}
where $a$ is the lattice spacing, $S_G$ is the Wilson gauge action~\cite{Wilsongauge,book2010} where $\beta=2N_c/g^2_{\rm YM}$ with $g_{\rm YM}$ the coupling strength of the gauge fields to the quarks, $m$ is the fermion mass, $U_{\mu}=e^{iaA_{\mu}}$, $V_{\mu}=e^{ia e A^{\rm EM}_{\mu}}$, $\eta _{\mu}(n)=(-1)^{\sum _{\nu<\mu}n_{\nu}}$ and $U_{-\mu}(n)=U_{\mu}^{\dagger}(n-\mu)$, $V_{-\mu}(n)=V_{\mu}^*(n-\mu)$, $\eta _{-\mu}=-\eta _{\mu}(n-\mu)$.
A twisted boundary condition is applied to ensure gauge invariance~\cite{u1phaseisimaginary3,twist1,twist2,twist3}, therefore we use
\begin{equation}
\begin{split}
&f=Q_qa^2F=\frac{2k\pi}{L_{\mu}L_{\nu}},\;\;\;k\in \mathbb{Z},\\
&V_{\nu}=e^{i f n_{\mu}},\;\;V_{\mu}(n_{\mu}=L_{\mu})=e^{-ifL_{\mu}n_{\nu}}.\\
\end{split}
\label{eq.2.5}
\end{equation}
where $L_{\mu}$ is the extent at direction $\mu$.

In lattice simulations, the origin of the axis is set to be the middle of the spatial volume and at $n_{\tau}=1$.
With twisted boundary condition, $S_q$ is $U(1)$ gauge invariant, i.e. free of gauge choice~(but different gauge choice will result in different twisted boundary conditions), and therefore the results do not depend on the gauge choice.
For ${\bf E}$ at the ${\bf z}$-direction and for axial gauge, $V_{\mu}(n)=1$ except for
\begin{equation}
\begin{split}
&V_{\tau}(n)=e^{-i a Q_q e E_z z},\;\;V_z(a^{-1}z=L_z/2-1)=e^{i a Q_q e E_z L_z \tau},\\
\end{split}
\label{eq.2.6}
\end{equation}
with quantized electric field $a^2 E_z=6k \pi / L_{\tau}L_z$ so that $f$ satisfies $2k \pi / L_{\tau}L_z$ with $|Q_q|=1/3$.
In the simulation, we use $L_x\times L_y\times L_z\times L_{\tau} =12\times 12\times 12 \times 6$, therefore $a^{-1}z=-6,-5,\ldots,5$, $a^{-1}\tau =0,1,\ldots,5$ and $a^2E_z=k\times \pi / 12$.
In this work, we use $k=0,1,\ldots , 12$.
The case of $k=12+n$ is equivalent as $k=12-n$ with an electric field on the opposite direction, where $n$ are integers.
Note that, for $a^2eE_z\sim \mathcal{O}(1)$ or larger, the results suffer from strong discretization errors.
In this case, the lattice action can no longer approximate the presence of external electric field well.
Therefore, in sections.~\ref{sec3.2.3}, \ref{sec3.2.4} and \ref{sec3.3.3} which are closely related to physical phenomena, we only consider the results with $0\leq k \leq 4$.

\section{\label{sec3}Numerical results}

\subsection{\label{sec3.1}Matching}

\begin{table}[!htbp]
\begin{center}
\begin{tabular}{c|c|c|c||c|c|c|c}
%$\beta$ & $r_0/a$ & $a^{-1}\;({\rm MeV})$ & $\langle \bar{\psi}\psi\rangle_l$ &
%$\beta$ & $r_0/a$ & $a^{-1}\;({\rm MeV})$ & $\langle \bar{\psi}\psi\rangle_l$ \\
%\hline
%$5.30$ & $3.079(30)$ & $1215(12)$ & $0.08773(9)$ &
%$5.48$ & $4.906(46)$ & $1936(18)$ & $0.06799(5)$ \\
%$5.32$ & $3.277(42)$ & $1293(16)$ & $0.08421(9)$ &
%$5.50$ & $5.189(36)$ & $2048(14)$ & $0.06665(3)$ \\
%$5.34$ & $3.484(50)$ & $1375(20)$ & $0.08113(9)$ &
%$5.52$ & $5.435(60)$ & $2145(24)$ & $0.06554(5)$ \\
%$5.36$ & $3.677(31)$ & $1451(12)$ & $0.07857(8)$ &
%$5.54$ & $5.579(51)$ & $2202(20)$ & $0.06455(4)$ \\
%$5.38$ & $3.924(45)$ & $1549(18)$ & $0.07619(9)$ &
%$5.56$ & $5.829(31)$ & $2300(12)$ & $0.06351(2)$ \\
%$5.40$ & $4.113(40)$ & $1623(16)$ & $0.07418(7)$ &
%$5.58$ & $6.133(47)$ & $2420(19)$ & $0.06255(3)$ \\
%$5.42$ & $4.382(58)$ & $1730(23)$ & $0.07224(9)$ &
%$5.60$ & $6.386(46)$ & $2520(18)$ & $0.06169(2)$ \\
%$5.44$ & $4.561(30)$ & $1800(12)$ & $0.07063(4)$ &
%$5.62$ & $6.545(39)$ & $2583(15)$ & $0.06092(2)$ \\
%$5.46$ & $4.808(61)$ & $1898(24)$ & $0.06924(8)$ &
%$5.64$ & $6.957(55)$ & $2746(22)$ & $0.06009(2)$ \\
$\beta$ & $r_0/a$ & $a^{-1}\;({\rm MeV})$ &
$\beta$ & $r_0/a$ & $a^{-1}\;({\rm MeV})$ \\
\hline
$5.30$ & $3.079(30)$ & $1215(12)$ &
$5.48$ & $4.906(46)$ & $1936(18)$ \\
$5.32$ & $3.277(42)$ & $1293(16)$ &
$5.50$ & $5.189(36)$ & $2048(14)$ \\
$5.34$ & $3.484(50)$ & $1375(20)$ &
$5.52$ & $5.435(60)$ & $2145(24)$ \\
$5.36$ & $3.677(31)$ & $1451(12)$ &
$5.54$ & $5.579(51)$ & $2202(20)$ \\
$5.38$ & $3.924(45)$ & $1549(18)$ &
$5.56$ & $5.829(31)$ & $2300(12)$ \\
$5.40$ & $4.113(40)$ & $1623(16)$ &
$5.58$ & $6.133(47)$ & $2420(19)$ \\
$5.42$ & $4.382(58)$ & $1730(23)$ &
$5.60$ & $6.386(46)$ & $2520(18)$ \\
$5.44$ & $4.561(30)$ & $1800(12)$ &
$5.62$ & $6.545(39)$ & $2583(15)$ \\
$5.46$ & $4.808(61)$ & $1898(24)$ &
$5.64$ & $6.957(55)$ & $2746(22)$ \\
\end{tabular}
\end{center}
\caption{\label{tab:matching1}The coupling constant $\beta$ and the lattice spacing matched by using $r_0=0.5\;{\rm fb}$.}
\end{table}

The lattice simulation is performed with the help of \verb"Bridge++" package~\cite{bridgepp}.
To study the effect of external electric field, the simulation is carried out with $N_f=1+1$, where $u$ and $d$ quarks carrying different electric charges.
The bare mass is chosen as $m_q=0.1a^{-1}$ for both fermion fields where $a$ is lattice spacing, when the electric field is not presented, the two fermion fields degenerate and $N_f=2$.
The coupling constant of gauge field $\beta$, and the corresponding $a$ are listed in Table~\ref{tab:matching1}.
The lattice spacing is matched by measuring static quark potential $V(r)$~\cite{r0a,r0b,r0c} and matching the `Sommer scale' $r_{0}$ to $r_{0}=0.5\; {\rm fm}$~\cite{sommer,r1a,r1b} at low temperature~(at $L_{\tau}=48$).
Throughout this paper, the statistical error is estimated as $\sigma=\sigma _{\rm jk} \sqrt{2\tau _{\rm ind}}$~\cite{book2010}, where $\sigma _{\rm jk}$ is statistical error calculated using `jackknife' method, and $2\tau _{\rm ind}$ is the separation of molecular dynamics time units~(T.U.) such that the two configurations can be regarded as independent, which is calculated by using `autocorrelation' with $S=1.5$~\cite{autocorrelation} on the bare chiral condensation of quark~($u$ quark in the case of $N_f=1+1$).
When matching, we use $200$ trajectories as thermalization, and $1000$ configurations are measured for each $\beta$.
In the following, for each $\beta$, $200+3000\times 13$ trajectories are simulated.
The first $200$ trajectories are discarded for thermalization, then $3000\times 13$ trajectories are simulated with sequentially growing $a^2eE_z=k\pi/12$ for $k=0,1,2,\ldots, 12$.
The first $100$ trajectories of the $3000$ are discarded for thermalization and $2900$ configurations are measured for each $\beta$.

\begin{figure}
\begin{center}
\includegraphics[width=0.6\textwidth]{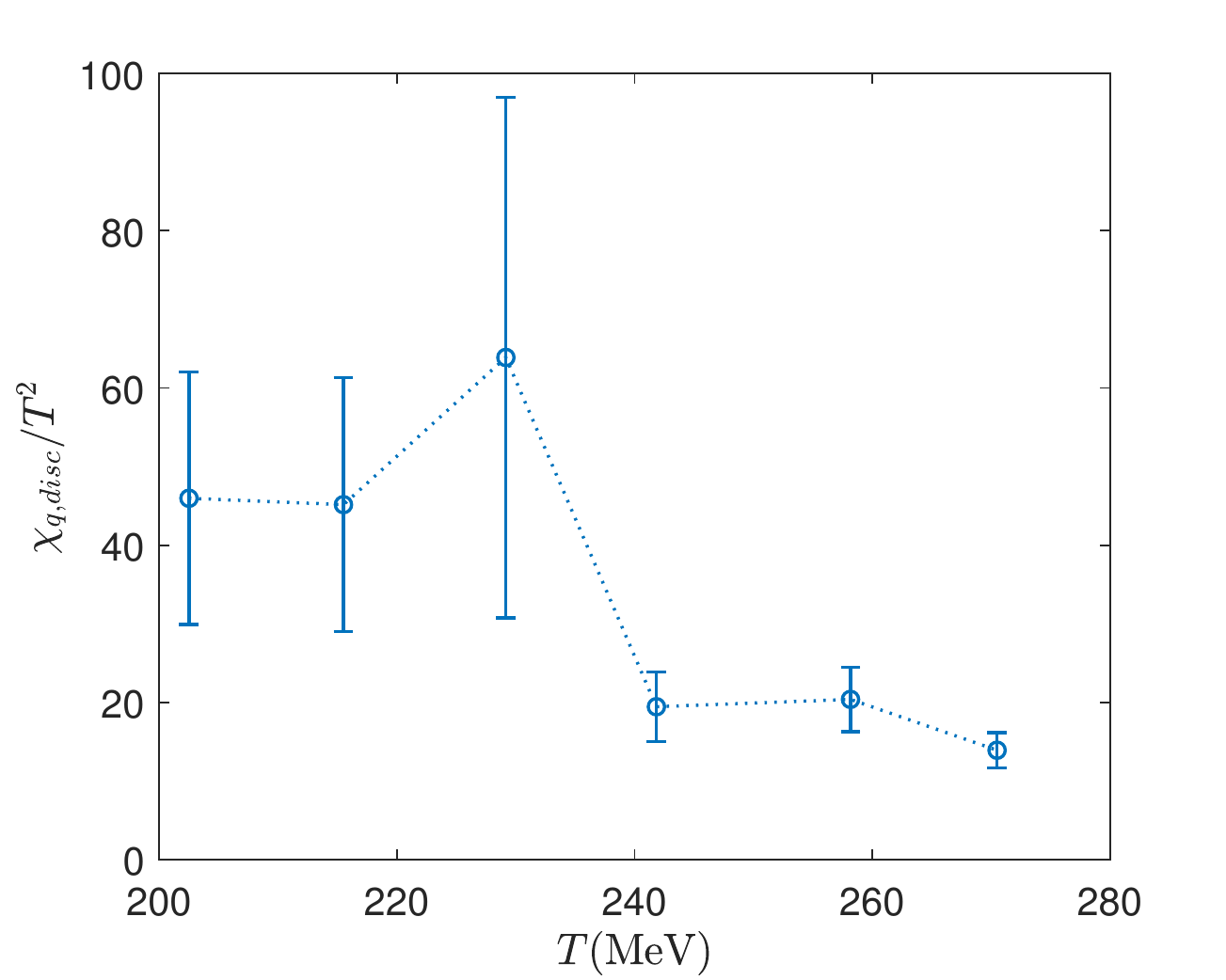}
\caption{\label{fig:phase0}$\chi _{q,disc}/T^2$ as a function of $T$.}
\end{center}
\end{figure}

The pseudo critical temperature is determined at $E_z=0$ by using disconnected susceptibility of chiral condensation defined as~\cite{chiralsusp}
\begin{equation}
\begin{split}
&\chi _{q,disc}=\frac{N_f^2}{16L_x^3L_{\tau}}\left(\langle {\rm tr}\left[D_q^{-1}\right]^2\rangle-\langle {\rm tr}\left[D_q^{-1}\right]\rangle^2\right),
\end{split}
\label{eq.suspchiraldefine}
\end{equation}
which is depicted in Fig.~\ref{fig:phase0}.
It can be found that $\beta_c = 5.34$ and $T_c = 229\;{\rm MeV}$.
Note that, for different $\beta$, $m_q$ is different.

\subsection{\label{sec3.2}Chiral condensation}

Since the bare mass $m_q$ is different for different lattice spacing, we directly use $c_q=\langle \bar{\psi}_q\psi_q \rangle/V$ where $V$ is volume.
Such a definition has the problem of renormalization and is not suitable for comparison between different temperatures~(there would have been difficulties to compare between different temperatures since $m_q$ is different at different temperatures in our simulations), but can show the pattern of $c_q$ with different $E_z$.
Other quantities of interest are charge density defined as $c^4_q=\langle \bar{\psi}_q\gamma _4 \psi_q \rangle/V$, and current density $c^3_q=\langle \bar{\psi}_q\gamma _3 \psi_q \rangle/V$.
In terms of staggered fermion field, they are
\begin{equation}
\begin{split}
&c_q=\frac{1}{4L_xL_yL_zL_{\tau}}\frac{2}{a^3}\langle \sum_n\bar{\chi }(n)\chi (n)\rangle,\\
&c_q^{\mu}=\frac{1}{4L_xL_yL_zL_{\tau}}\frac{1}{a^3}\langle \sum _n\eta _{\mu}(n)\sum _{\delta=\pm \mu}\bar{\chi }(n)U_{\delta}(n)V_{\delta}(n)\chi (n+\delta)\rangle.\\
\end{split}
\label{eq.cqdefine}
\end{equation}
In order to study the influence of the chemical potential as the coordinate $z$ changes, we also define $c_q(z)$, as $c_q(z)=\langle \sum _{n_z=z}\bar{\psi}_q(n)\psi_q(n) \rangle / \left(L_xL_yL_{\tau}\right)$, which is the chiral condensation of a $z$-slice.
$c^{3,4}_q(z)$ are defined similarly.
In the following, $c_q$ and $c_q^3$ are treated as real numbers, and $c_q^4$ is treated as complex.

\subsubsection{\label{sec3.2.1}The Z distribution of chiral condensation}

\begin{figure}
\begin{center}
\includegraphics[width=0.48\textwidth]{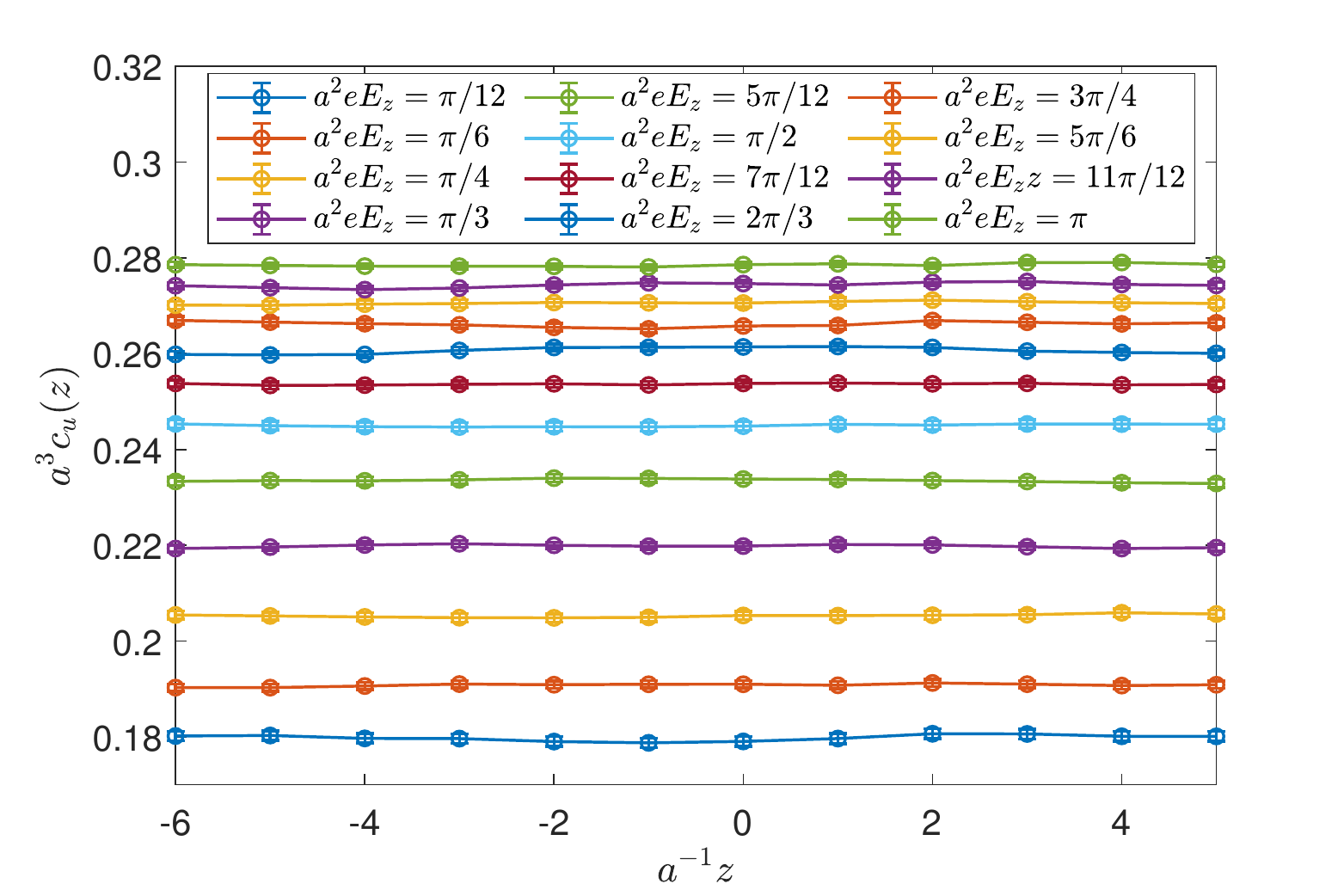}
\includegraphics[width=0.48\textwidth]{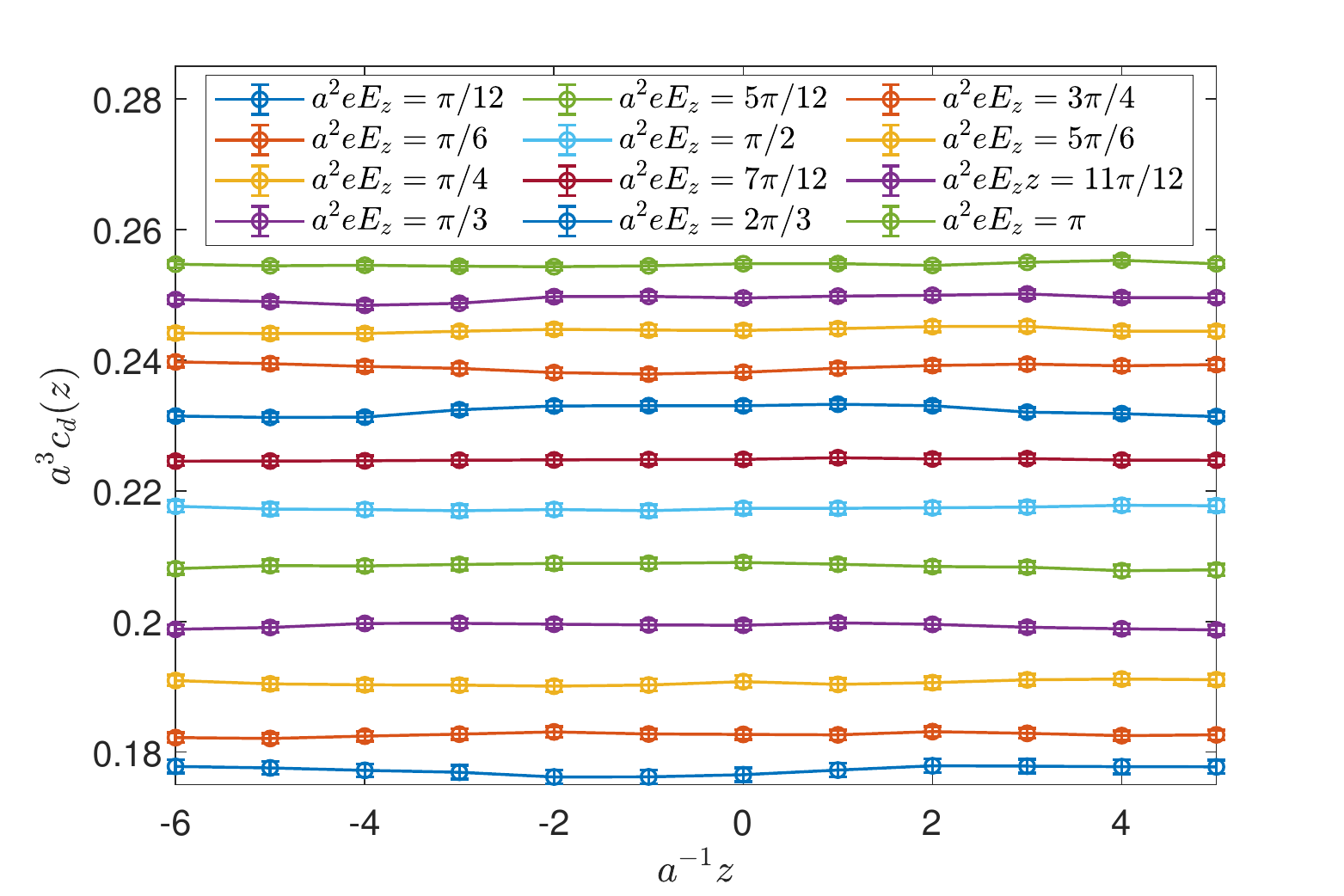}
\caption{\label{fig:cq530}$c_u(z)$~(the left panel) and $c_d(z)$~(the right panel) at $\beta=5.3$.}
\end{center}
\end{figure}

\begin{figure}
\begin{center}
\includegraphics[width=0.48\textwidth]{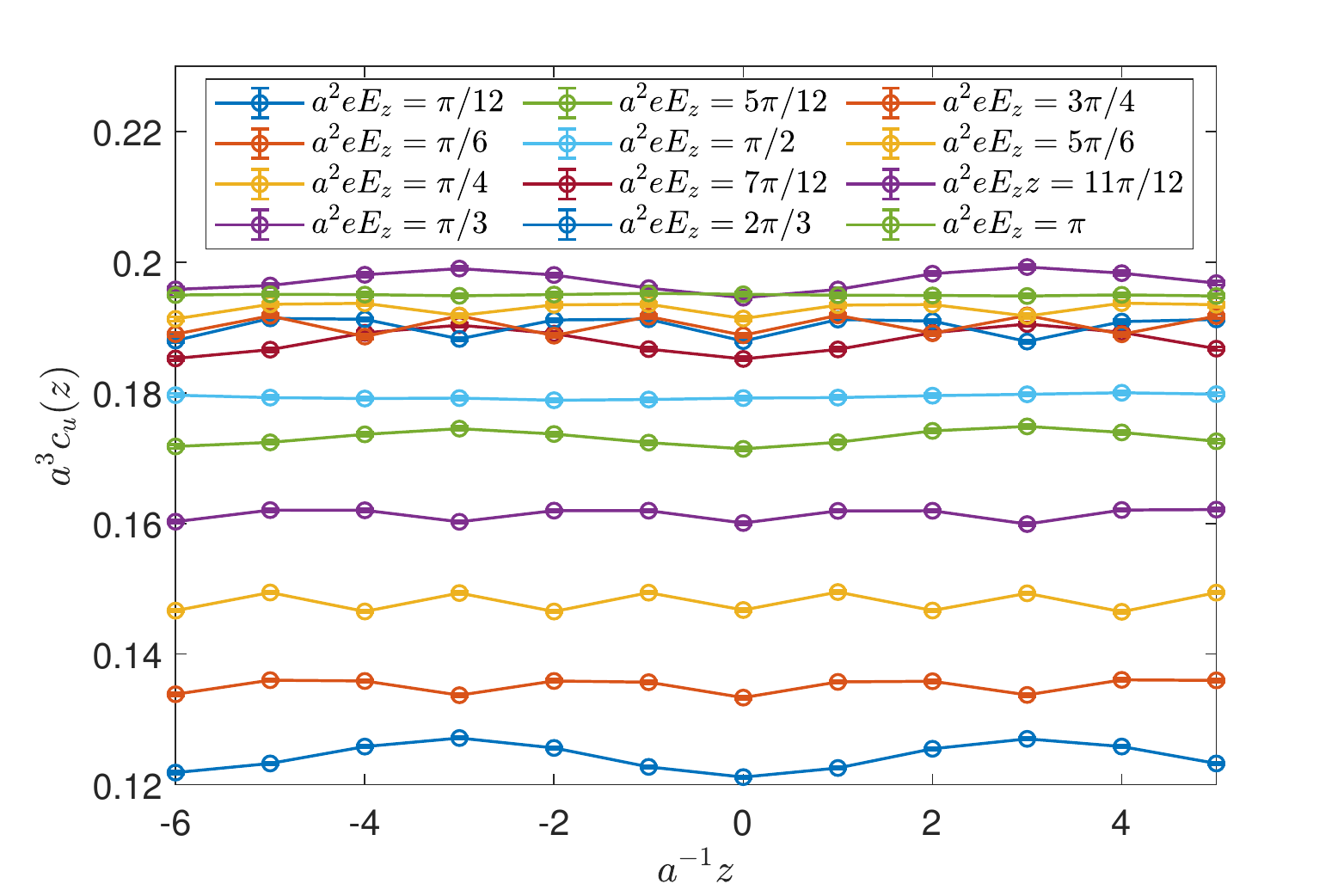}
\includegraphics[width=0.48\textwidth]{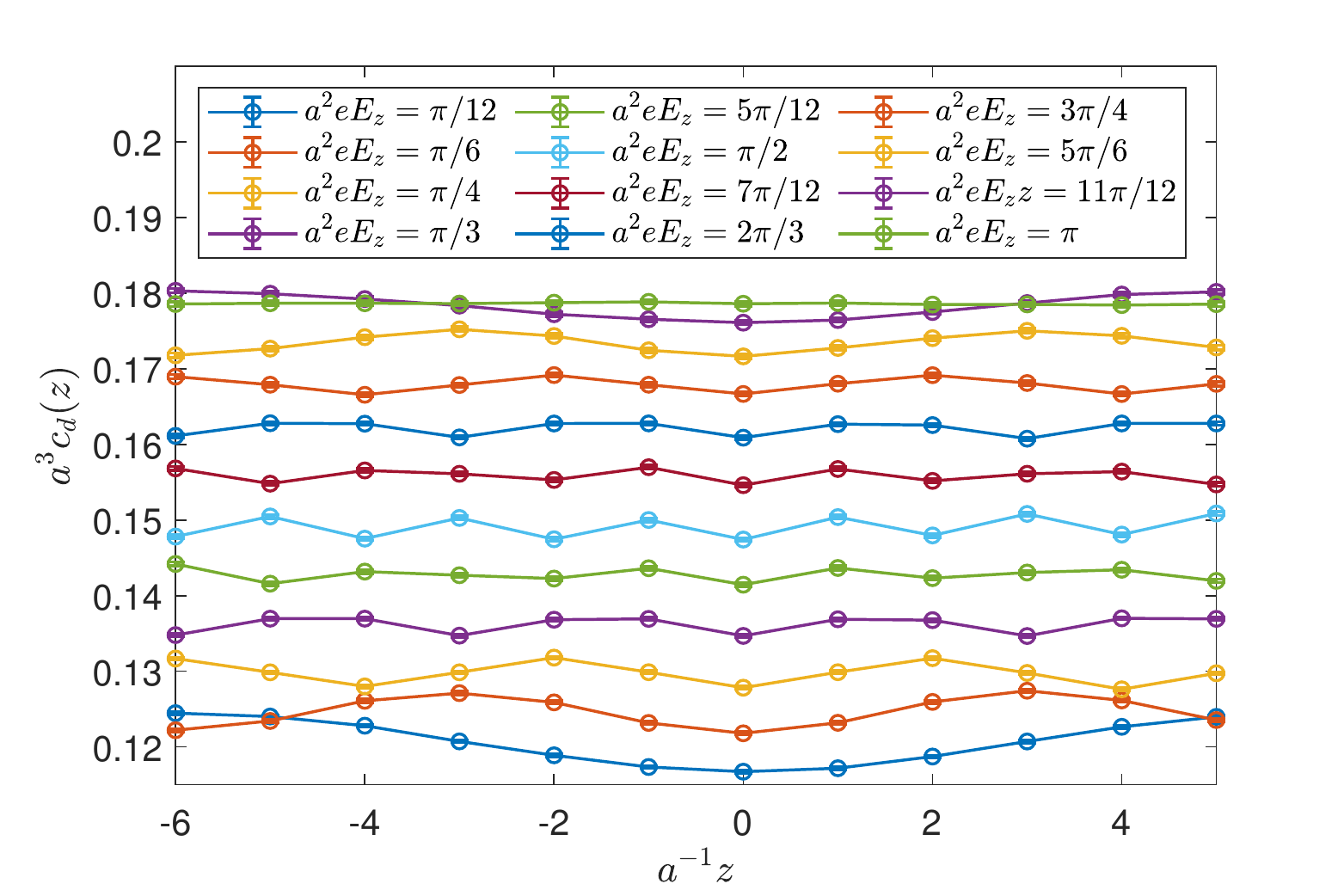}
\caption{\label{fig:cq564}Same as Fig.~\ref{fig:cq530} but for $\beta=5.64$.}
\end{center}
\end{figure}

As introduced, one of our main concerns is whether the imaginary chemical potential, which varies with coordinate $z$, brings about a distribution that varies with $z$ or whether it brings about an overall change.
We find that the results depend on the temperature at which the quark matter is located.
For $\beta=5.3$ and $\beta=5.64$, $c_q(z)$ are shown in Figs.~\ref{fig:cq530} and \ref{fig:cq564}.
As can be seen, the chiral condensation rises as the electric field strength increases.
The difference is that for lower temperatures, the chiral condensation does not appear to vary with the coordinate $z$.
For high temperatures, it is clearly an oscillatory function of $z$.
The shape of the function is consistent with a trigonometric function.

\begin{figure}
\begin{center}
\includegraphics[width=0.48\textwidth]{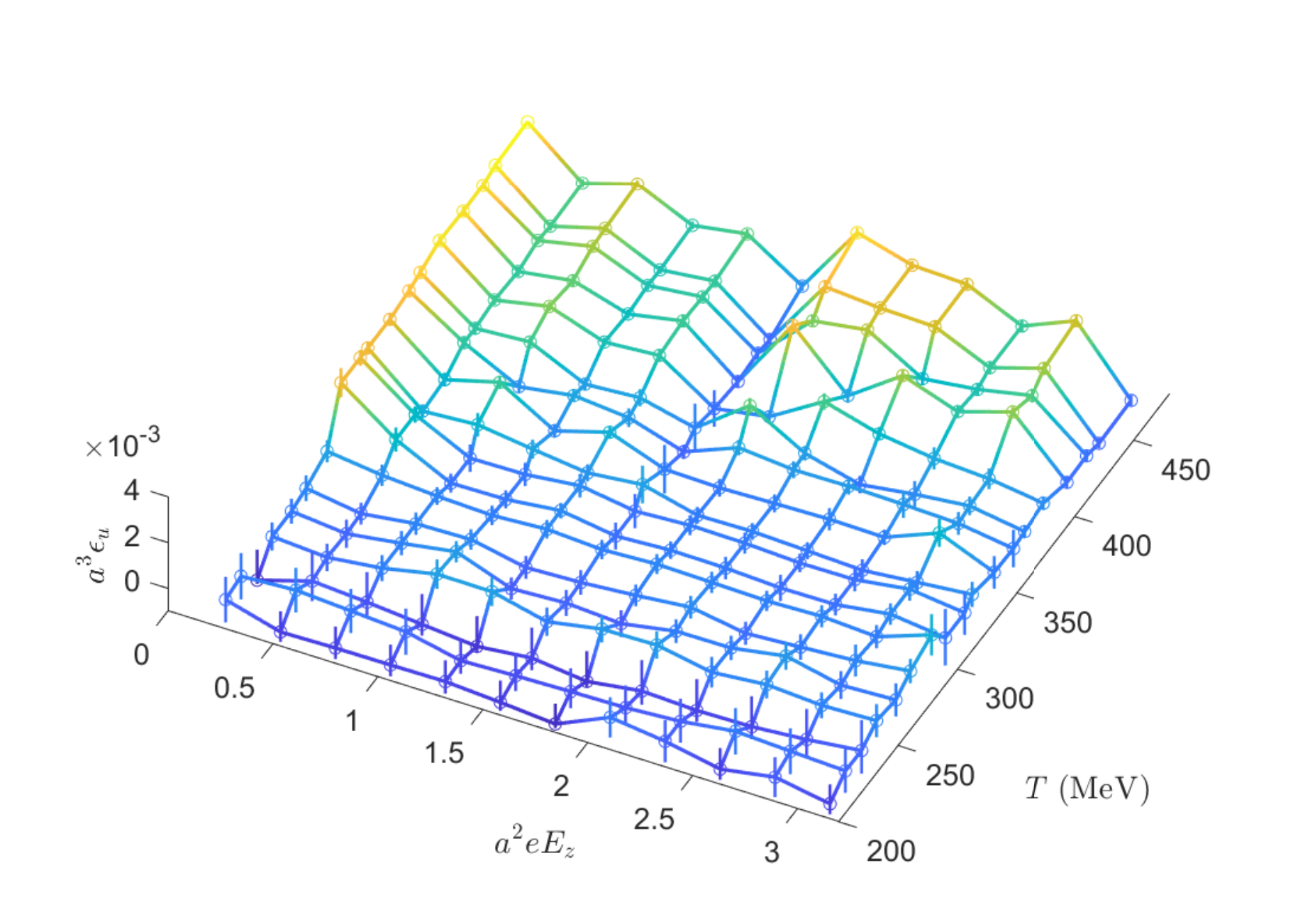}
\includegraphics[width=0.48\textwidth]{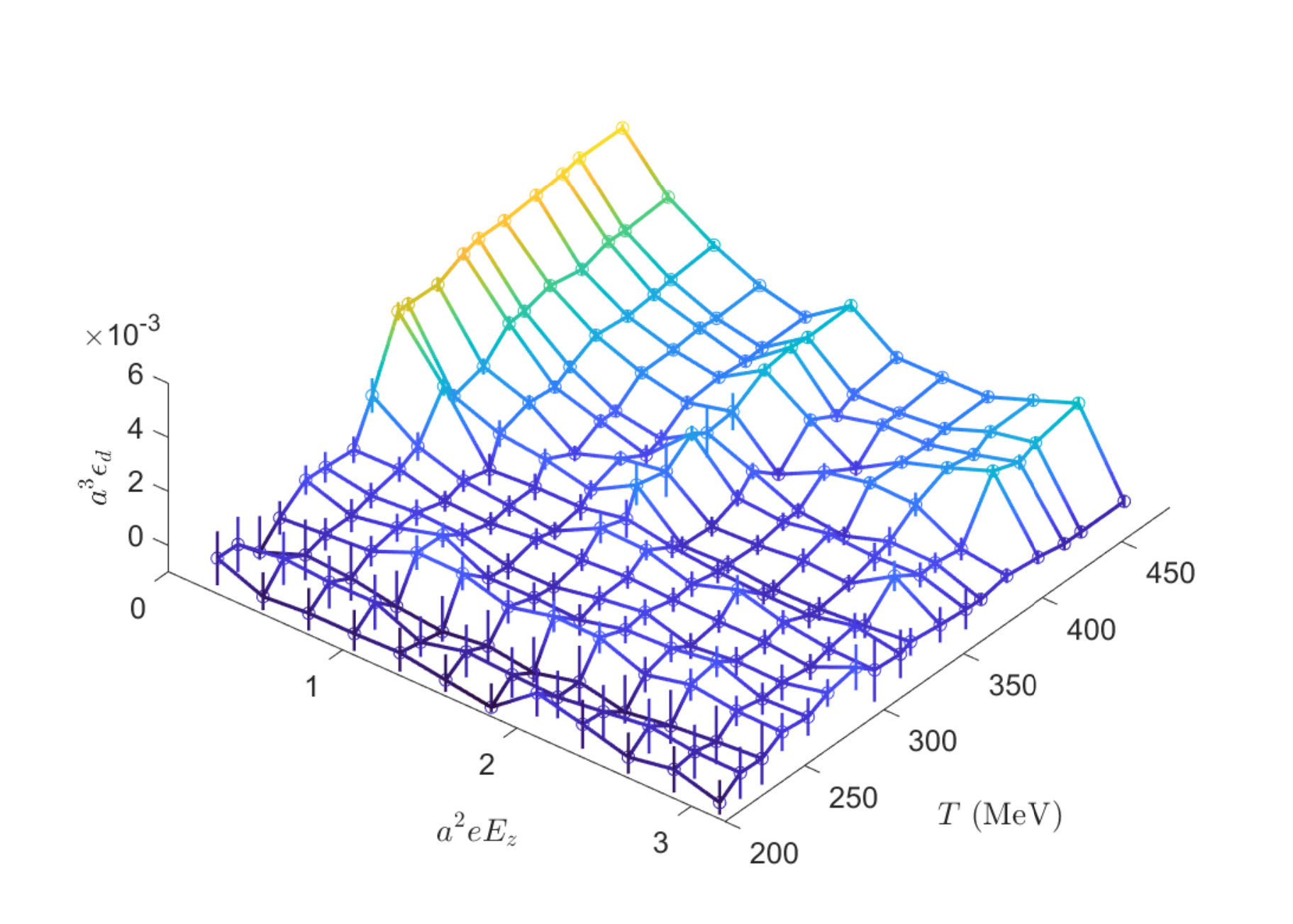}
\caption{\label{fig:stdcq}$\epsilon _u$~(the left panel) and $\epsilon _d$~(the left panel) as functions of temperature $T$ and $E_z$.}
\end{center}
\end{figure}
As a signature of whether the chiral condensation varies with $z$, we borrow the definition of standard deviation and define the magnitude of the oscillation of the chiral condensation over $z$, as
\begin{equation}
\begin{split}
&\varepsilon _q(T,E_z) = \sqrt{\frac{1}{L_z-1}\sum _z\left(c_q^{T, E_z}(z)-\frac{1}{L_z}\sum _{z'} c_q^{T, E_z}(z')\right)^2},
\end{split}
\label{eq.epsilonq}
\end{equation}
where $c_q^{T, E_z}(z)$ is $c_q(z)$ at temperature $T$ and electric field strength $E_z$.
When $E_z=0$, there is no imaginary chemical potential that varies with $z$, therefore $\varepsilon _q(T,0)$ are set as baselines, and we define $\epsilon _q(T,E_z)=\varepsilon _q(T,E_z)-\varepsilon _q(T,0)$.
$\epsilon _q$ are shown in Fig.~\ref{fig:stdcq}.
Generally, the amplitude of oscillation grows with temperature.
It can also be observed that, for the case of $d$ quark whose electric charge is $-1/3$, the oscillation disappears at $a^2eE_z=\pi$, for the case of $u$ quark whose electric charge is $2/3$, the oscillation disappears at $a^2eE_z=\pi/2$ and $a^2eE_z=\pi$.
This can be explained when the frequency of the oscillation is investigated.

Since the linearly changed imaginary chemical potential leads to a periodic change in the partition function, it can be speculated that the chiral condensation that oscillates periodically with the $z$-direction is a reflection of the linearly varying imaginary chemical potential.
That is, at high temperatures, the whole system looks more inclined to be a simple combination of $z$-slices corresponding to different imaginary chemical potentials that reach equilibrium independently and then come together.
This implies that the chiral condensation located at a certain place does not feel the imaginary chemical potential far away perhaps due to some screening effect and the effect of imaginary chemical potential is therefore a short-range effect on the chiral condensation.
On the contrary, when the oscillation disappears, the whole system must be considered as a whole, so that when equilibrium is reached, the chiral condensation located at a certain place can feel the overall distribution of the imaginary chemical potential, and the effect of imaginary chemical potential is therefore a long-range effect on the chiral condensation.

\subsubsection{\label{sec3.2.2}The fitting of chiral condensation}

\begin{figure}
\begin{center}
\includegraphics[width=0.99\textwidth]{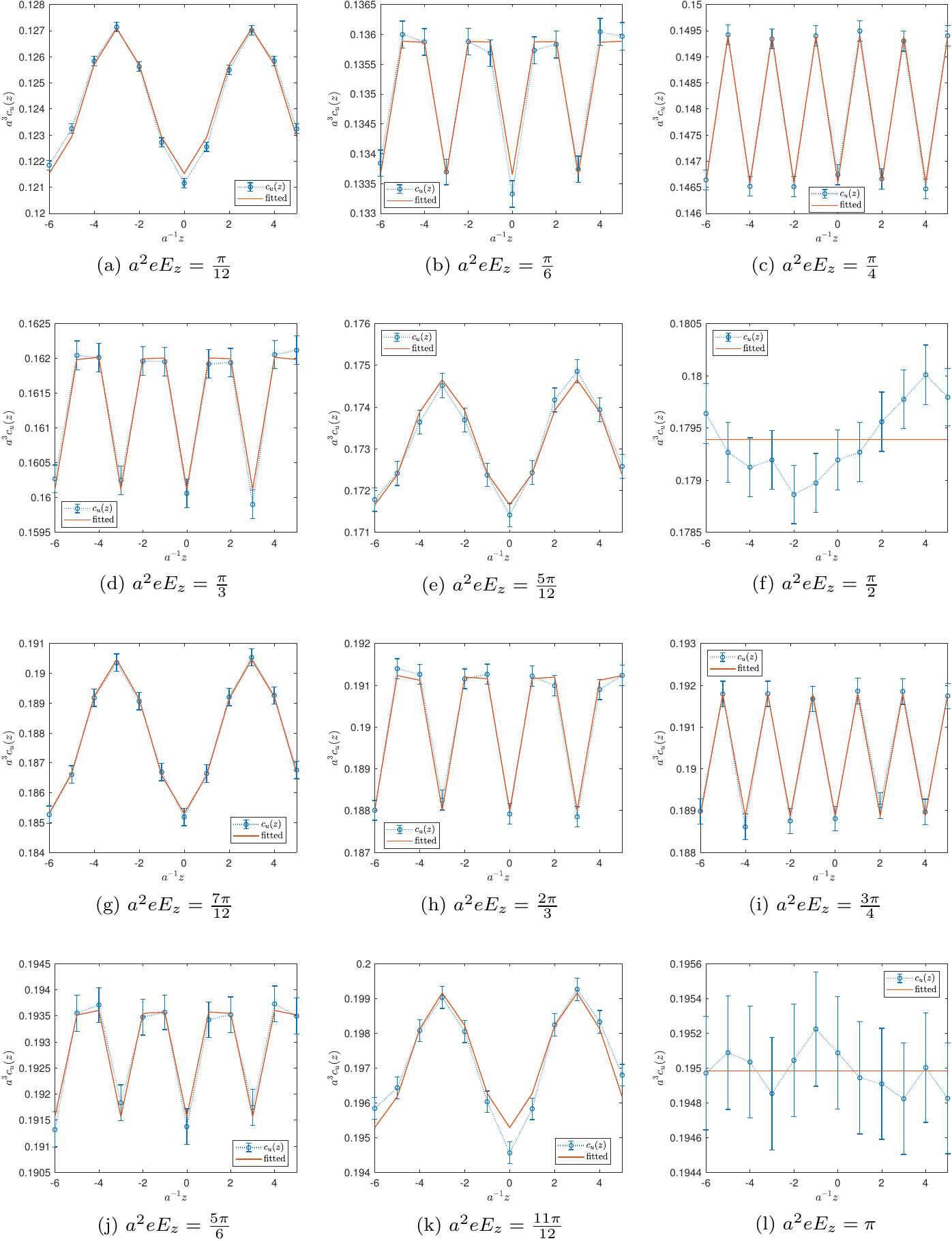}
\caption{\label{fig:cu564fit}$c_u(z)$ at $\beta=5.64$ compared with the results of fittings. Note that, the dashed and solid lines are only shown for visual guidance, but not the images of $c_u(z)$.}
\end{center}
\end{figure}

\begin{figure}
\begin{center}
\includegraphics[width=0.99\textwidth]{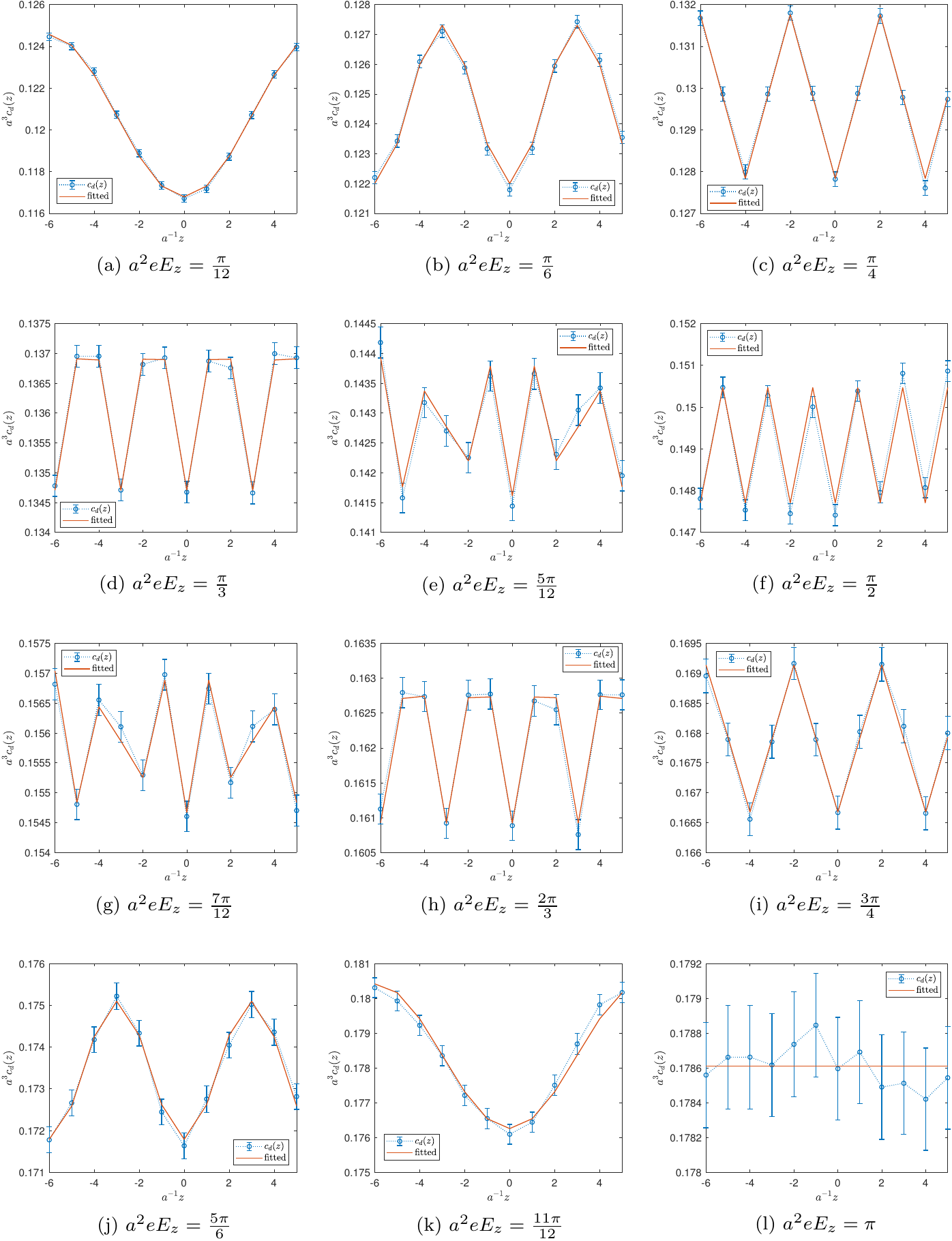}
\caption{\label{fig:cd564fit}Same as Fig.~\ref{fig:cu564fit} but for $c_d(z)$ at $\beta=5.64$.}
\end{center}
\end{figure}

\begin{figure}
\begin{center}
\includegraphics[width=0.48\textwidth]{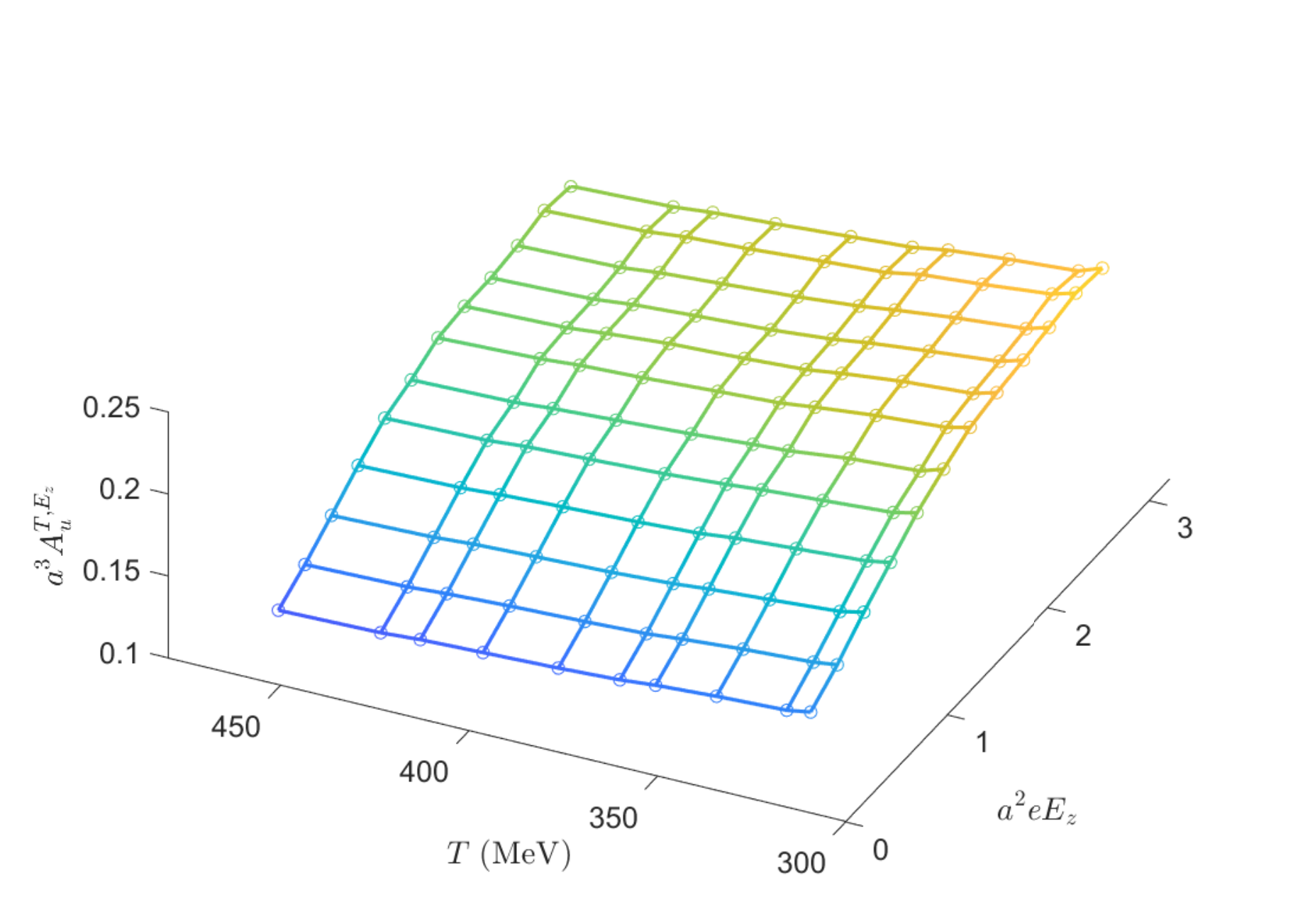}
\includegraphics[width=0.48\textwidth]{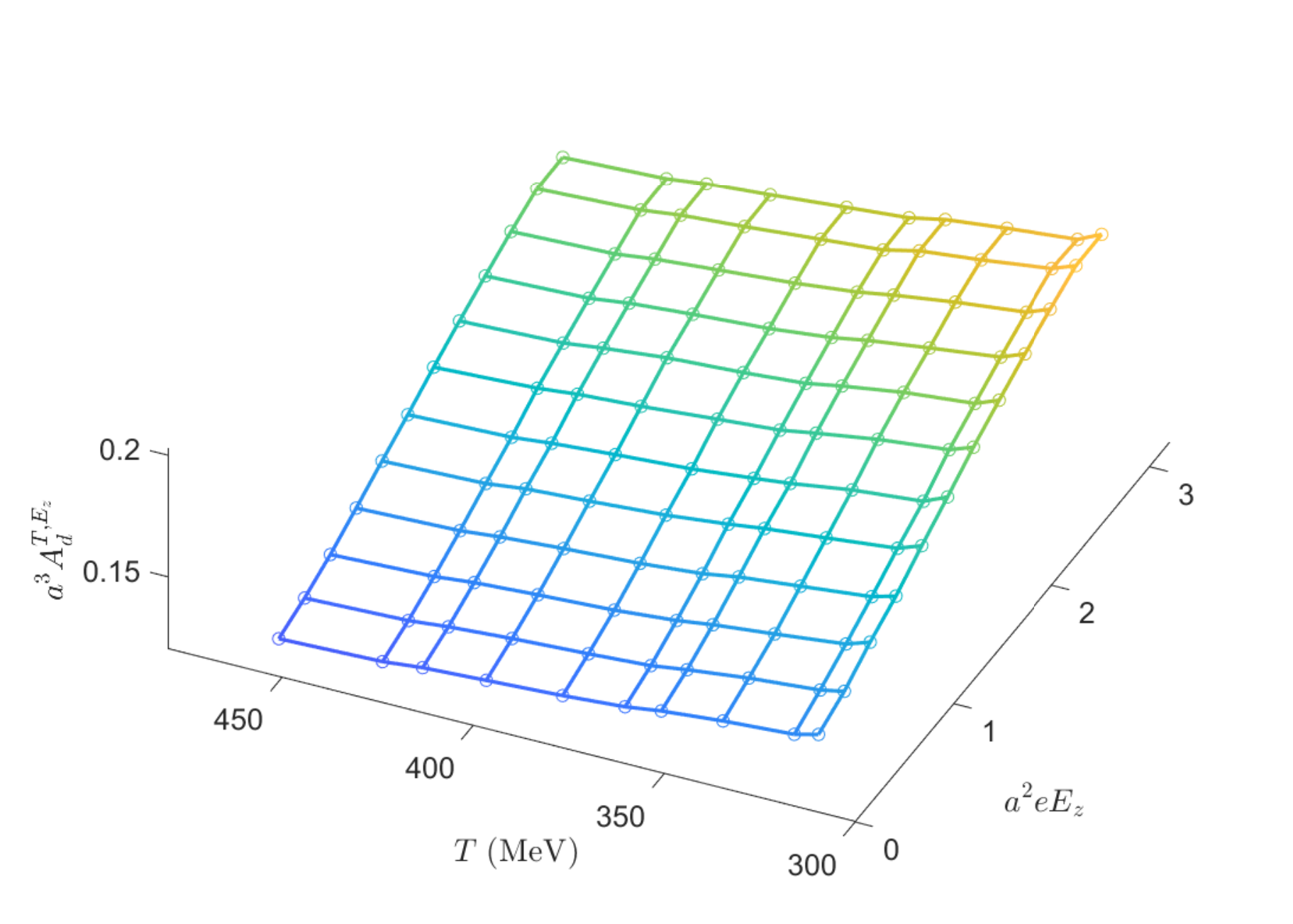}
\includegraphics[width=0.48\textwidth]{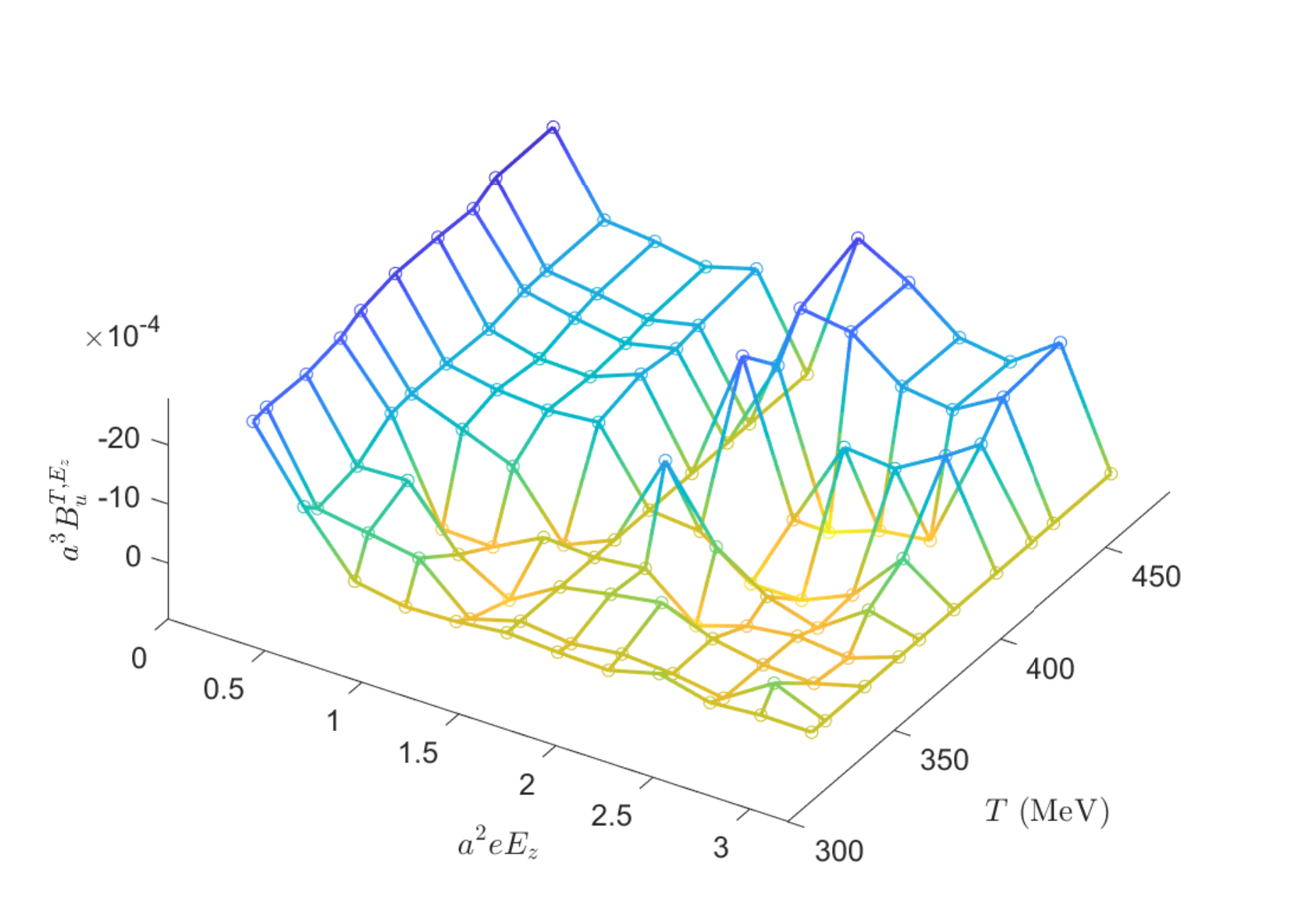}
\includegraphics[width=0.48\textwidth]{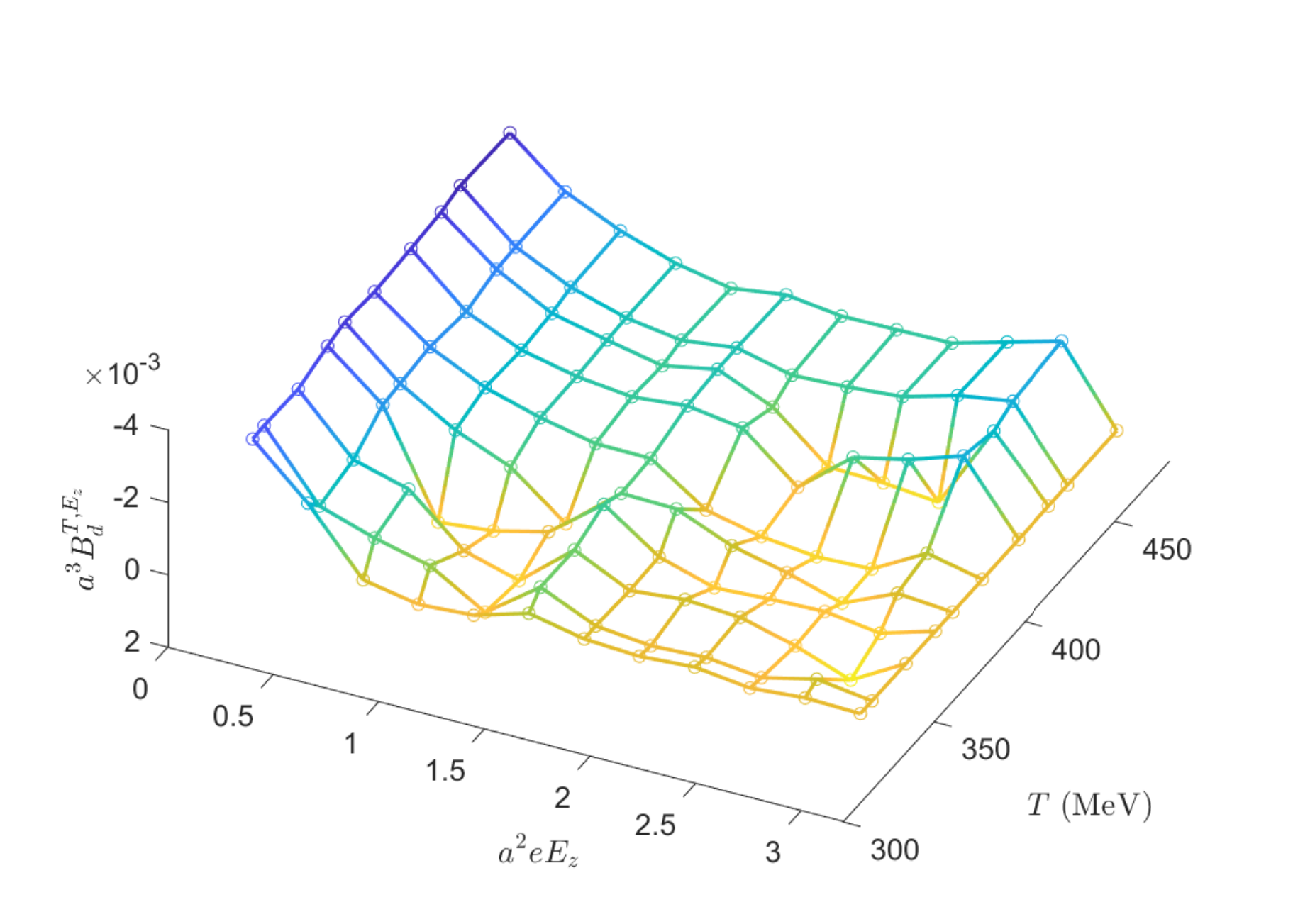}
\caption{\label{fig:aqbq}$A_q$ and $B_q$ as functions of temperature and $E_z$.}
\end{center}
\end{figure}

\begin{figure}
\begin{center}
\includegraphics[width=0.6\textwidth]{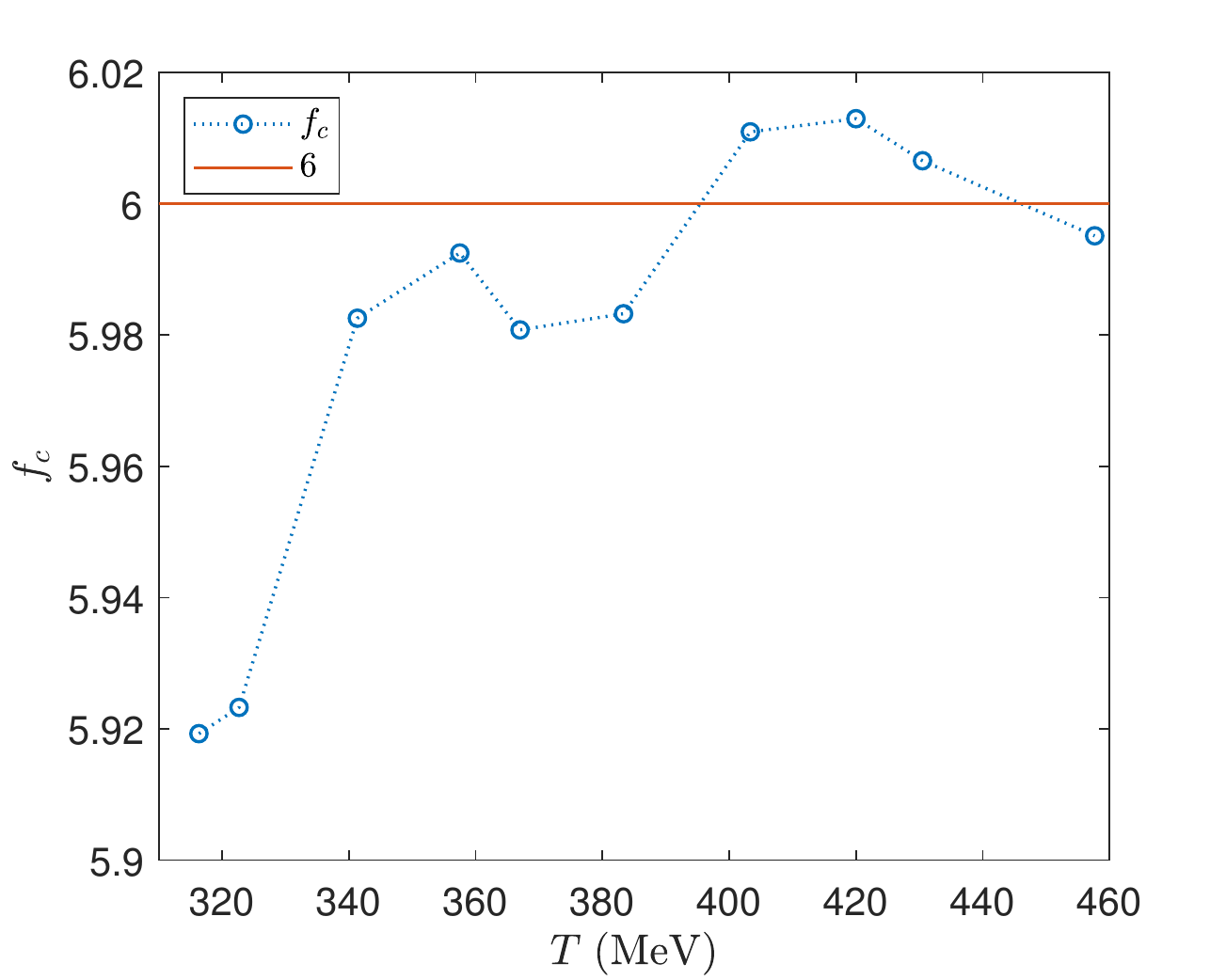}
\caption{\label{fig:fc}$f_c$ as a function of temperature $T$.}
\end{center}
\end{figure}
It has been shown , when oscillating, $c_q^{T,E_z}(z)$ is approximately a trigonometric function.
We find that, $c_q^{T,E_z}(z)$ can be fitted as
\begin{equation}
\begin{split}
&c_q^{T,E_z}(z) = A_q^{T,E_z}+B_q^{T,E_z}\cos(a f_c z Q_q e E_z),\\
\end{split}
\label{eq.fitansatz}
\end{equation}
where $A_q^{T,E_z}$ can be viewed as an overall change of the chiral condensation, $B_q^{T,E_z}$ is the amplitude of oscillation, $f_c$ is the frequency of the oscillation.

$c_q^{T,E_z}(z)$ is fitted according to the ansatz in Eq.~(\ref{eq.fitansatz}) using the following steps.
\begin{enumerate}
\item For a fixed temperature~(a fixed $\beta$), choose an initial value for $f_c$.
\item Find $A_q^{E_z}$ and $B_q^{E_z}$ which minimizes the error $\sum _{z} \delta^2(q,k,z)$ where $\delta (q,k,z)=A_q^{E_z^{(k)}}+B_q^{E_z^{(k)}}\cos(a f_c z Q_q e E_z^{(k)})-c_q^{E_z^{(k)}}(z)$, and $E_z^{(k)}=k \pi/12 a^{-2}$.
\item For $A_q^{E_z}$ and $B_q^{E_z}$ obtained in step $2$, find $f_c$ which minimize the error $\sum _{q,k,z} \delta^2 (q,k,z) $.
\item Repeat step $2$ and $3$ until $f_c$ converges.
\item Apply step $2$ for the last time and finish the fit.
\end{enumerate}

The results of the fit depend on the initial value chosen for $f_c$, therefore, we fit the case of $a^2E_z=\pi/12$ for $u$ quark according to the ansatz in Eq.~(\ref{eq.fitansatz}) first to set the initial value for $f_c$.
To ensure reliability, $c_u(z)$ at $a^2e E_z=\pi/2$ and $a^2e E_z=\pi$, $c_d(z)$ at $a^2e E_z=\pi$ are excluded in the fit.

Taking the case of $\beta=5.64$ for example, $c_q^{E_z}(z)$ and fitted $c_q^{E_z}(z)$ are shown in Figs.~\ref{fig:cu564fit} and \ref{fig:cd564fit}.
Note that the dashed and solid lines are only shown for visual guidance, but not the images of $c_q(z)$.
In this paper, we use $\chi ^2 / d.o.f.$ to estimate the goodness of fits, which is defined as
\begin{equation}
\begin{split}
&\chi ^2 / d.o.f. = \frac{1}{M-N} \sum _{i}^M\frac{\left(f(x_i, \vec{\alpha } ) - y_i\right)^2}{\sigma _i ^2},\\
\end{split}
\label{eq.chisquare}
\end{equation}
where $M$ is the number of points $(x_i, y_i)$ participated in the fit, $f(x_i,\vec{\alpha})$ is the result of fit with $\vec{\alpha }=(\alpha _1,\alpha _2,\ldots,\alpha _N)$ representing the $N$ fit parameters, $\sigma _i$ is the statistical error for $y_i$.

To ensure reliability, we only fit $c_q(z)$ in the regions with strong oscillations.
Starting from $\beta =5.46$, $\epsilon _q^{T,E_z}$ at $a^2eE_z=\pi/12$ are one order of magnitude larger than the statistical errors of $\epsilon _q^{T,E_z}$.
In the case $\beta\geq 5.46$, we find $\chi ^2 / d.o.f. = 0.48 \sim 1.07$ for different temperatures.

$A_q^{T,E_z}$ and $B_q^{T,E_z}$ are shown in Fig.~\ref{fig:aqbq}, $f_c$ are shown in Fig.~\ref{fig:fc}.
It can be observed that, $A_q$ grows with $E_z$ and decrease with temperature.
The overall effect of the electric field on the chiral condensation will be investigated later.
Meanwhile, $B_q$ is consistent with $\epsilon_q$ in Fig.~\ref{fig:stdcq}.
The amplitude of oscillation grows with the temperature and decreases with the electric field strength.

Another interesting and noteworthy conclusion is that $f_c\approx 6$ is a constant integer for different flavors at different temperatures and different electric field strengths.
This also explained the phenomena that the oscillation disappears for $c_u(z)$ at $a^2e E_z=\pi/2$ and $a^2e E_z=\pi$, for $c_d(z)$ at $a^2e E_z=\pi$.
Since $a^2 f_c Q_u e E_z = 2\pi$ for $a^2eE_z=\pi/2$ and $4\pi$ for $a^2eE_z=\pi$, $a^2 f_c Q_d E_z = -2\pi$ for $a^2e E_z = \pi$.

The frequency $f_c$ actually responds to the $U(1)$ gauge field in Eq.~(\ref{eq.2.6}).
When $k = 1$ and $f=6\pi/(L_zL_{\tau})=\pi/12$, the flux through the $z-\tau$ plane is $6 \pi$, and for quarks, the periods of oscillations along the $z$ coordinate are exactly $Q_q \times 6 \pi$.
Thus, the frequency along $z$ coordinate is actually $f_c =L_{\tau}=6$.

It should be pointed out that, so far it is not able to exclude other cases, for example, $f_c$ may not be $L_{\tau}$, but $2N_c$, or $-2/Q_d$, as well as possibly $L_z/2$.
It will be necessary to use other simulations to finally verify $f_c=L_{\tau}$.
But we prefer $f_c$ to be $L_{\tau}$, and in the following, we use $L_{\tau}$ directly instead of $f_c$ to denote the frequency.

\subsubsection{\label{sec3.2.3}The chiral condensation as a function of electric field strength}

\begin{figure}
\begin{center}
\includegraphics[width=0.48\textwidth]{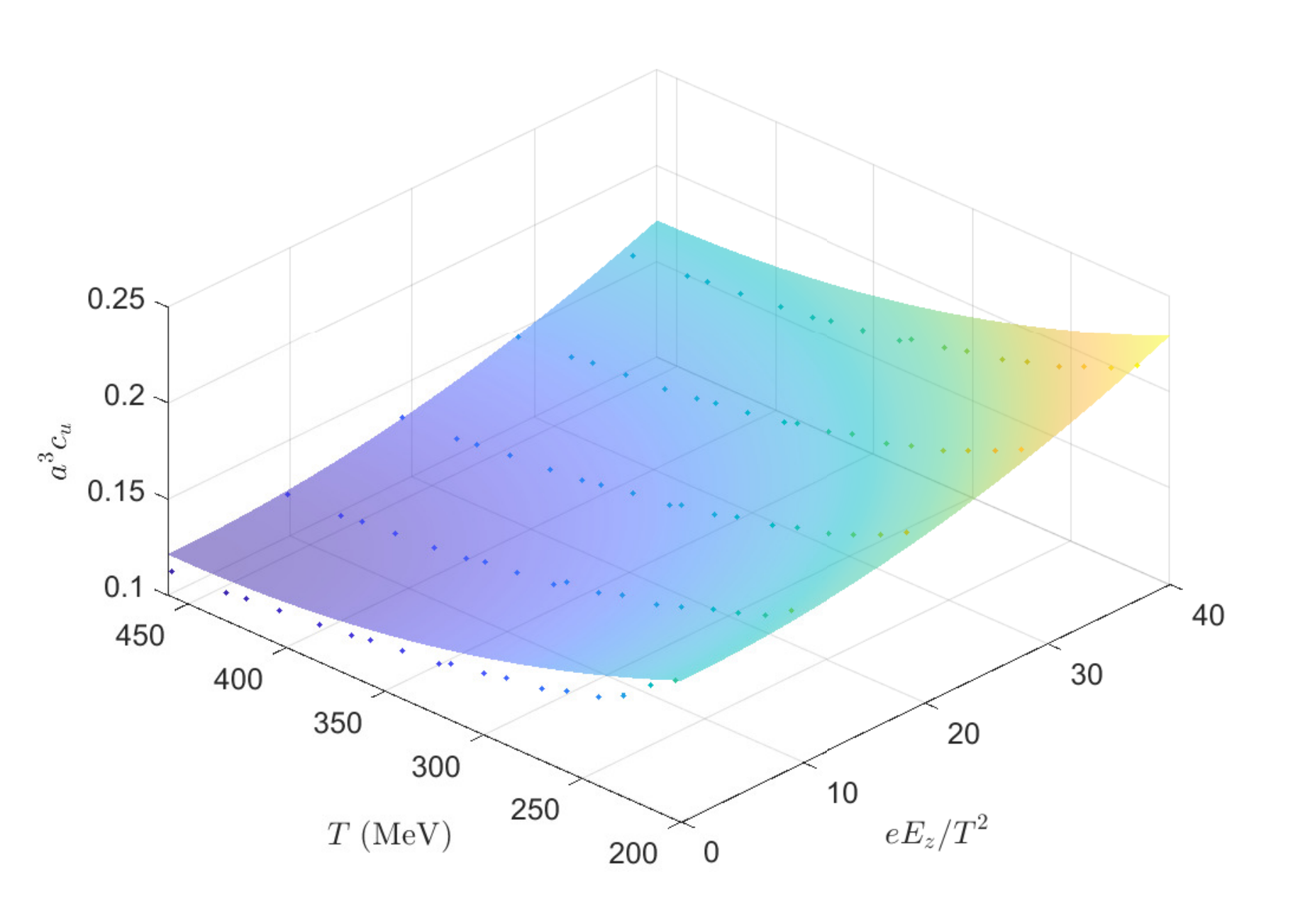}
\includegraphics[width=0.48\textwidth]{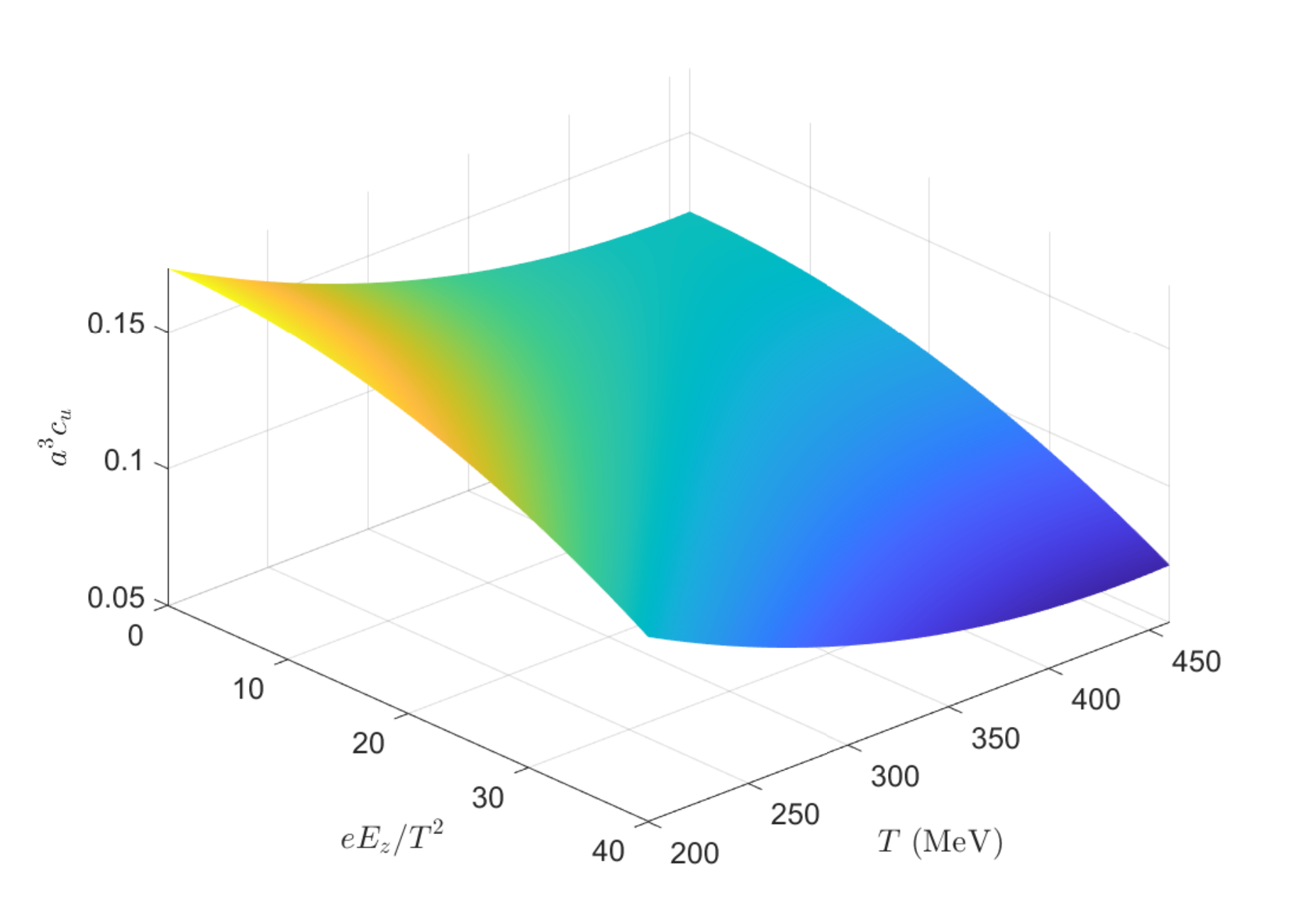}
\caption{\label{fig:cusq}$c_u^{T,E_z}$ and $c_u^{T,E_z}$ fitted using Eq.~(\ref{eq.fitansatcqe2})~(the left panel) and analytical extension of fitted $c_u^{T,E_z}$~(the right panel).}
\end{center}
\end{figure}

\begin{figure}
\begin{center}
\includegraphics[width=0.48\textwidth]{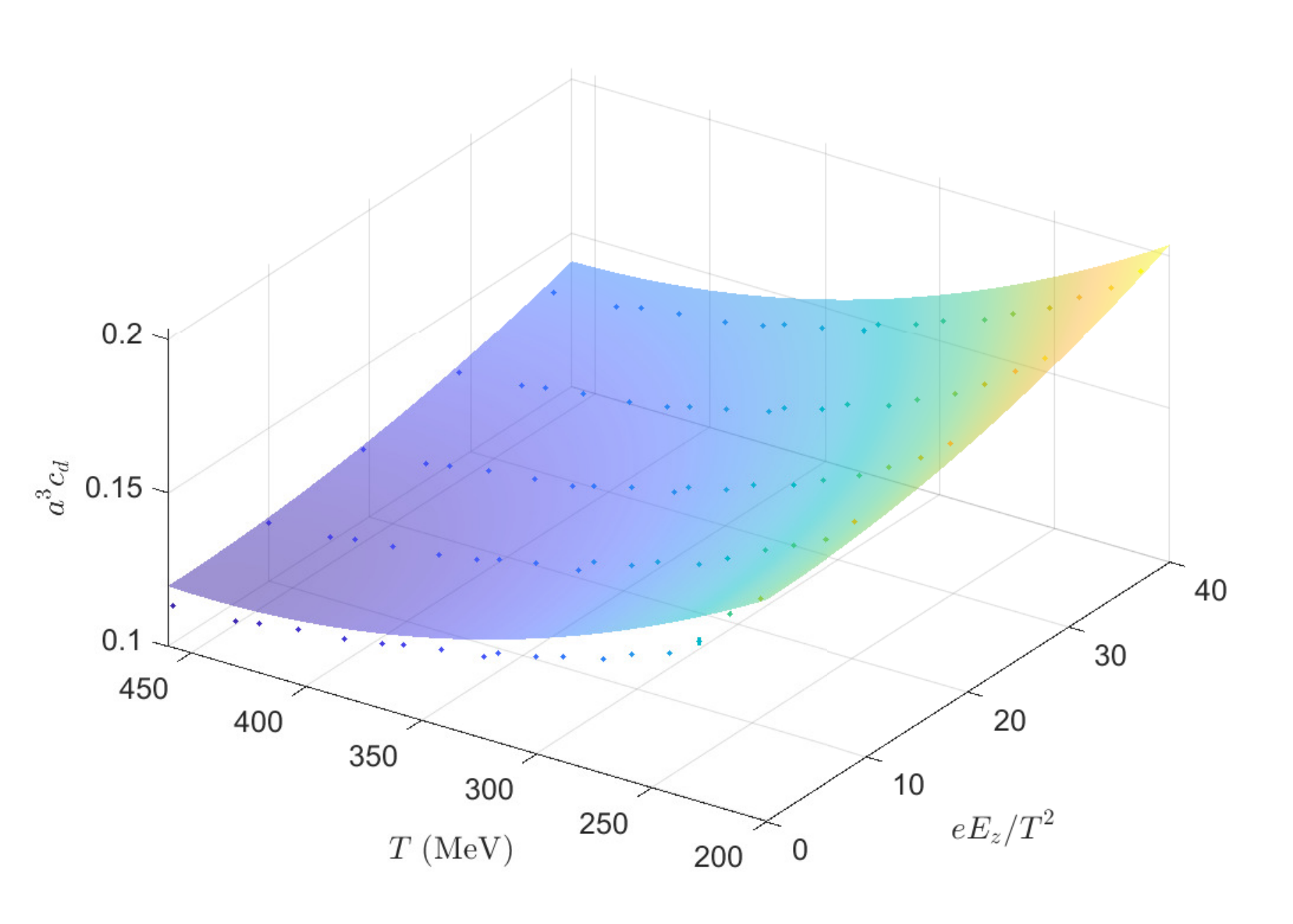}
\includegraphics[width=0.48\textwidth]{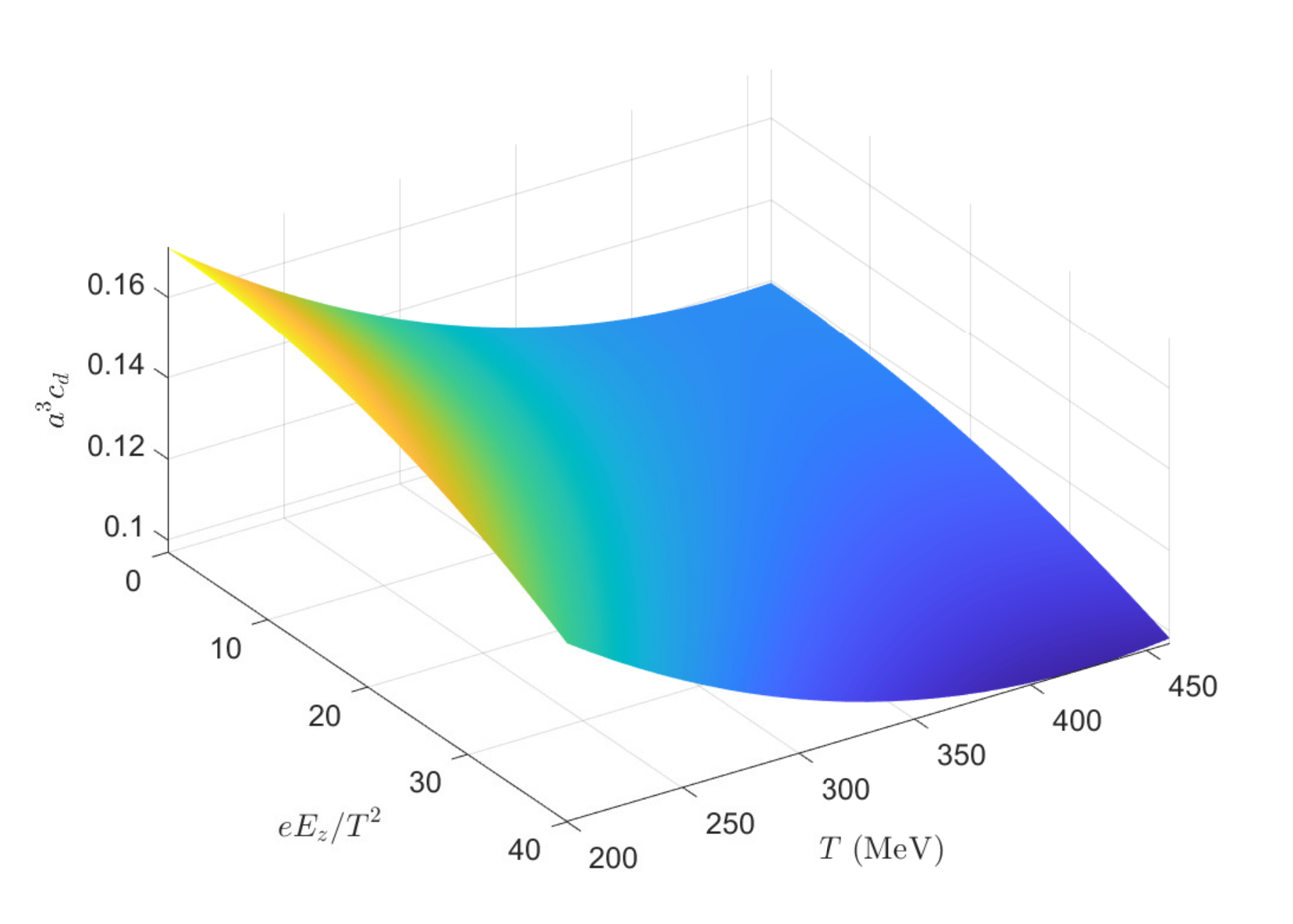}
\caption{\label{fig:cdsq}Same as Fig.~\ref{fig:cusq} but for $c_d^{T,E_z}$.}
\end{center}
\end{figure}

\begin{figure}
\begin{center}
\includegraphics[width=0.48\textwidth]{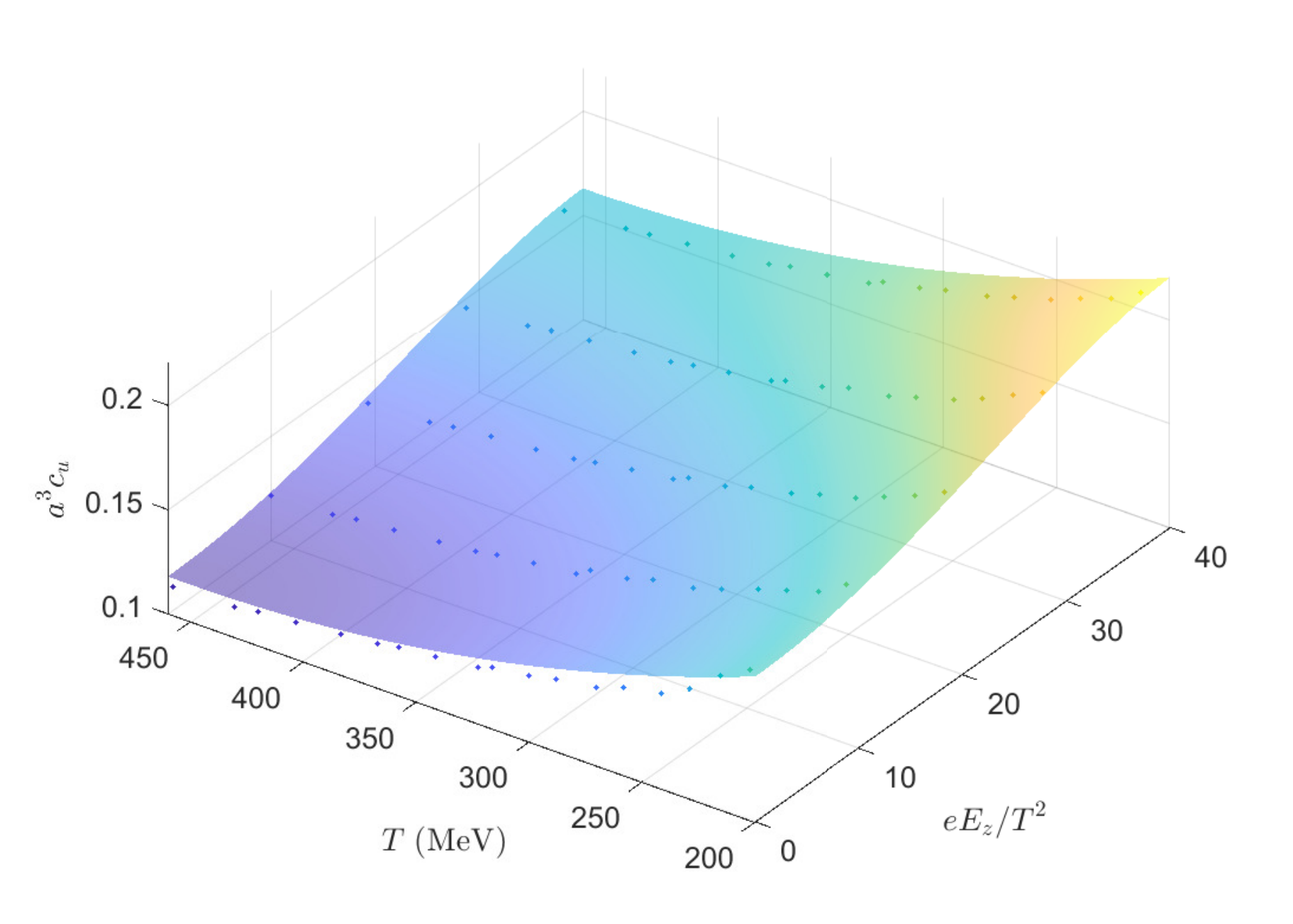}
\includegraphics[width=0.48\textwidth]{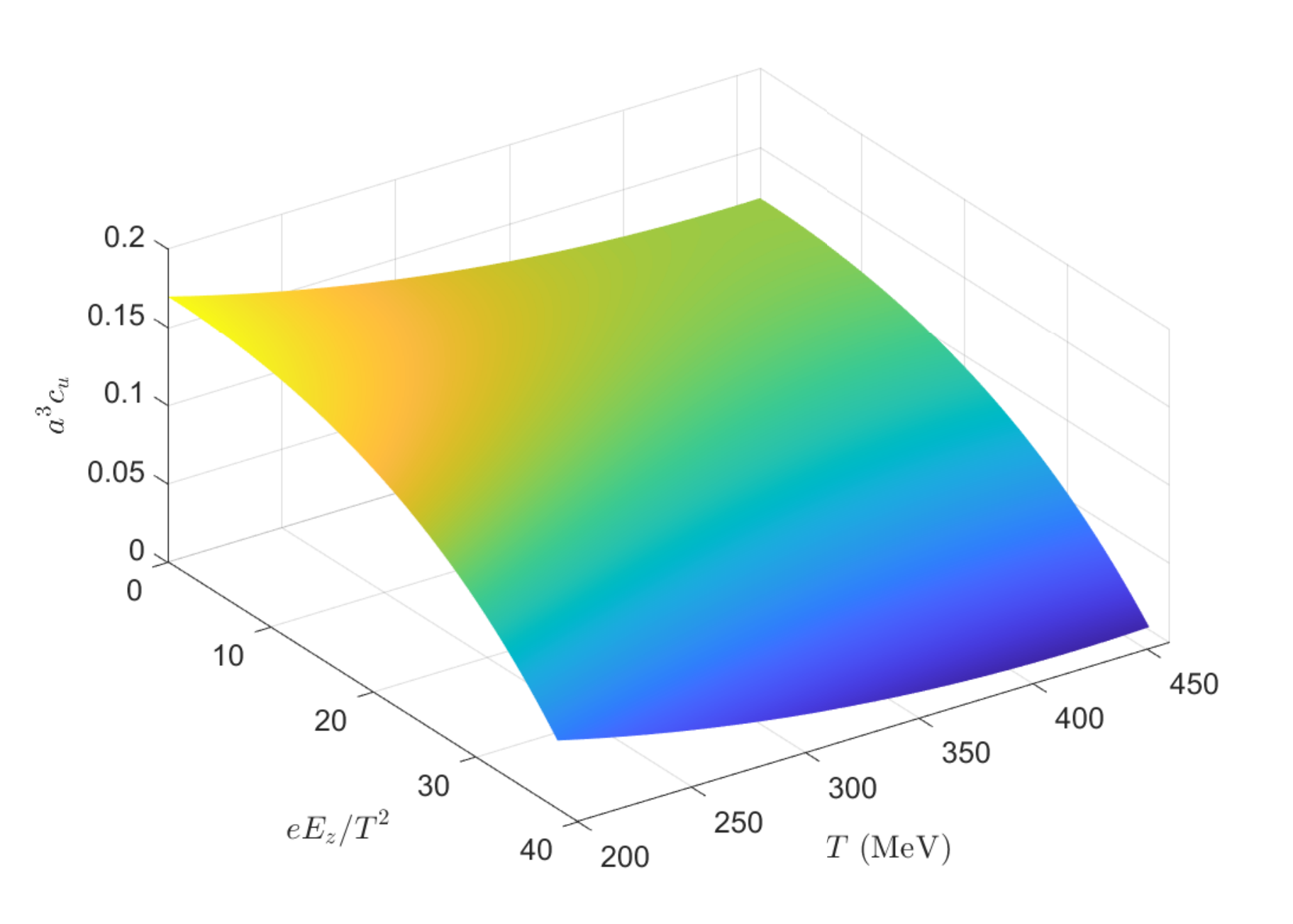}
\caption{\label{fig:cuquad}Same as Fig.~\ref{fig:cusq} but fitted using Eq.~(\ref{eq.fitansatcqe4}).}
\end{center}
\end{figure}

\begin{figure}
\begin{center}
\includegraphics[width=0.48\textwidth]{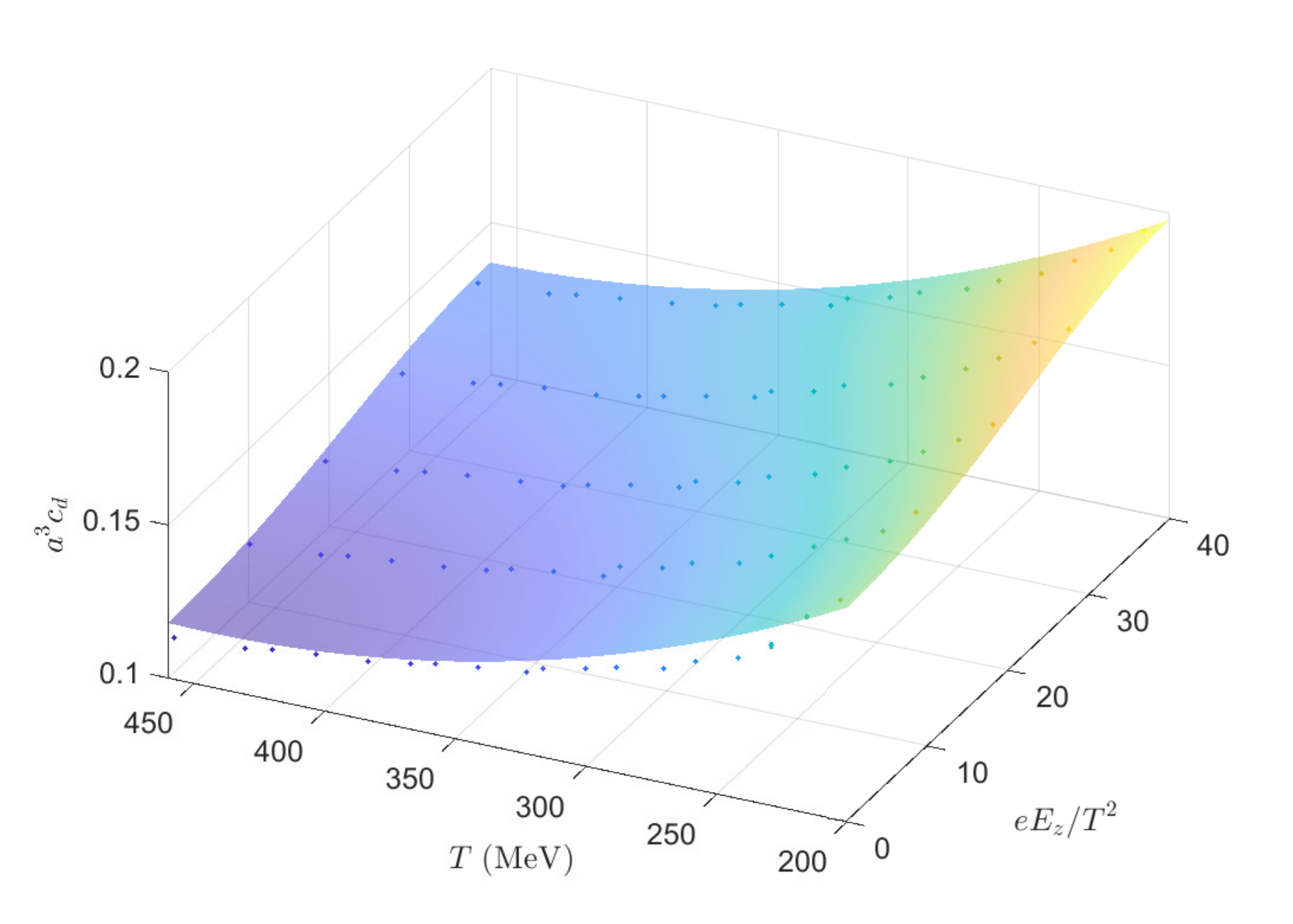}
\includegraphics[width=0.48\textwidth]{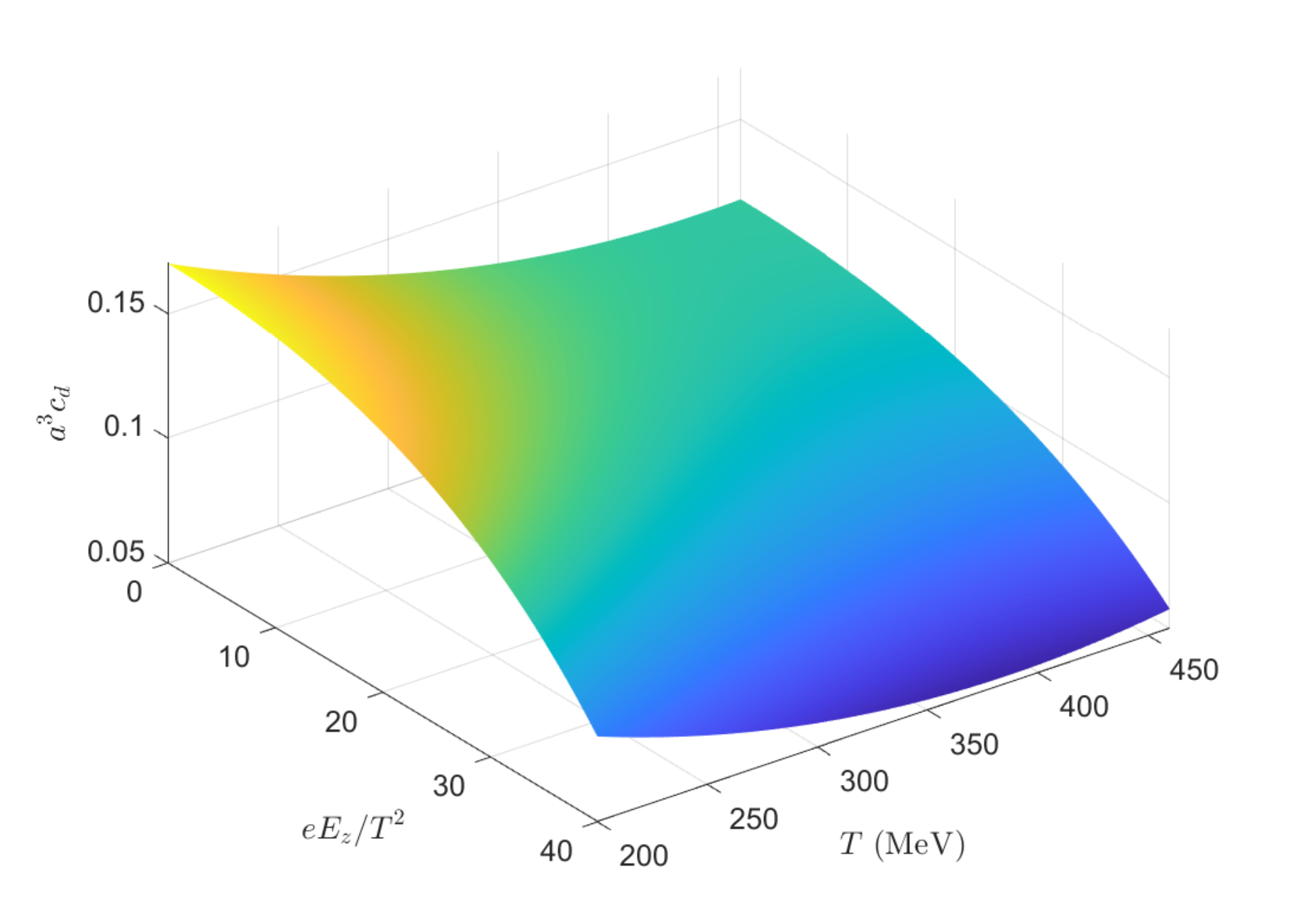}
\caption{\label{fig:cdquad}Same as Fig.~\ref{fig:cdsq} but fitted using Eq.~(\ref{eq.fitansatcqe4}).}
\end{center}
\end{figure}

The relationship between chiral condensation and the electric field strength is one of the questions of interest in this work.
When the system is considered as a whole, $c_q$ as functions of $T$ and $E_z$ is calculated.
The results suffer from strong discretization errors at large $E_z$, therefore only the results for $0\leq a^2 eE_z \leq \pi / 3$ are presented.

It needs to be kept in mind that the relationship between $c_q$ and temperature is not simple, as the renormalization is not applied, it is affected by the relationship between $c_q$ and lattice spacing $a$, in addition, as $am_q$ is a constant, it is also messed up by the relationship between $c_q$ and $m_q$ at the same time.
Apart from those, the electric field in physical units are different for different $a$, therefore in this subsection, the results are present as functions of $eE_z/T^2$ instead of $a^2eE_z$.

Although further verification is needed, it is useful to assume that chiral condensation is still perturbable as the electric field strength varies.
In other words, we assume $c_q^{T,E_z}\approx a_0(T)+a_1(T) E_z^2 + a_2(T) E_z^4+\ldots $.
With this assumption, analytical extension can be applied to obtain the relationship between the chiral condensation and the real electric field strength.
The ansatz of $c_q^{T,E_z}$ up to the order of $E_z^2$ and $E_z^4$ are
\begin{equation}
\begin{split}
&c_q^{T,E_z} = \left(a_0+a_1T+a_2T^2\right)+\left(b_0+b_1T\right)E_z^2,\\
\end{split}
\label{eq.fitansatcqe2}
\end{equation}
\begin{equation}
\begin{split}
&c_q^{T,E_z} = \left(a_0+a_1T+a_2T^2\right)+\left(b_0+b_1T\right)E_z^2+\left(c_0+c_1T\right)E_z^4.\\
\end{split}
\label{eq.fitansatcqe4}
\end{equation}

The results of $c_u$ and $c_d$ keeping up to $E_z^2$ are shown in the left panels of Figs.~\ref{fig:cusq} and \ref{fig:cdsq}, respectively.
The results of $c_u$ and $c_d$ keeping up to $E_z^4$ are shown in the left panels of Figs.~\ref{fig:cuquad} and \ref{fig:cdquad}, respectively.
The $\chi ^2 / d.o.f.$ are $7.3$ and $3.2$ for $c_u$ and $c_d$ keeping up to $E_z^2$, $2.5$ and $1.9$ for $c_u$ and $c_d$ keeping up to $E_z^4$.
In our simulation, we use an imaginary electric field strength.
To obtain the relationship between $c_q$ and $E_z$, an analytical extension is applied to rotate the electric field strength back to the real axis.
The results of $c_u$ and $c_d$ keeping up to $E_z^2$ after analytical extension are shown in the right panels of Figs.~\ref{fig:cusq} and \ref{fig:cdsq}, respectively.
The results of $c_u$ and $c_d$ keeping up to $E_z^4$ after analytical extension are shown in the right panels of Figs.~\ref{fig:cuquad} and \ref{fig:cdquad}, respectively.
Both the ansatz in Eqs.~(\ref{eq.fitansatcqe2}) and (\ref{eq.fitansatcqe4}) support the conclusion that the external electric field restores the chiral symmetry as predicted by previous works.

\begin{figure}
\begin{center}
\includegraphics[width=0.7\textwidth]{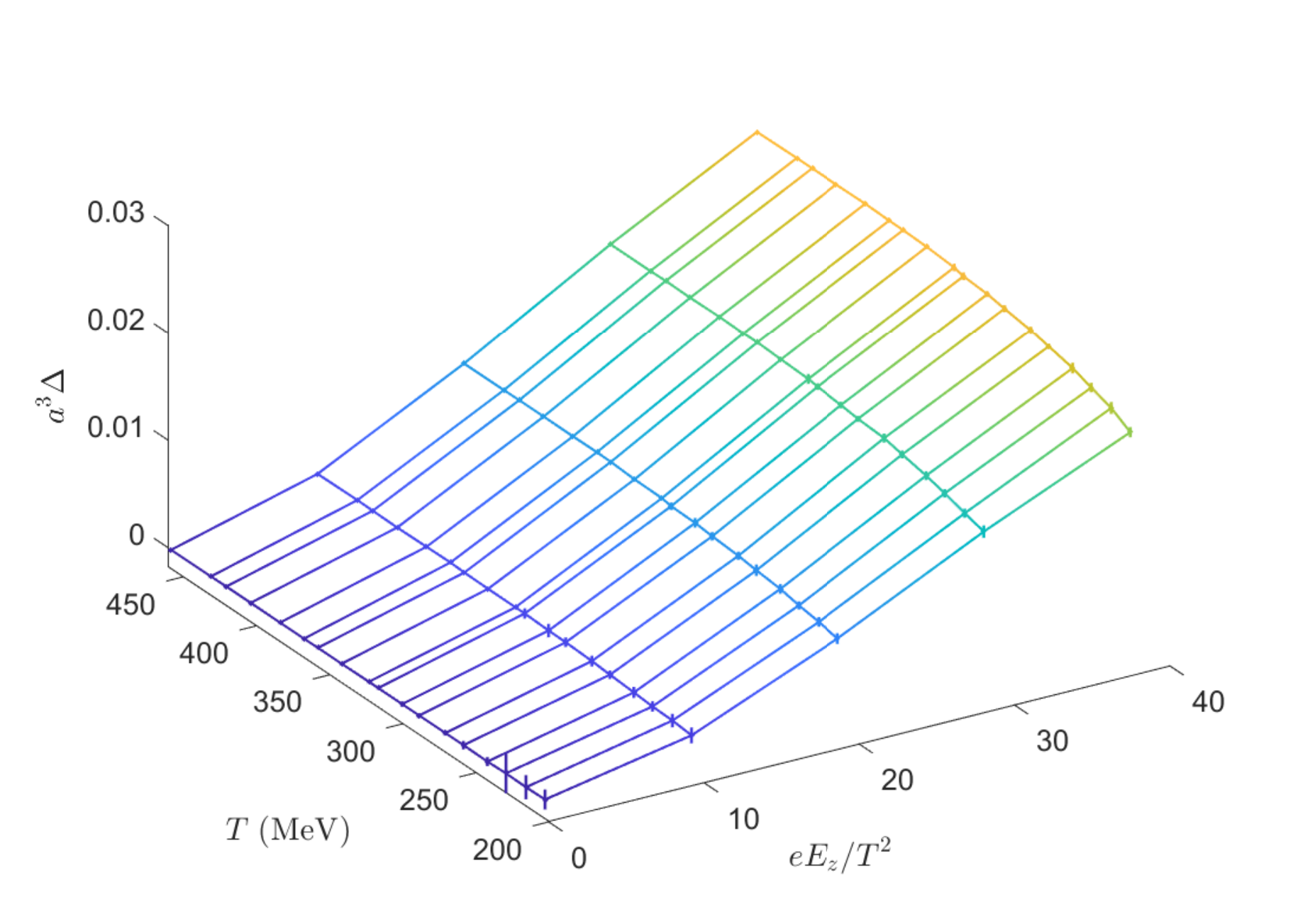}
\caption{\label{fig:delta}$\Delta =c_u-c_d$ as a function of $T$ and $eE_z/T^2$.}
\end{center}
\end{figure}
Moreover, since $\left|Q_u\right|>\left|Q_d\right|$, the effect of the electric field is larger for $c_u$ than $c_d$.
The difference defined as $\Delta =c_u-c_d$~($\Delta$ is also $\langle \bar{\psi} \tau _3 \psi \rangle$ where $\tau _3$ is the Pauli matrix) is also calculated and depicted in Fig.~\ref{fig:delta}.
It can be seen that, $a^3\Delta$ is insensitive to temperature.

\subsubsection{\label{sec3.2.4}The charge distribution}

\begin{figure}
\begin{center}
\includegraphics[width=0.7\textwidth]{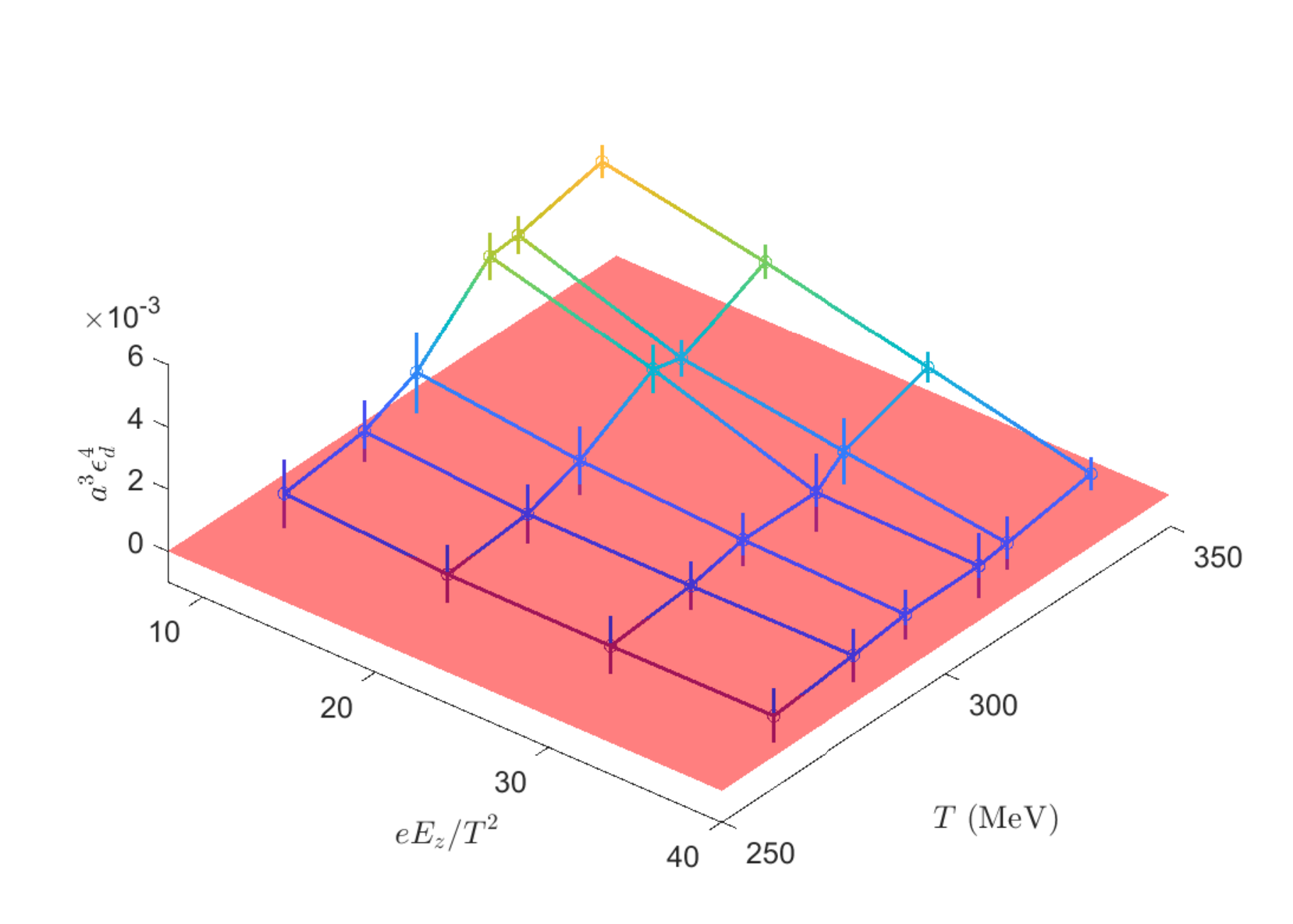}
\caption{\label{fig:c4oscillation}$\epsilon _d^4$ as a function of temperature $T$ and $eE_z/T^2$.}
\end{center}
\end{figure}

It has been pointed out that, in equilibrium, the electric and diffusion forces acting on quarks balance each other, which happens when there is a non-constant charge distribution~\cite{chargedistrib1,chargedistrib2}.
The charge density can be defined as $Q_q e\bar{q} \gamma _0 q$, which can be related with $c_q^4$ defined in Eq.~(\ref{eq.cqdefine}).
We find that, at high temperatures, the imaginary of $c_q^4$ shows nontrivial dependence on $z$.
As will be explained later, we concentrate on ${\rm Im}[c_d^4]$.
To show the oscillation of ${\rm Im}[c_d^4]$, $\epsilon _d^4(T, E_z)$ is defined as same as $\epsilon _q(T, E_z)$ but with $c_q^{T,E_z}$ replaced by ${\rm Im}[c_d^4]$.
$\epsilon _d^4$ is shown in Fig.~\ref{fig:c4oscillation}.
Starting from $T\approx 300\;{\rm MeV}$, the imaginary of $c_q^4$ starts to oscillate over $z$ coordinate for $\pi / 12 \leq a^2 eE_z \leq \pi / 3$.

\begin{figure}
\begin{center}
\includegraphics[width=1\textwidth]{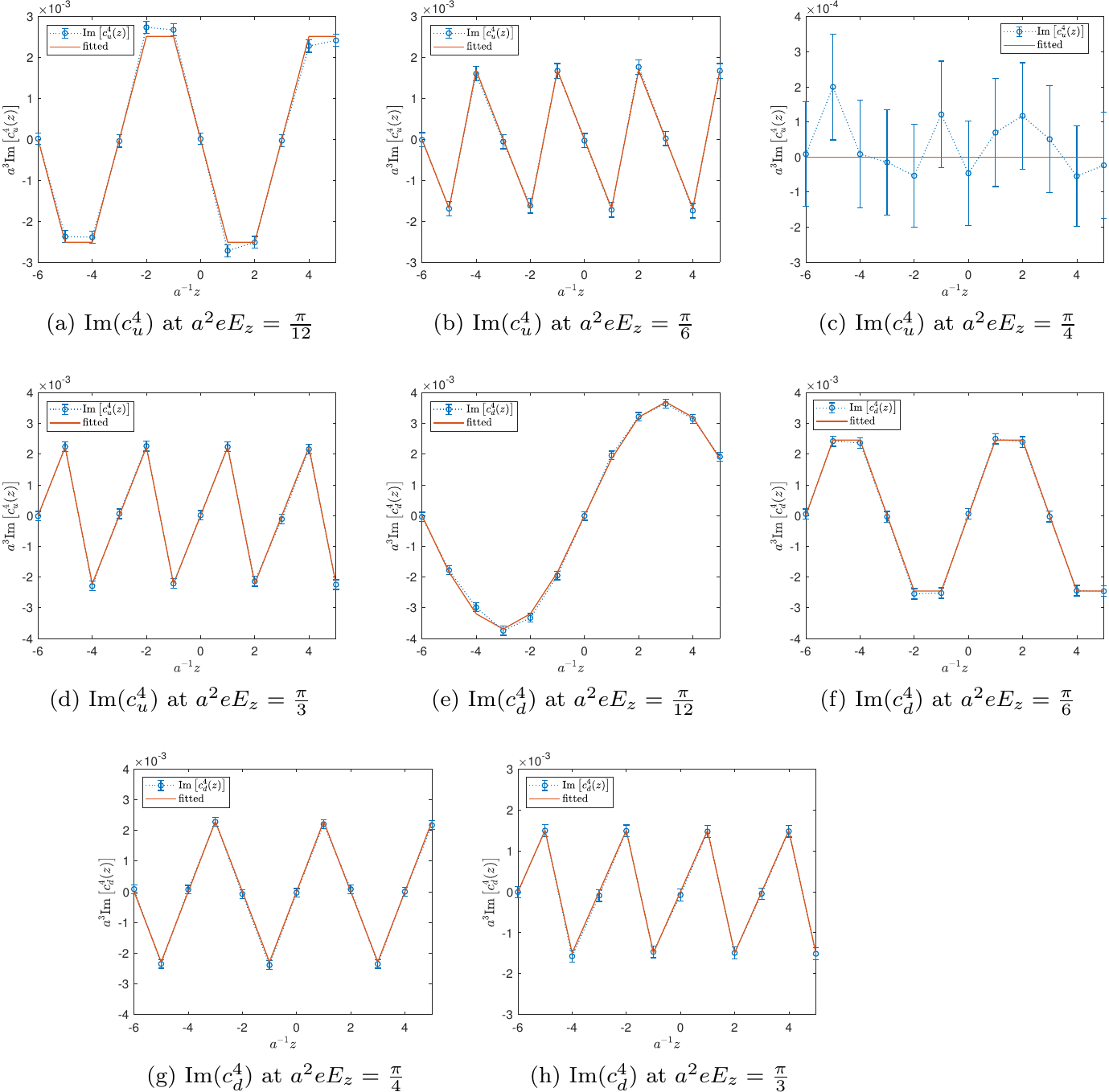}
\caption{\label{fig:c4564fit}Same as Fig.~\ref{fig:cu564fit} but for ${\rm Im}\left[c_q^4(z)\right]$ at $\beta=5.64$.}
\end{center}
\end{figure}

We find that, the frequencies of the oscillation are as same as those of $c_q(z)$, i.e., the ${\rm Im}[c_q^{4,T,E_z}]$ can be fitted with the ansatz,
\begin{equation}
\begin{split}
&{\rm Im}\left[c_q^{4,T,E_z}(z)\right] = A_q^{4,T,E_z}\sin ( a L_{\tau}zQ_q eE_z).\\
\end{split}
\label{eq.fitansatzcq4}
\end{equation}
For $\beta=5.64$, ${\rm Im}\left[c_q^4(z)\right]$ are shown in Fig.~\ref{fig:c4564fit}.
$\chi^2 /d.o.f.$ are found to be $0.086\sim 0.118$.
For $c_u^4$, the case of $a^2 eE_z=\pi / 4$ corresponds to no oscillation, that is the reason why $\epsilon _d^4$ is used to show the oscillation.
In all cases, $A_q^4<0$.
We also find that $|A_d^4|$ decrease with $E_z$.

\begin{figure}
\begin{center}
\includegraphics[width=0.7\textwidth]{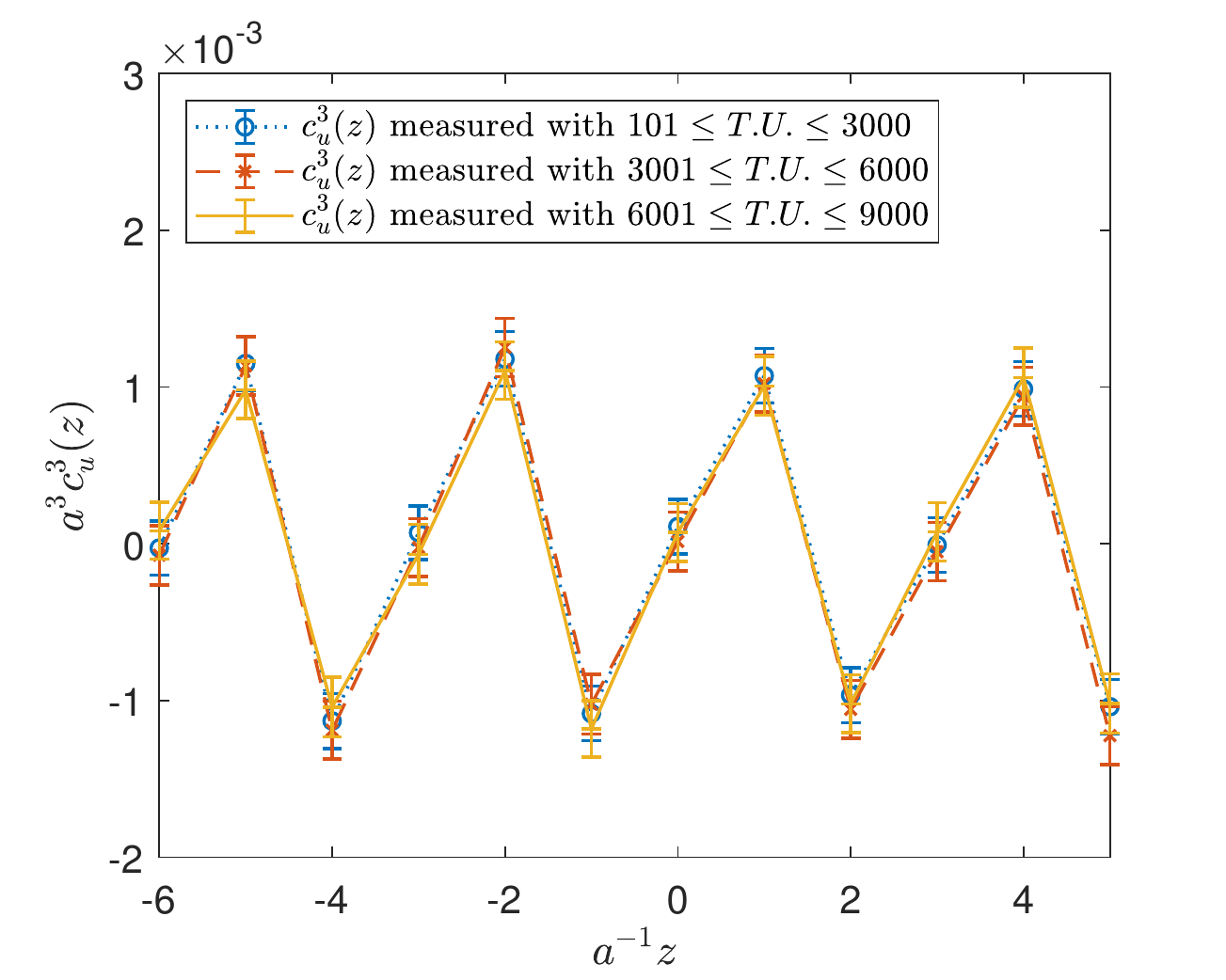}
\caption{\label{fig:cq3564}The oscillations of $c_u^3(z)$ along $z$ measured using different configurations.}
\end{center}
\end{figure}
Another interesting quantity is $c_q^3$.
We find that, at high temperatures and at large $E_z$~($a^2eE_z=\pi / 3$), $c_u^3(z)$ also oscillates over $z$~($c_d^3(z)$ starts to oscillate at a larger $E_z$ beyond our scope due the large discretization errors).
To verify that the oscillations of $c_u^3(z)$ along $z$ are not caused by not reaching the equilibrium, for $\beta=5.64$ and $a^2eE_z=\pi / 3$, additional $6000$ trajectories are simulated starting from $T.U.=3000$.
The results are shown in Fig.~\ref{fig:cq3564}.
Although $c_u^3(z)$ has a none trivial distribution, $c_u^3=0$.
The none trivial $c_u^3(z)$ may indicate a none trivial meson condensation caused by Schwinger mechanism~\cite{isospincharge}.
Also, since the case of a large $E_z$ is affected by large discretization errors, it is possible that this is a fake phenomenon from discretization errors.

\subsection{\label{sec3.3}Polyakov loop}

The signal of R-W transition can be observed by the phase of Polyakov loop, the latter is defined as
\begin{equation}
\begin{split}
&P ({\bf n})=\prod _{n_{\tau}}U_{\tau}({\bf n}, n_{\tau}) ,\;\;P (z) = \frac{1}{L_xL_y}\sum _{n_x,n_y} P({\bf n}=(n_x,n_y,a^{-1}z)),\;\;P=\frac{1}{L_z}\sum _{z} P(z),\\
\end{split}
\end{equation}
where the product in the definition of $P ({\bf n})$ is performed sequentially from $n_{\tau} =0$ to $n_{\tau} =L_{\tau}-1$, and $P (z)$ is an average of $P({\bf n})$ over a $z$-slice, $P$ is an average of $P({\bf n})$ over the whole spatial volume.

\begin{figure}
\begin{center}
\includegraphics[width=0.48\textwidth]{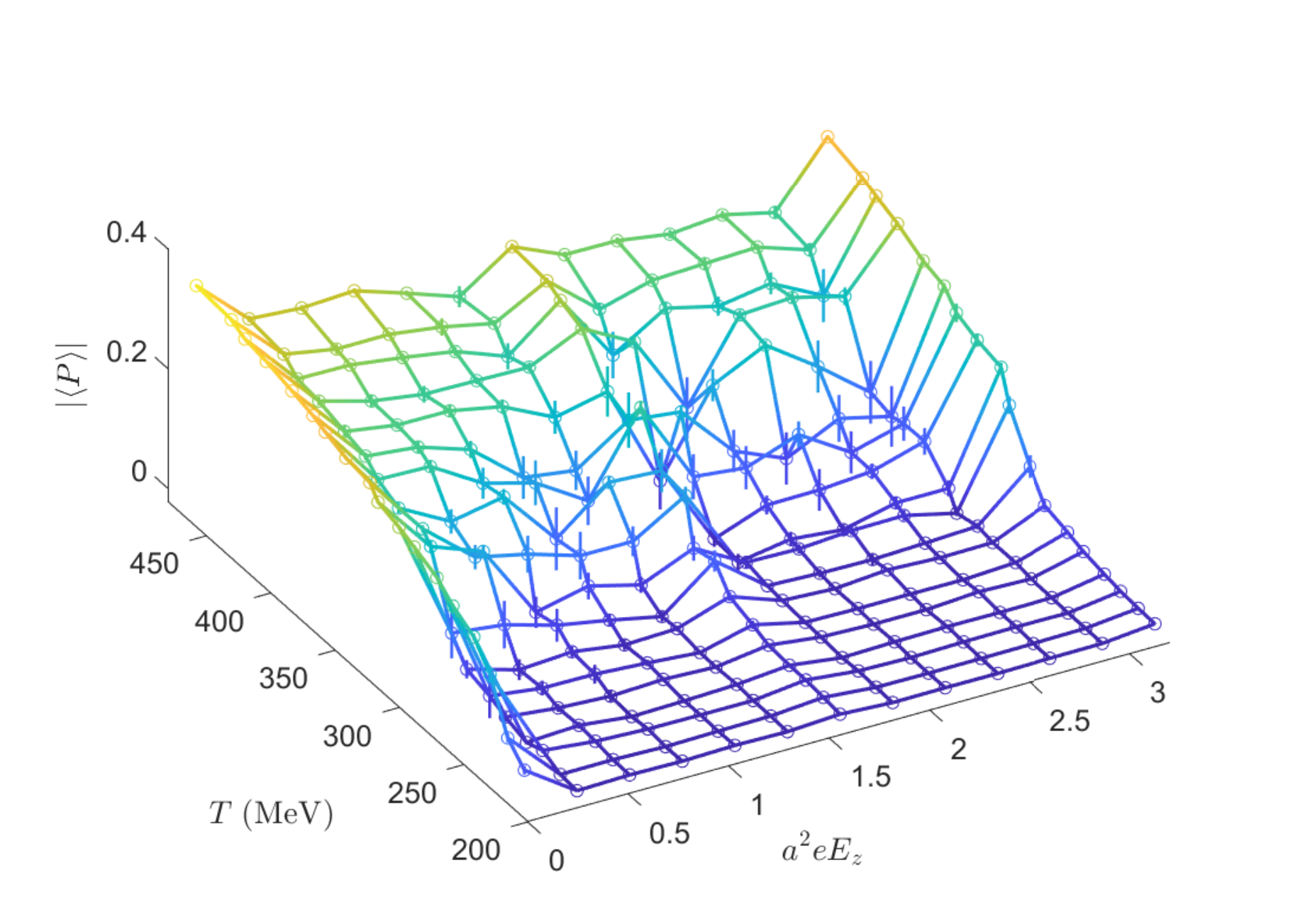}
\includegraphics[width=0.48\textwidth]{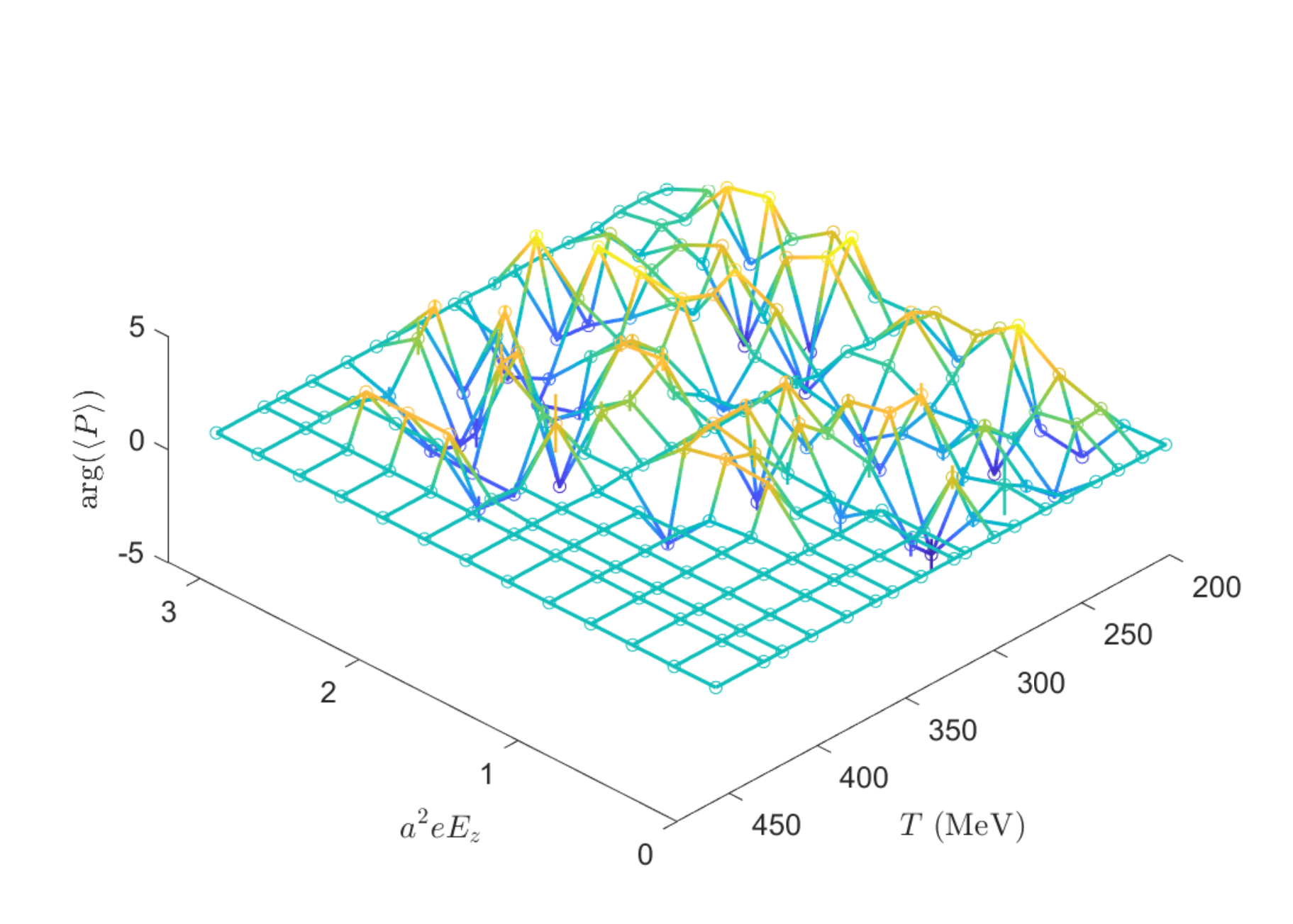}
\caption{\label{fig:polya}$\left|\langle P\rangle \right|$~(the left panel) and $\arg \left(\langle P\rangle \right)$~(the right panel) as functions of $T$ and $E_z$.}
\end{center}
\end{figure}

For different $\beta$ and $a^2eE_z$, $\left|\langle P\rangle \right|$ and $\arg \left(\langle P\rangle \right)$ are shown in Fig.~\ref{fig:polya}.
The R-W transition is clearly presented according to the none-zero $\arg \left(\langle P\rangle \right)$ at lower temperatures.

\subsubsection{\label{sec3.3.1}The properties of Polyakov loop}

\begin{figure}
\begin{center}
\includegraphics[width=0.48\textwidth]{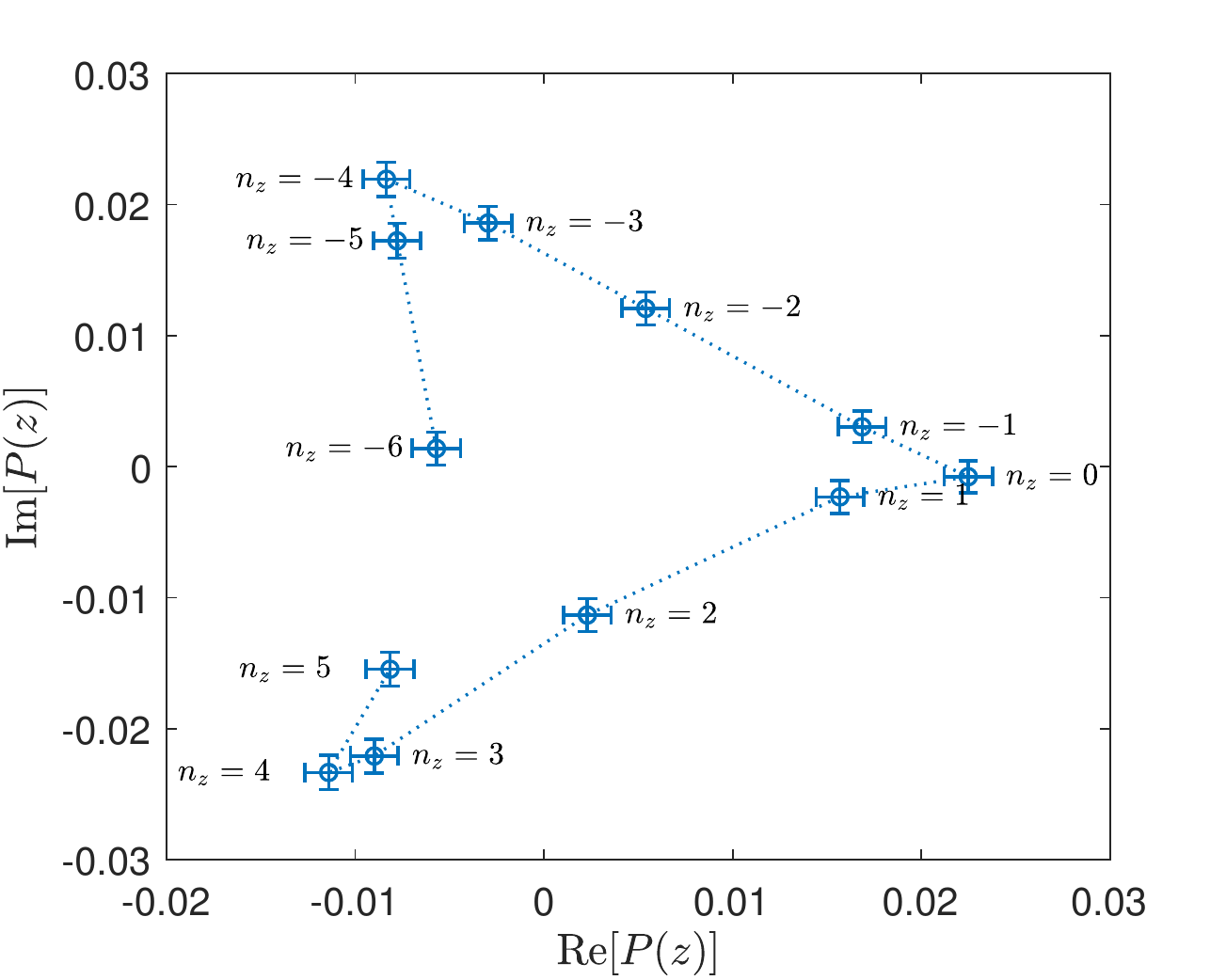}
\includegraphics[width=0.48\textwidth]{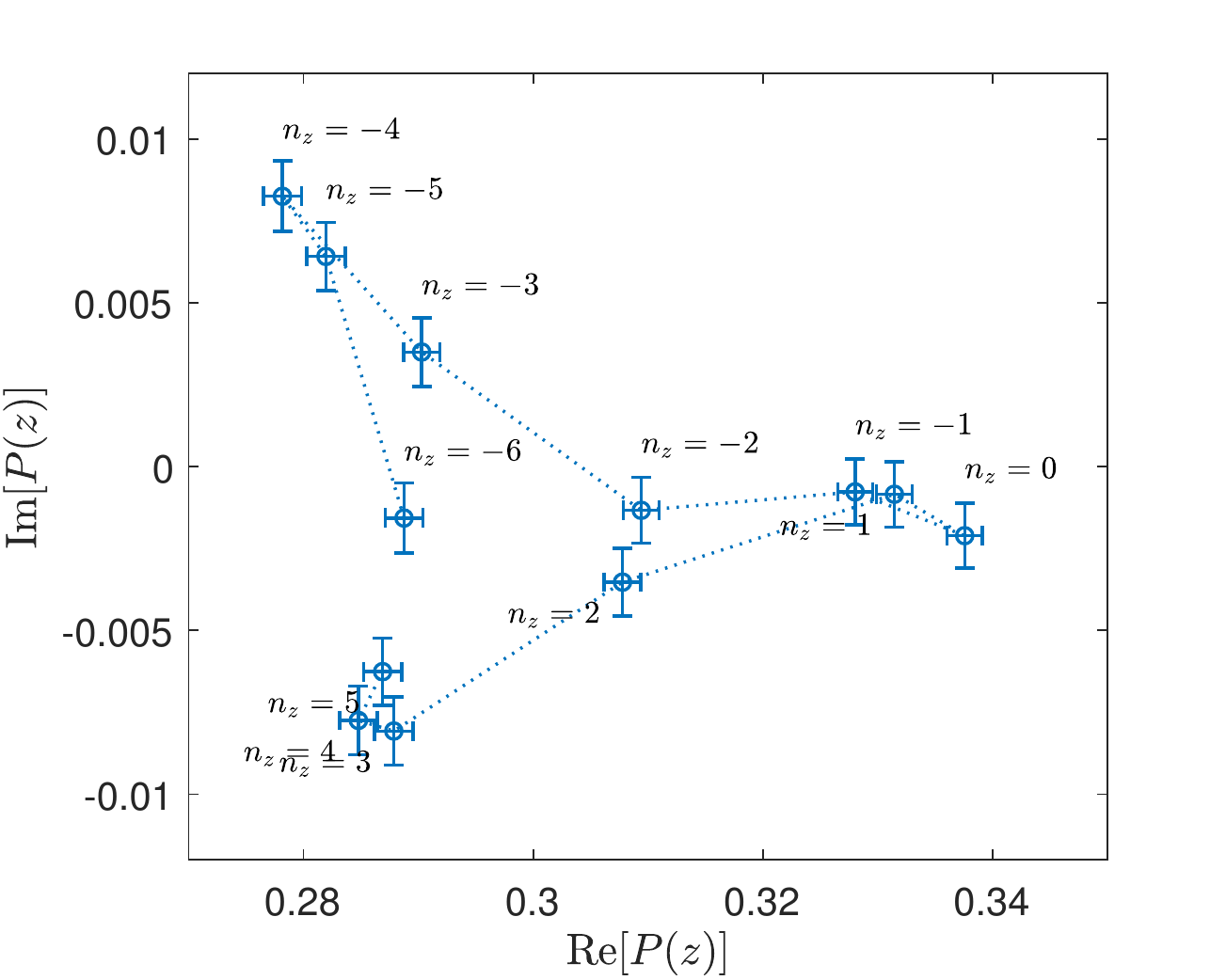}
\caption{\label{fig:complex530564}$\langle P(z)\rangle$ at $a^2eE_z = \pi/12$ shown in the complex plane for $\beta=5.3$~(the left panel) and $\beta=5.64$~(the right panel).}
\end{center}
\end{figure}
The phase of the Polyakov loop is important to study the R-W transition.
It is observed that the phase of the Polyakov loop is also oscillating with $z$, but quite different from the case of chiral condensation, the phase of Polyakov loop is oscillating at lower temperatures.
Since a large $E_z$ corresponds to a high frequency of oscillation and is not suitable for presentation, we show $P(z)$ only for $a^2eE_z = \pi/12$, in the complex plane, similar to Ref.~\cite{pzcomplexplane}.
The case of $\beta=5.30$ and $\beta=5.64$ are shown in Fig.~\ref{fig:complex530564}, $\langle P(z)\rangle$ in the complex plane exhibits the shape of a pointed triangle for both low temperature and high temperature.

\begin{figure}
\begin{center}
\includegraphics[width=0.7\textwidth]{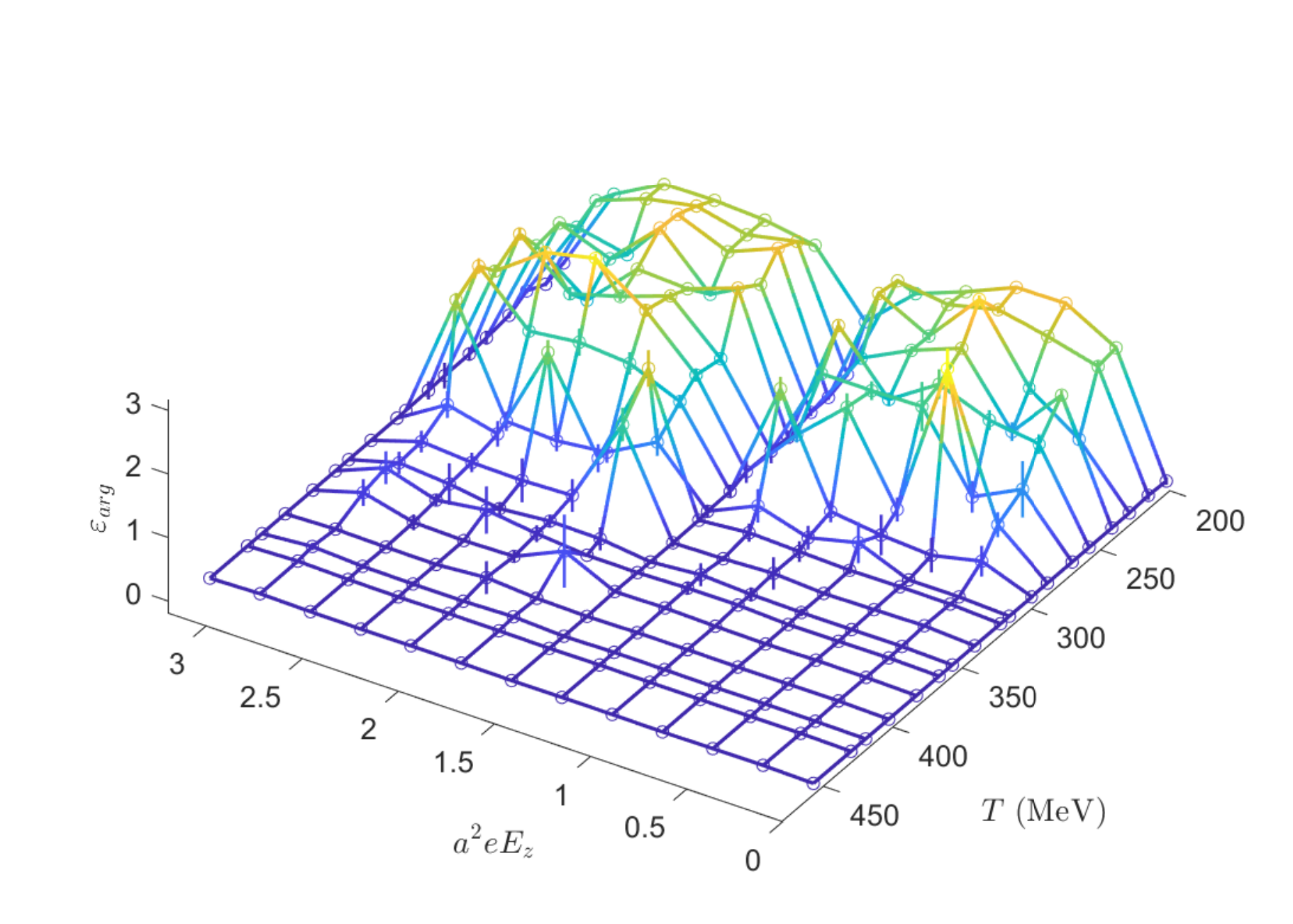}
\caption{\label{fig:stdarg}$\varepsilon _{\rm arg}$ as a function of $T$ and $E_z$.}
\end{center}
\end{figure}

\begin{figure}
\begin{center}
\includegraphics[width=0.48\textwidth]{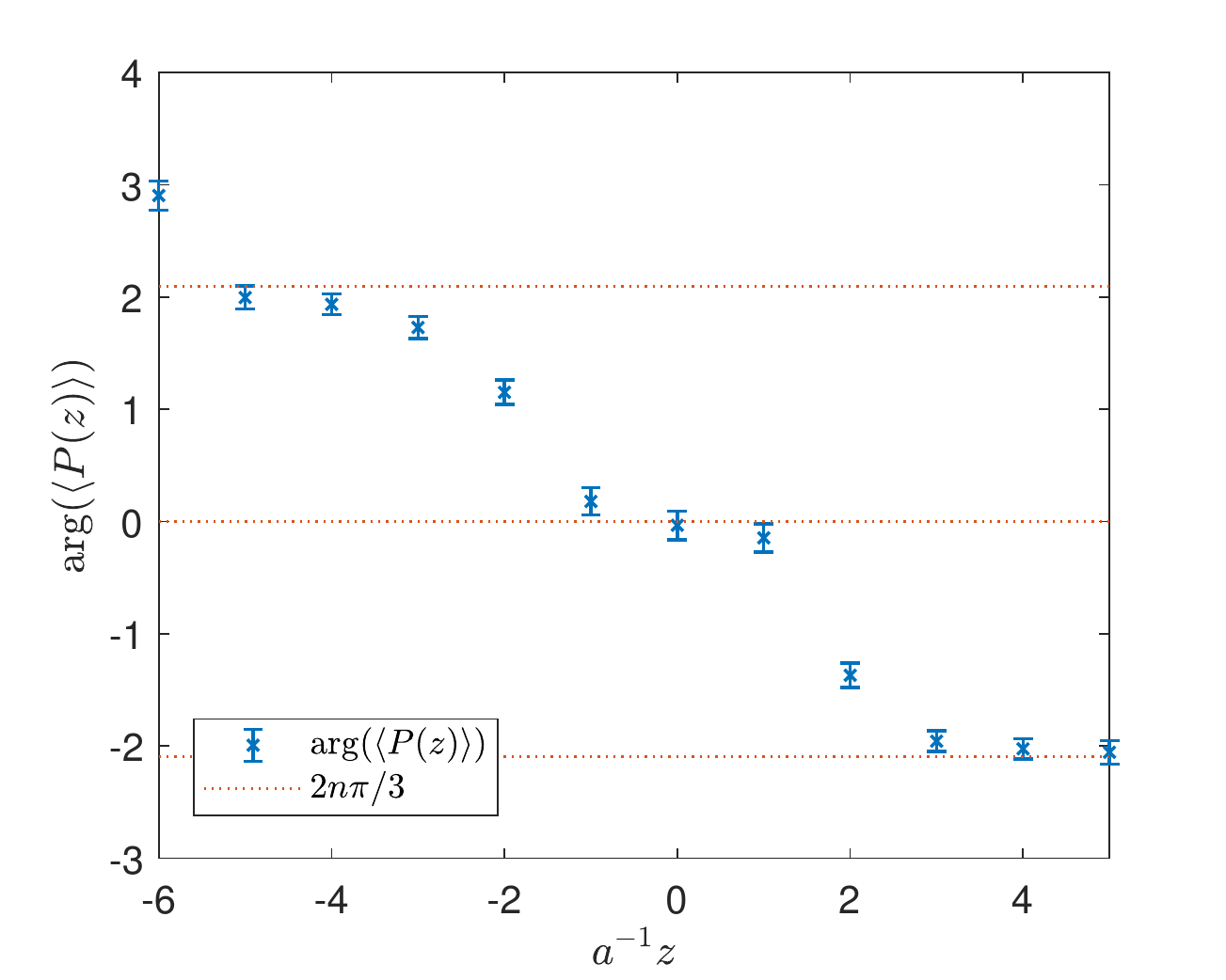}
\includegraphics[width=0.48\textwidth]{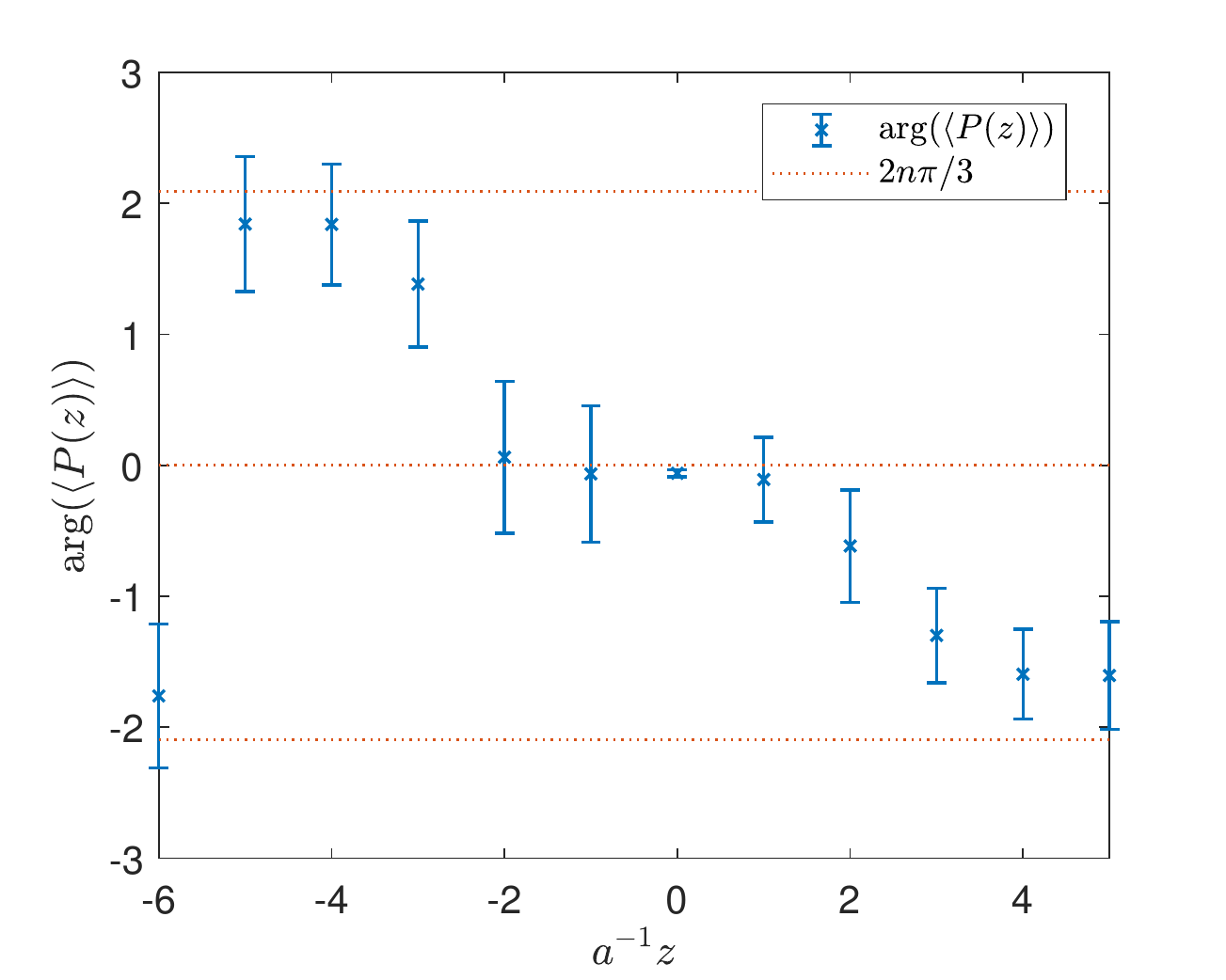}
\caption{\label{fig:arglowez1}$\arg\left(\langle P(z)\rangle\right)$ at $a^2eE_z=\pi/12$ for $\beta=5.3$~(the left panel) and $\beta=5.38$~(the right panel).}
\end{center}
\end{figure}

\begin{figure}
\begin{center}
\includegraphics[width=0.48\textwidth]{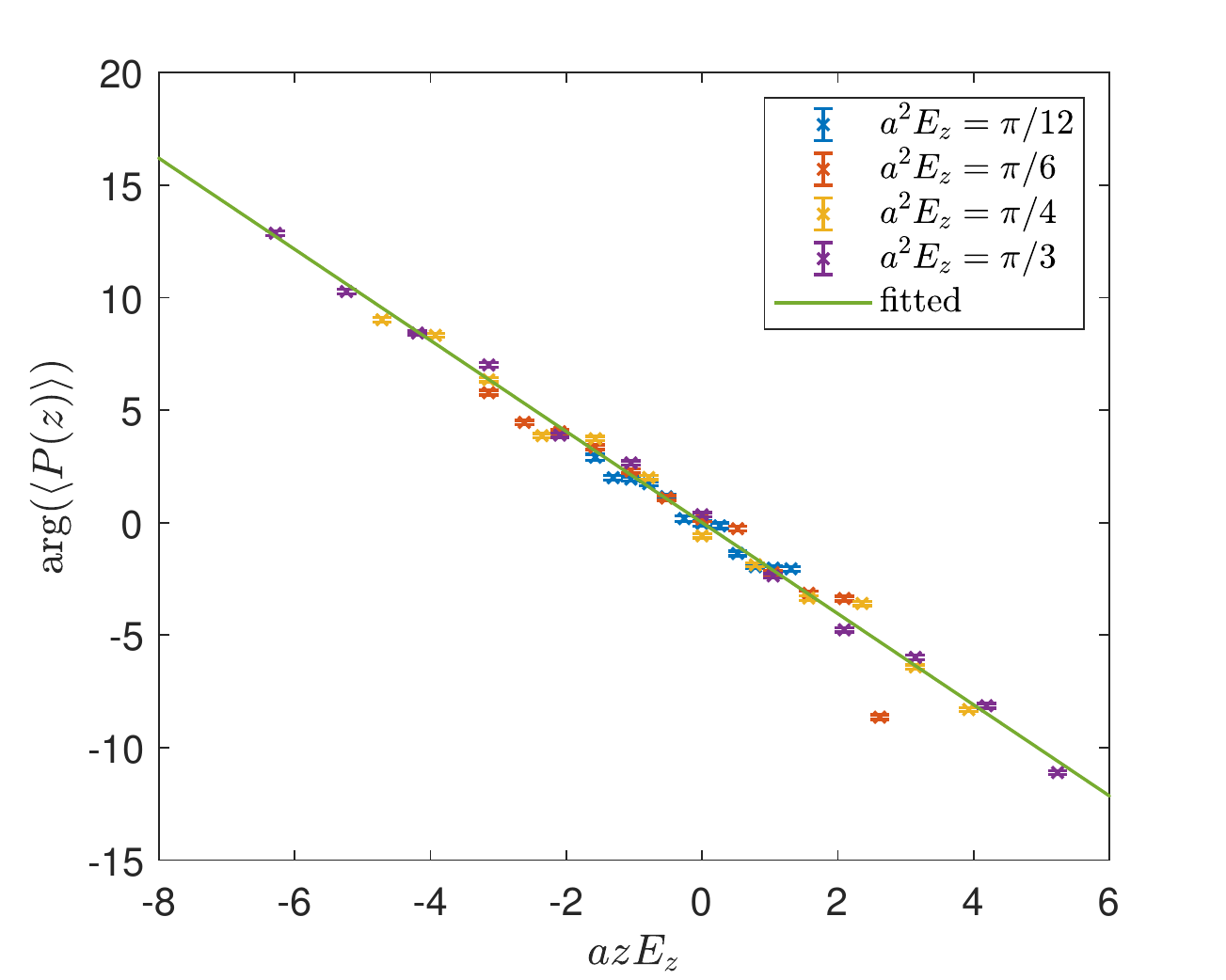}
\includegraphics[width=0.48\textwidth]{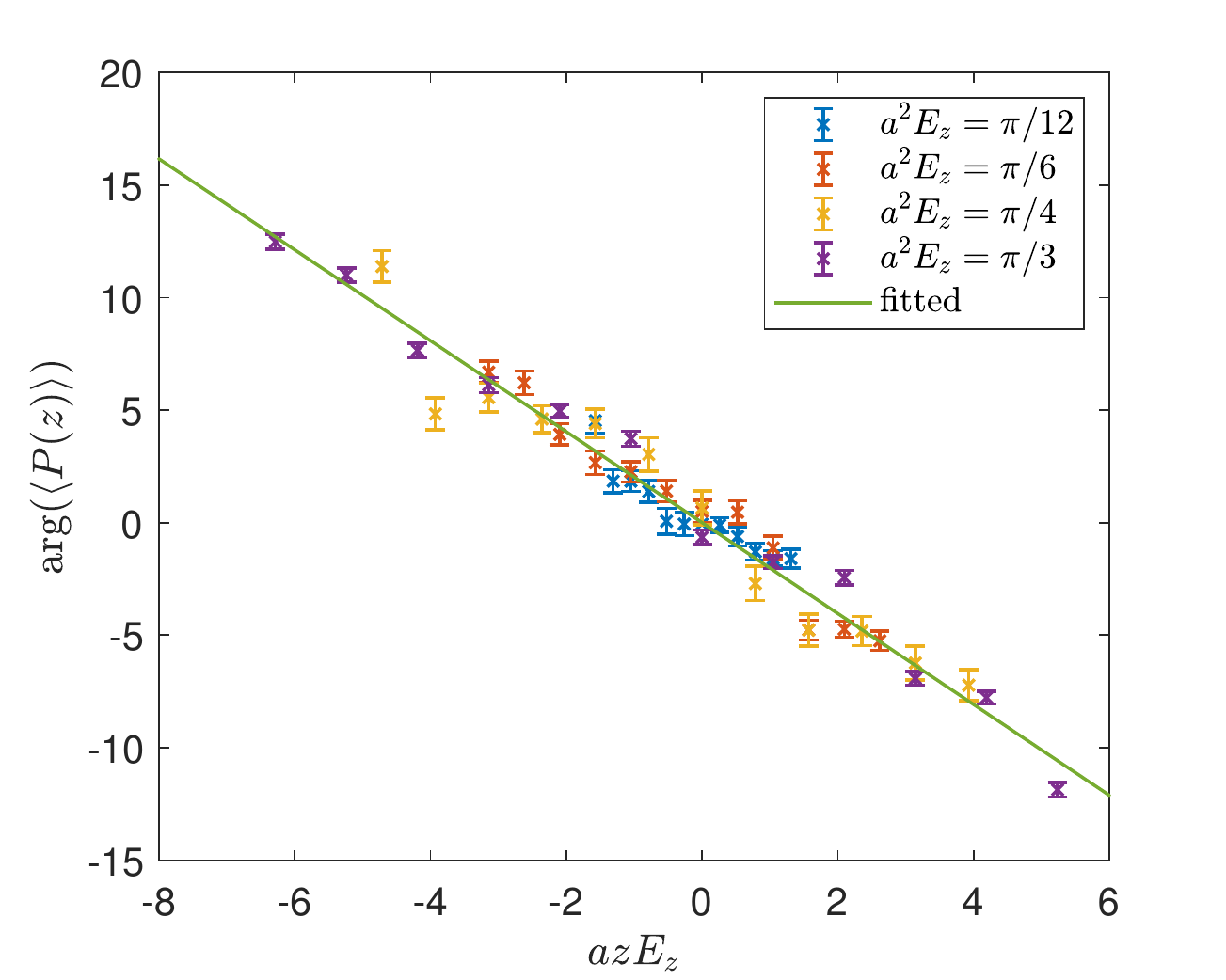}
\caption{\label{fig:arglowfit}$\arg\left(\langle P(z)\rangle\right)$ and fitted $\arg\left(\langle P(z)\rangle\right)$ according to the ansatz in Eq.~(\ref{eq.phaseansatz}) for $\beta=5.3$~(the left panel) and $\beta=5.38$~(the right panel).}
\end{center}
\end{figure}

\begin{figure}
\begin{center}
\includegraphics[width=0.6\textwidth]{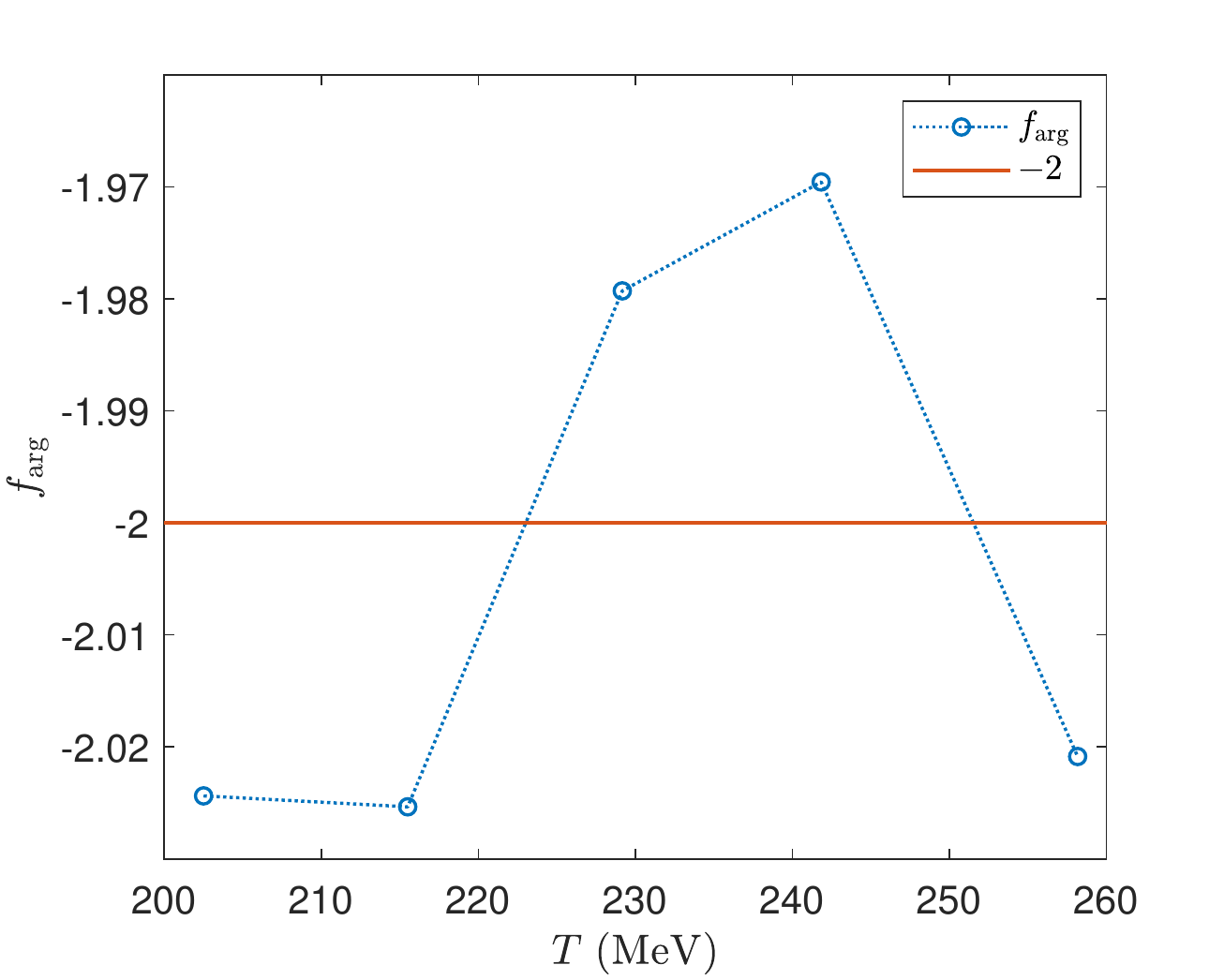}
\caption{\label{fig:dt}$f_{\rm arg}$ in Eq.~(\ref{eq.phaseansatz}) as a function of temperature.}
\end{center}
\end{figure}

Similar as $\varepsilon _q$ in Eq.~(\ref{eq.epsilonq}), we define
\begin{equation}
\begin{split}
&\varepsilon _{\rm arg}(T,E_z) = \sqrt{\frac{1}{L_z-1}\sum _z\left(\arg\left(\langle P^{T, E_z}(z)\rangle \right)-\frac{1}{L_z}\sum _{z'} \arg\left(\langle P^{T, E_z}(z')\rangle \right)\right)^2}.
\end{split}
\label{eq.epsilonarg}
\end{equation}
$\varepsilon _{arg}$ is shown in Fig.~\ref{fig:stdarg}.
Note that, for high temperatures, there are cases that $\arg \left(\langle P\rangle\right)$ is large while $\varepsilon _{\rm arg}$ is small.
For $a^2 eE_z = \pi / 2$, $\varepsilon _{\rm arg}$ is small.
Those properties will be explained later.

For a lower temperature and a small external electric field, the phase of the Polyakov loop is consistent with the R-W transition.
The cases of $\beta=5.3$ and $\beta=5.38$ at $a^2eE_z=\pi/12$ are shown in Fig.~\ref{fig:arglowez1}.
It can be seen that there are plateaus in the $\arg \left(\langle P(z)\rangle\right)$ located at $2n\pi/3$ where $n$ are integers.

Since $a^2e\Delta E_z=\pi/12$ is large, the widths of the plateaus are narrow.
For larger $a^2eE_z$, the widths of plateaus can be neglected, and $\arg \left(\langle P(z)\rangle\right)$ tends to be a linear function of $zE_z$, therefore we assume
\begin{equation}
\begin{split}
&\arg\left(\langle P (z) \rangle \right)=a f_{\rm arg} z e E_z.\\
\end{split}
\label{eq.phaseansatz}
\end{equation}
By adding integer times of $2\pi$, $\arg\left(\langle P (z) \rangle \right)$ is fitted according to Eq.~(\ref{eq.phaseansatz}).
The results of the fit depend on the manually added $2n\pi$ where $n$ are integers, and to minimize the effect of the manually added $2n\pi$, we fit only for the case of $a^2eE_z\leq \pi /3$.
Taking $\beta =5.3$ and $\beta =5.38$ as examples, $\arg\left(\langle P (z) \rangle \right)$ and fitted $\arg\left(\langle P (z) \rangle \right)$ are shown in Fig.~\ref{fig:arglowfit}.

In the region $5.3\leq \beta \leq 5.38$, $\arg\left(\langle P (z) \rangle \right)$ are fitted, and $f_{\rm arg}$ is shown in Fig.~\ref{fig:dt}.
The $\chi ^2 /d.o.f. = 6.54\sim 42.7$ for different $\beta$, the worst case is $\beta=5.3$.
We find that $f_{\rm arg}\approx -2$ which is a constant integer for different temperatures.
This integer also explains the phenomena $\arg\left(\langle P (z) \rangle \right)=0$ and $\varepsilon _{\rm arg}=0$ for $a^2eE_z=\pi $.
However, the case of $a^2eE_z=\pi/2$ cannot be explained, and will be postponed to section.~\ref{sec3.3.2}.

\begin{figure}
\begin{center}
\includegraphics[width=0.48\textwidth]{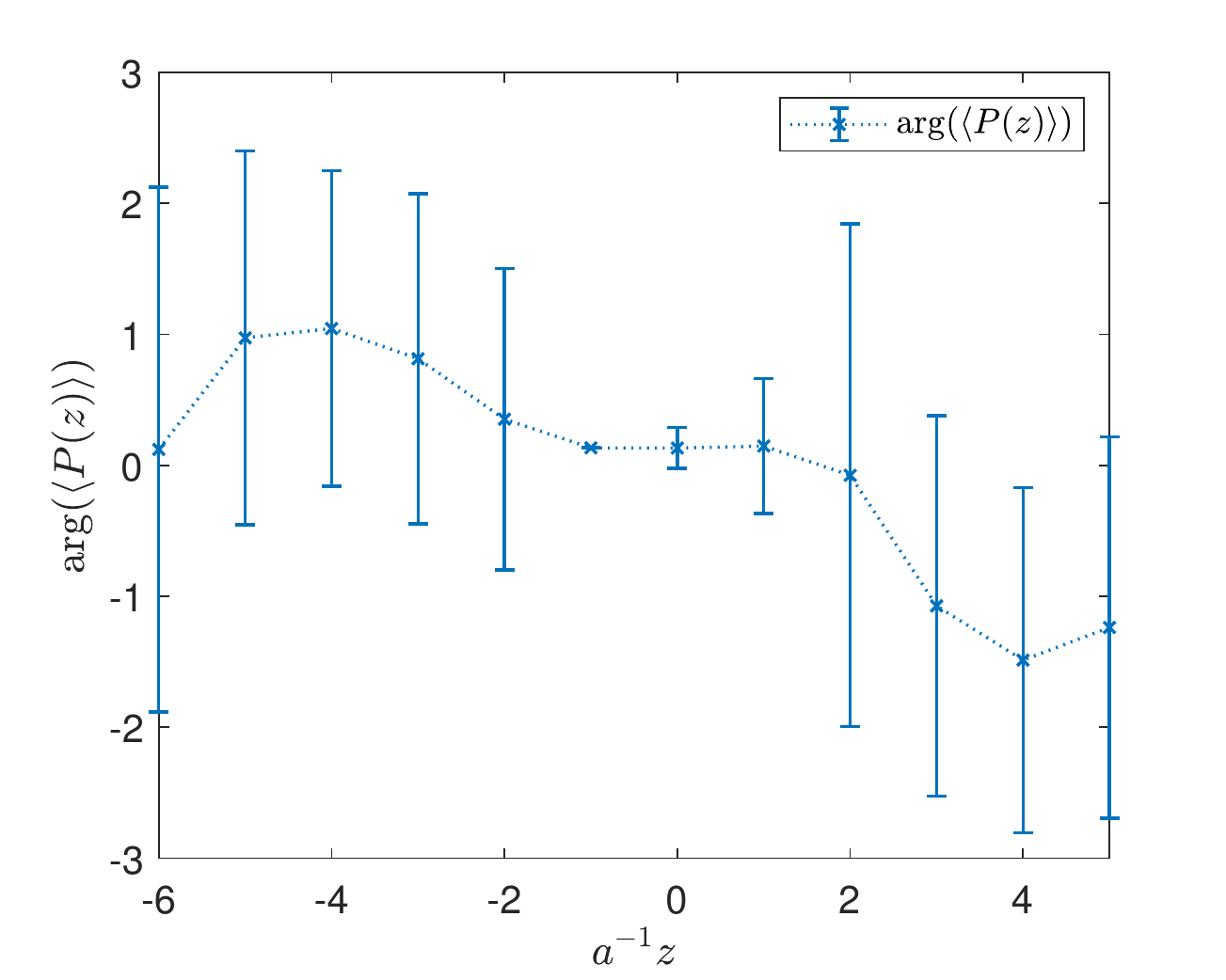}
\includegraphics[width=0.48\textwidth]{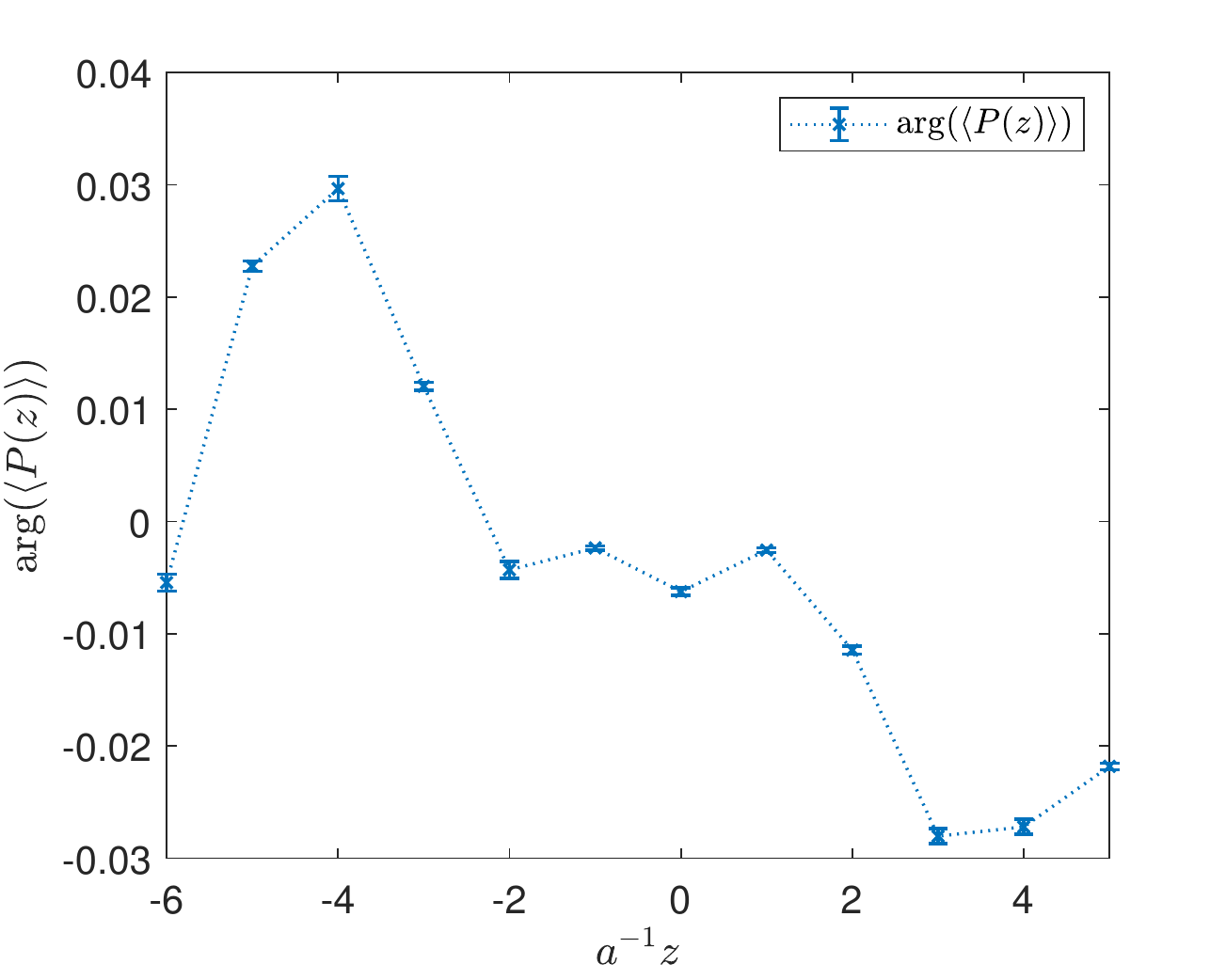}
\caption{\label{fig:arghighez1}Same as Fig.~\ref{fig:arglowez1} but for $\beta=5.4$~(the left panel) and $\beta=5.64$~(the right panel).}
\end{center}
\end{figure}

Such behavior fits when $\beta < 5.4$.
Starting from $\beta=5.4$, the behavior of the phase of Polyakov loop becomes different.
As examples, $\arg (\langle P(z)\rangle)$ at $\beta=5.4$ and $\beta =5.64$ and at $a^2eE_z=\pi /12$ are shown in Fig.~\ref{fig:arghighez1}.
It can be seen that, the variation of $\arg (\langle P(z)\rangle)$ is much smaller than the case in Fig.~\ref{fig:arglowez1}.

\begin{figure}
\begin{center}
\includegraphics[width=0.7\textwidth]{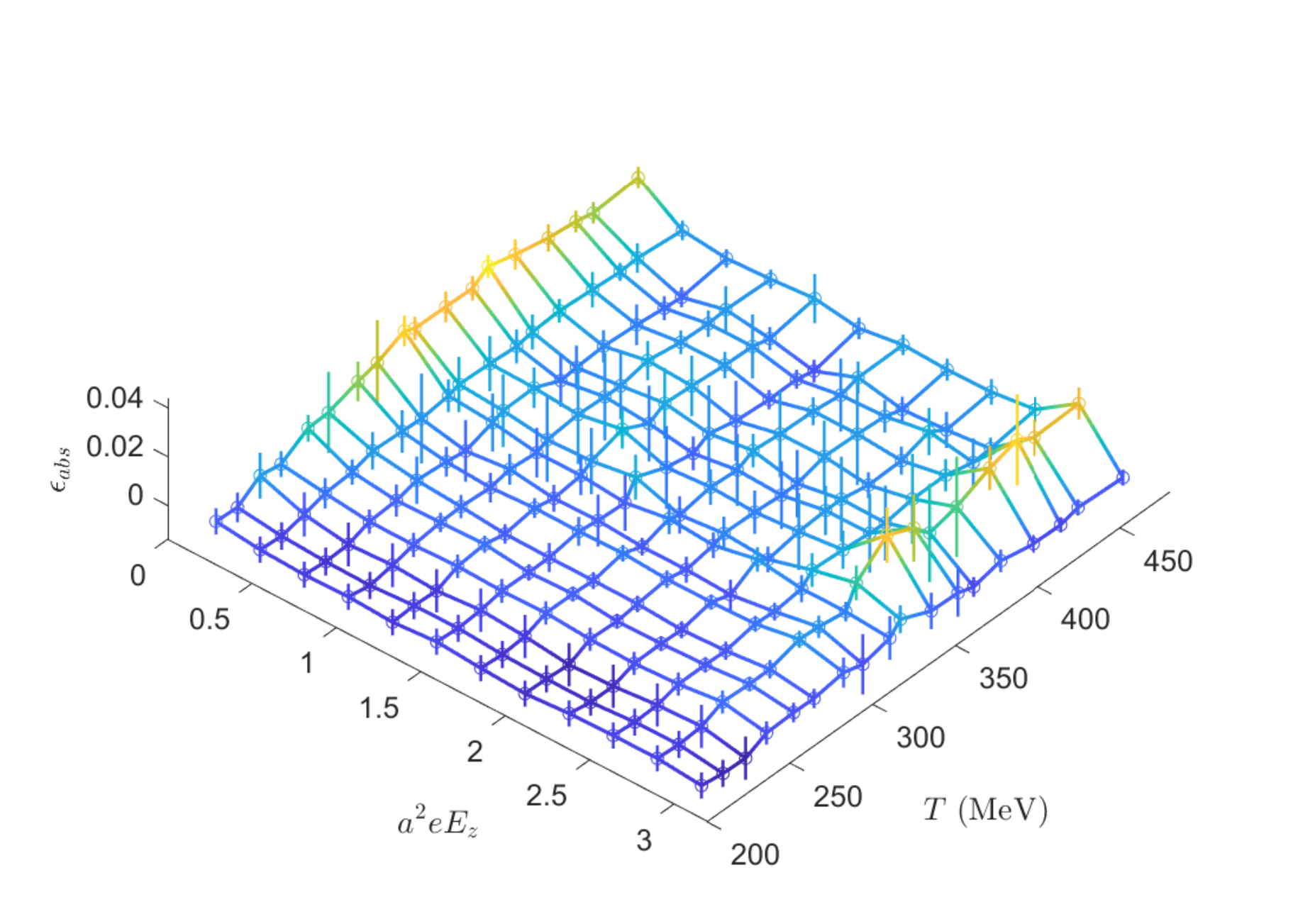}
\caption{\label{fig:stdabs}$\epsilon _{\rm abs}$ as a function of $T$ and $E_z$.}
\end{center}
\end{figure}

\begin{figure}
\begin{center}
\includegraphics[width=0.99\textwidth]{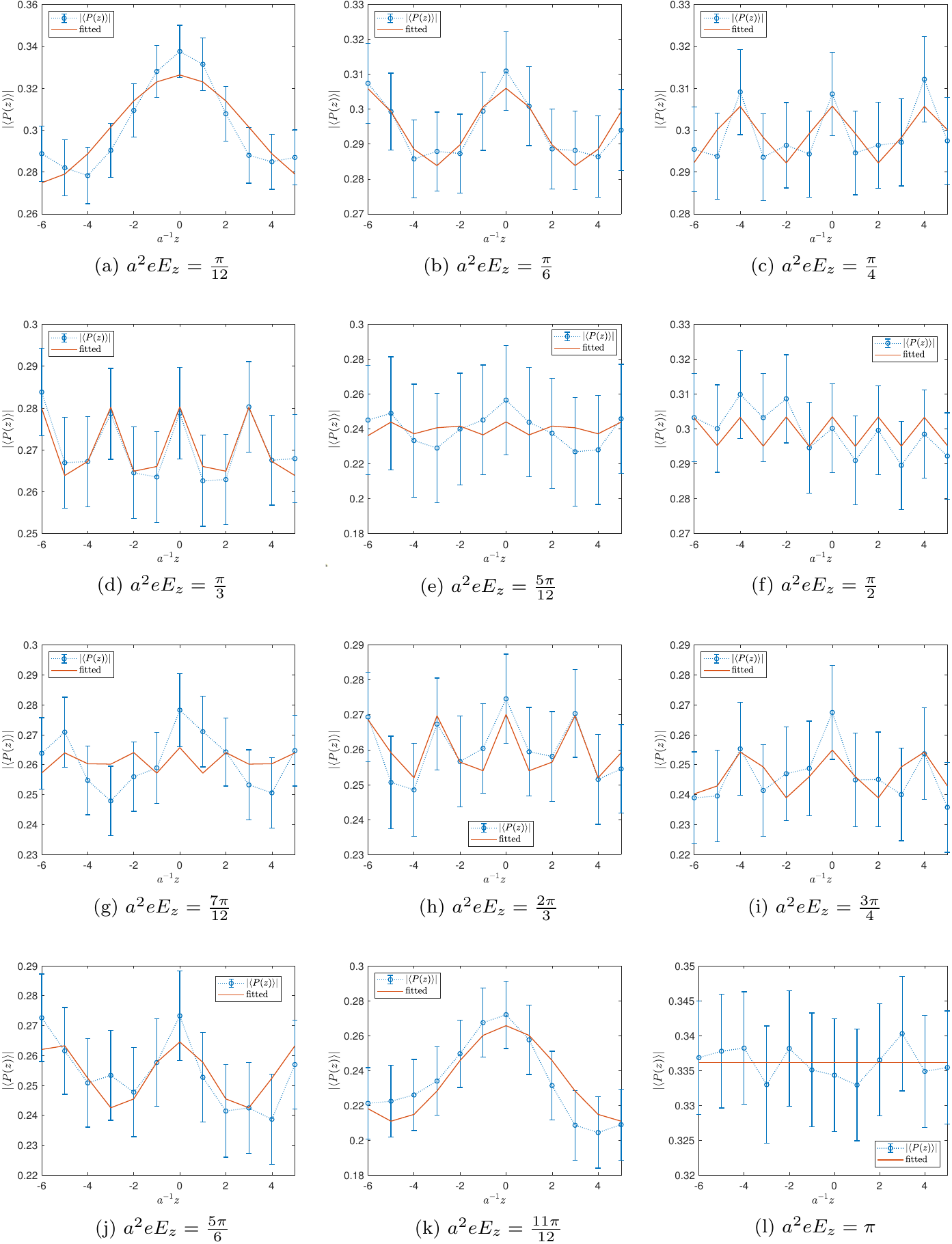}
\caption{\label{fig:polyaabs564fit}Same as Fig.~\ref{fig:cu564fit} but for $\left|\langle P(z)\rangle \right|$ at $\beta=5.64$.}
\end{center}
\end{figure}

\begin{figure}
\begin{center}
\includegraphics[width=0.6\textwidth]{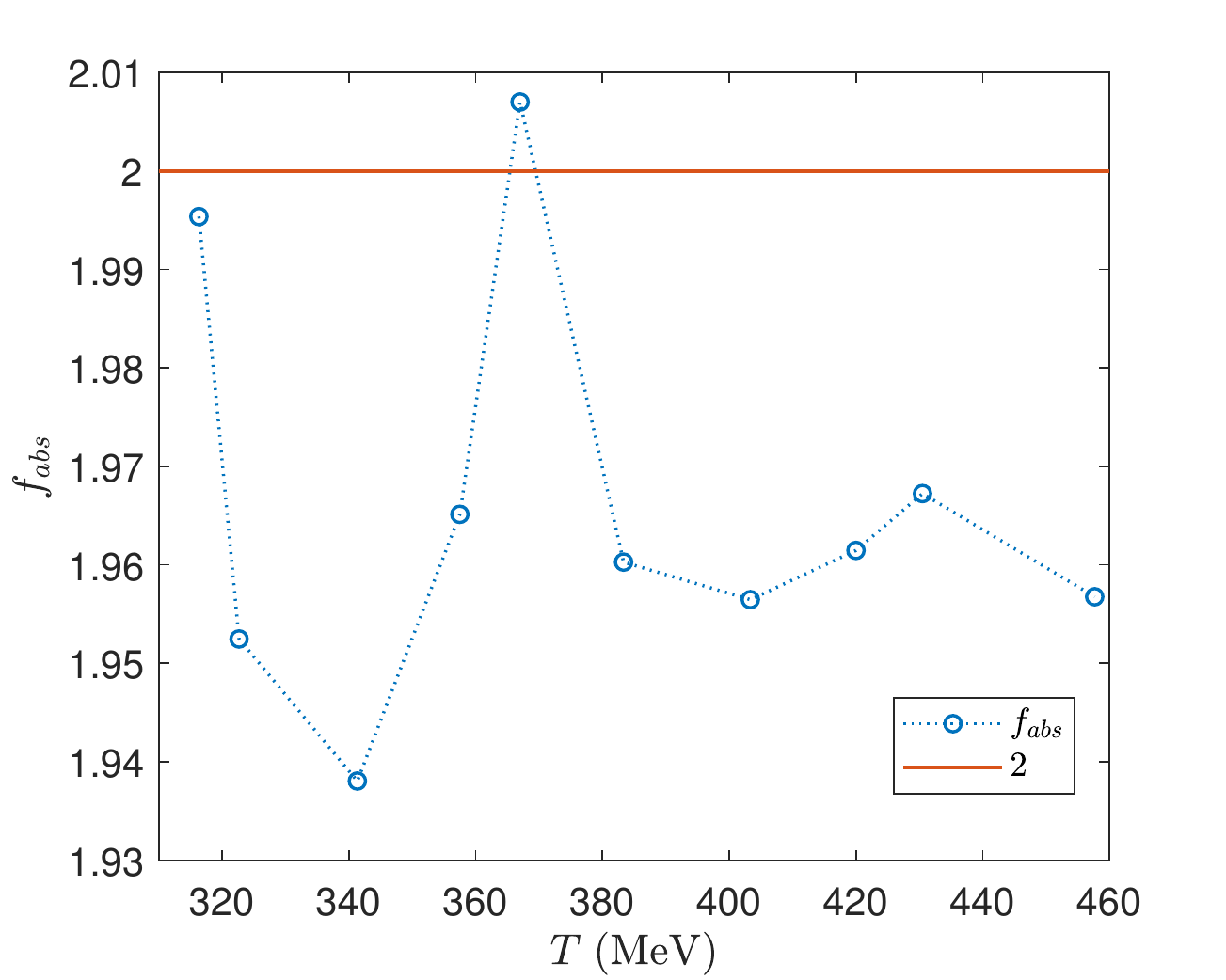}
\caption{\label{fig:fabs}$f_{\rm abs}$ as a function of temperature.}
\end{center}
\end{figure}

On the other hand, with growing temperatures the pattern of the $\left|\langle P(z)\rangle \right|$ gradually starts to become clear.
Similar as $\epsilon _q$, we define
\begin{equation}
\begin{split}
&\varepsilon _{\rm abs}(T,E_z) = \sqrt{\frac{1}{L_z-1}\sum _z\left(\left|\langle P^{T, E_z}(z)\rangle \right|-\frac{1}{L_z}\sum _{z'} \left|\langle P^{T, E_z}(z')\rangle \right|\right)^2},
\end{split}
\label{eq.epsilonabs}
\end{equation}
and $\epsilon _{\rm abs}(T,E_z)=\varepsilon _{\rm abs}(T,E_z)-\varepsilon _{\rm abs}(T,E_z=0)$.
$\epsilon _{\rm abs}$ is shown in Fig.~\ref{fig:stdabs}.

For $\left|\langle P(z)\rangle \right|$, we use the ansatz
\begin{equation}
\begin{split}
&\left|\langle P (z) \rangle \right|=A_{\rm abs}+B_{\rm abs}\cos \left(a f_{\rm abs} z eE_z\right).\\
\end{split}
\label{eq.absansatz}
\end{equation}
Using the method to fit for the chiral condensation, and using Eq.~(\ref{eq.absansatz}), $\left|\langle P (z) \rangle \right|$ is fitted.
As an example, the case for $\beta=5.64$ is shown in Fig.~\ref{fig:polyaabs564fit}, the $\chi ^2 /d.o.f. = 0.73$.
It can be seen that, except for the cases $a^2eE_z=5\pi/12$ and $a^2eE_z=7\pi/12$, Eq.~(\ref{eq.absansatz}) roughly describes the pattern of $|\langle P(z)\rangle|$.
Similar as the case of chiral condensation, the range of $5.46\leq \beta \leq 5.64$ is considered, and $f_{\rm abs}$ is shown in Fig.~\ref{fig:fabs}.
Again, we find $f_{\rm abs}\approx 2$ which is a constant integer.

At smaller $E_z$, another noteworthy interesting phenomenon is that $\epsilon _{u,d}$ in Fig.~\ref{fig:stdcq}, $\epsilon _{d}^4$ in Fig.~\ref{fig:c4oscillation} and $\epsilon _{\rm abc}$ in Fig.~\ref{fig:stdabs} decrease with $E_z$ in the small $E_z$ region.
This behavior implies that, the oscillation suddenly appears at a small none-zero $E_z$.
It has been pointed out in Refs.~\cite{chargedistrib1,chargedistrib2} that, physical observables exhibit a discontinuity between $E=0$ and any small $E>0$.
Our results can be seen as a support for the above conclusion.

\subsubsection{\label{sec3.3.2}An ansatz for the Polyakov loop}

After combining the analysis of the phase of Polyakov and the absolute value of Polyakov loop, we conclude that Polyakov loop is consistent with ansatz
\begin{equation}
\begin{split}
&\langle P (z) \rangle =A_p+\sum _{q=u,d}C_q\exp \left( L_{\tau} Q_q i a z eE_z\right),\\
\end{split}
\label{eq.polyaansatz}
\end{equation}
where $C_q$ are real numbers~(and $C_d>0$, $C_u\geq 0$ after fit) and $A_p$ is a complex number because there are cases that $\arg\left(\langle P\rangle\right)$ is large but $\varepsilon _{\rm arg}$ is small which corresponds to a complex $A_p$ and $|A_p|\gg C_q$.

\begin{figure}
\begin{center}
\includegraphics[width=0.48\textwidth]{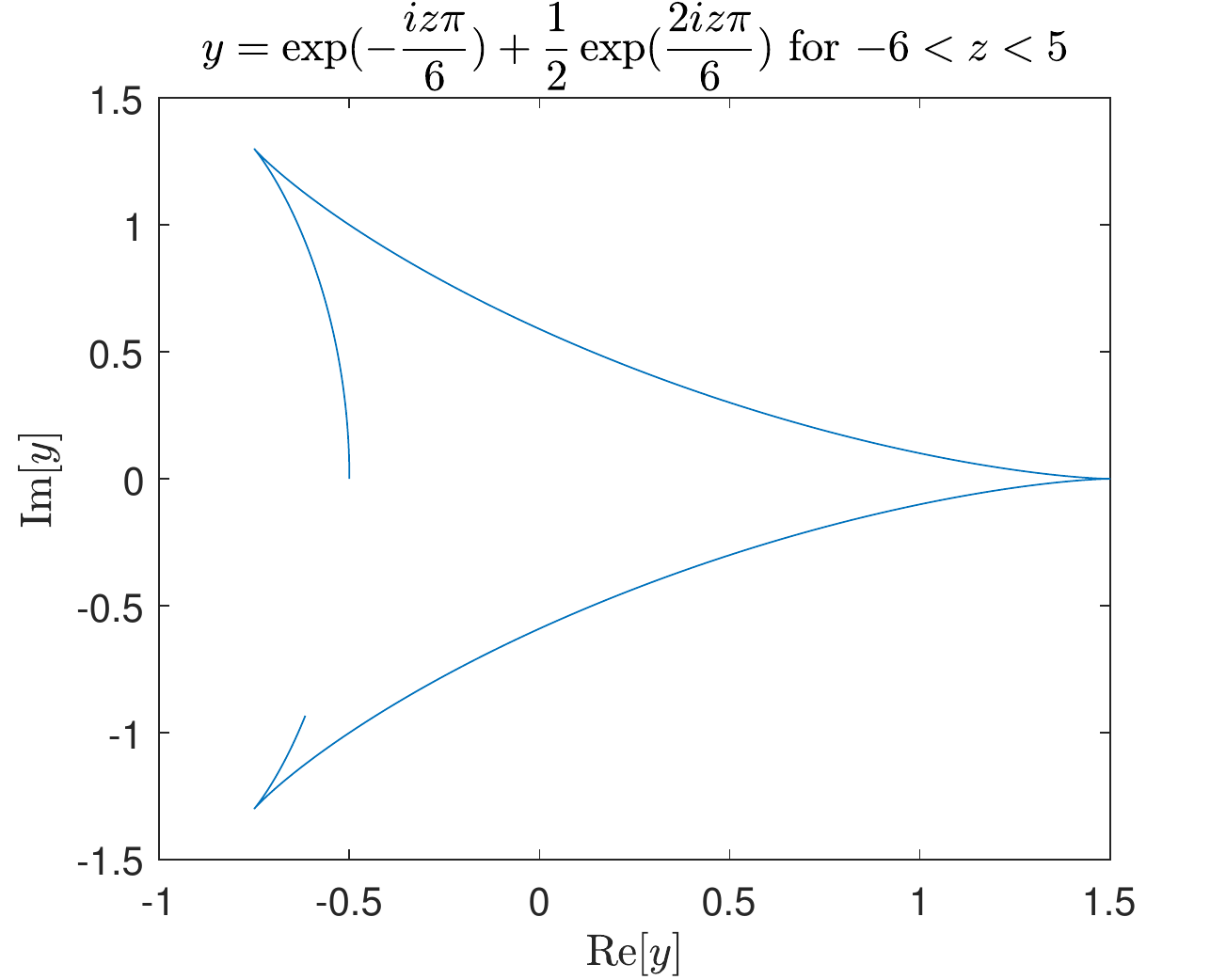}
\includegraphics[width=0.48\textwidth]{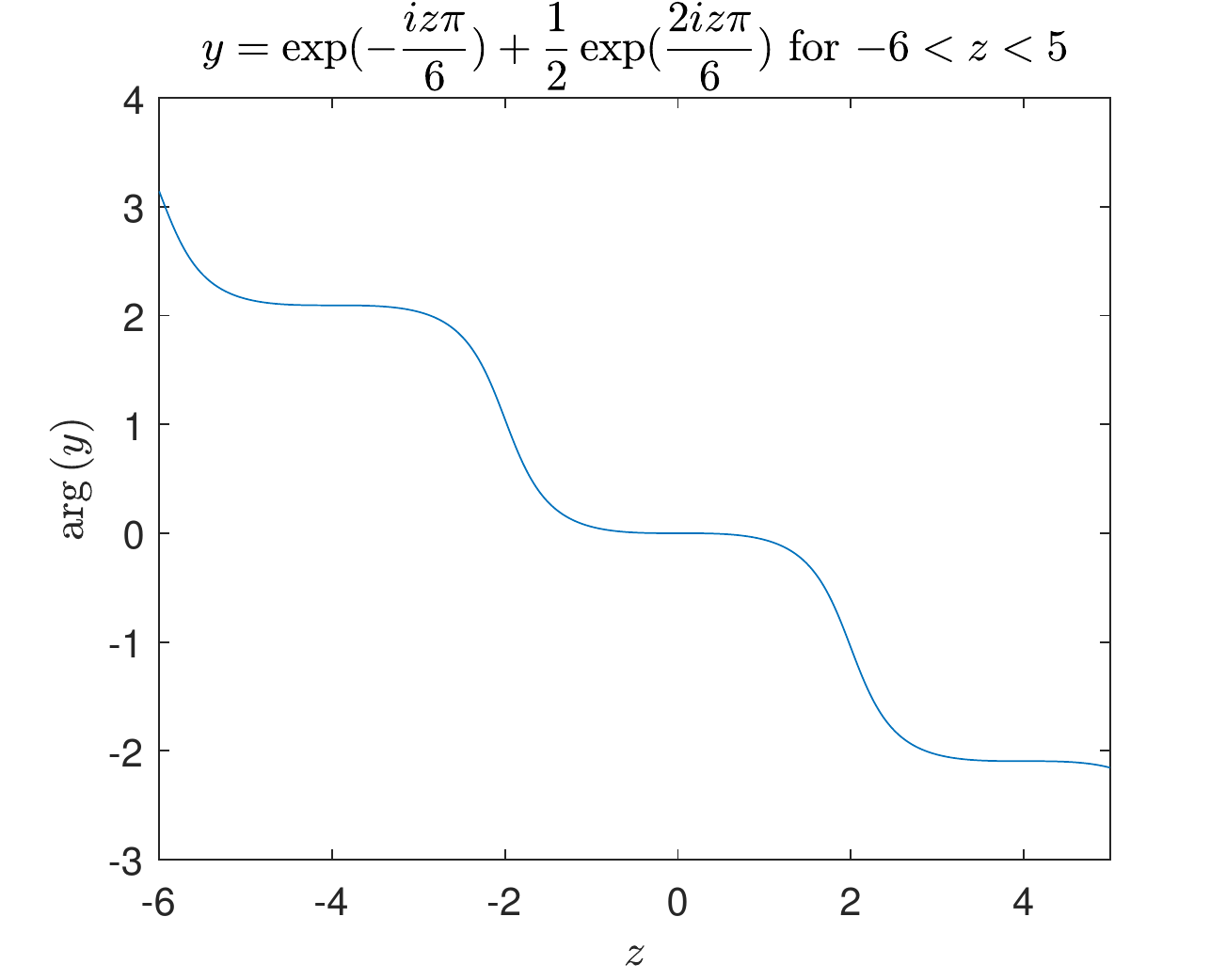}
\caption{\label{fig:example}An example when $|A_p|\ll C_q$ and $C_u<C_d$, the $\langle P(z)\rangle$ in ansatz Eq.~(\ref{eq.polyaansatz}) depicted in the complex plane~(the left panel) and $\arg(P(z))$~(the right panel).}
\end{center}
\end{figure}

\begin{figure}
\begin{center}
\includegraphics[width=0.7\textwidth]{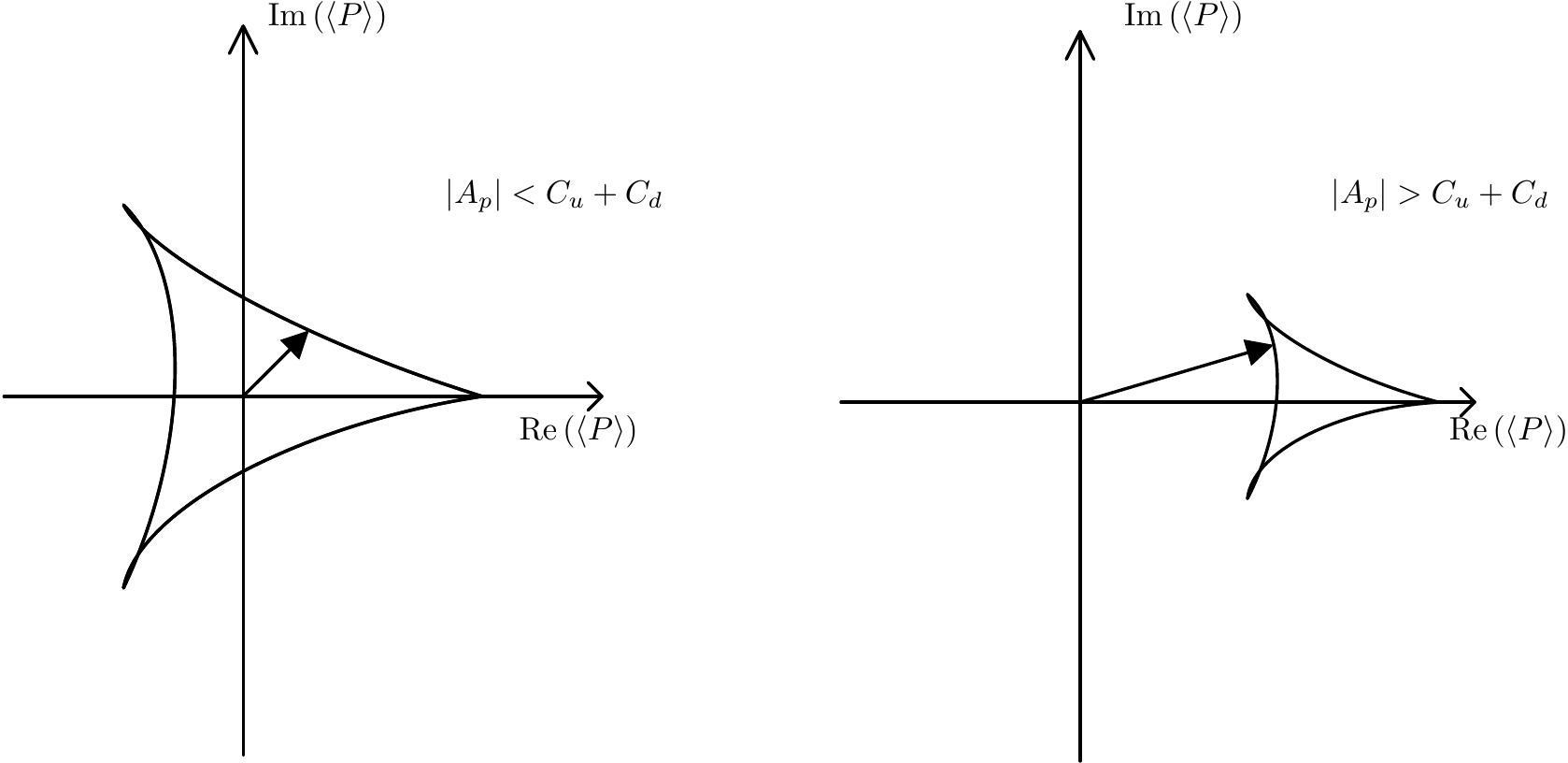}
\caption{\label{fig:complex}The difference patterns of the phase of the Polyakov loop when $|A_p|<C_u+C_d$~(the left panel) and $|A_p|>C_u+C_d$~(the right panel).}
\end{center}
\end{figure}

The ansatz in Eq.~(\ref{eq.polyaansatz}) counts both the effect from $u$ and $d$ quarks, and is able to describe the following phenomena.

\begin{itemize}
\item As shown in the left panel of Fig.~\ref{fig:example}, for $|A_p|\ll C_q$ and $C_u<C_d$ the ansatz describes $\langle P(z)\rangle$ in the complex plane in the left panel of Fig.~\ref{fig:complex530564}.
Similarly, when $|A_p|\gg C_q$, the ansatz describes the right panel of Fig.~\ref{fig:complex530564}.

\item As shown in the right panel of Fig.~\ref{fig:example}, for $|A_p|\ll C_q$ and $C_u<C_d$, Eq.~(\ref{eq.polyaansatz}) describes the phenomena in Fig.~\ref{fig:arglowez1}.
With $C_d>C_u$, the effect of $C_u$ term becomes small plateaus.
Ignoring the plateaus, $f_{\rm arg}=L_{\tau}\times Q_d=-2$, which explains the phenomena in Figs.~\ref{fig:arglowfit} and \ref{fig:dt}.

\item As shown in Fig.~\ref{fig:complex}, when $|A_p|>C_q$, the phase of the Polyakov loop ranges in $[-\pi, \pi)$, when $|A_p|<C_q$, the phase of the Polyakov loop ranges in $(-\pi / 2 , \pi / 2)$.
With the growth of temperature, $|A_p|>C_q$, the phase of Polyakov loop when $\beta\geq 5.4$ in Fig.~\ref{fig:arghighez1} can be explained.

\item When $|A_p|\gg C_q$ and $C_u\ll C_d$, Eq.~(\ref{eq.polyaansatz}) describes the oscillation in Fig.~\ref{fig:polyaabs564fit} and $f_{\rm abs}=|L_{\tau}\times Q_d|=2$ in Fig.~\ref{fig:fabs} can be understood.
A $C_u$ term also explains the discrepancy of the ansatz in Eq.~(\ref{eq.absansatz}) at $a^2eE_z=5\pi/12$ and $a^2eE_z=7\pi/12$.

\item Apart from that, Eq.~(\ref{eq.polyaansatz}) also explains the reason that $\varepsilon _{\rm arg}$ is small and $\arg\left(\langle P\rangle\right)\approx 0$ at $a^2eE_z=\pi / 2$.
This can be attributed to a real $A_p$ which is larger than $C_u+C_d$.
\end{itemize}

\begin{figure}
\begin{center}
\includegraphics[width=0.99\textwidth]{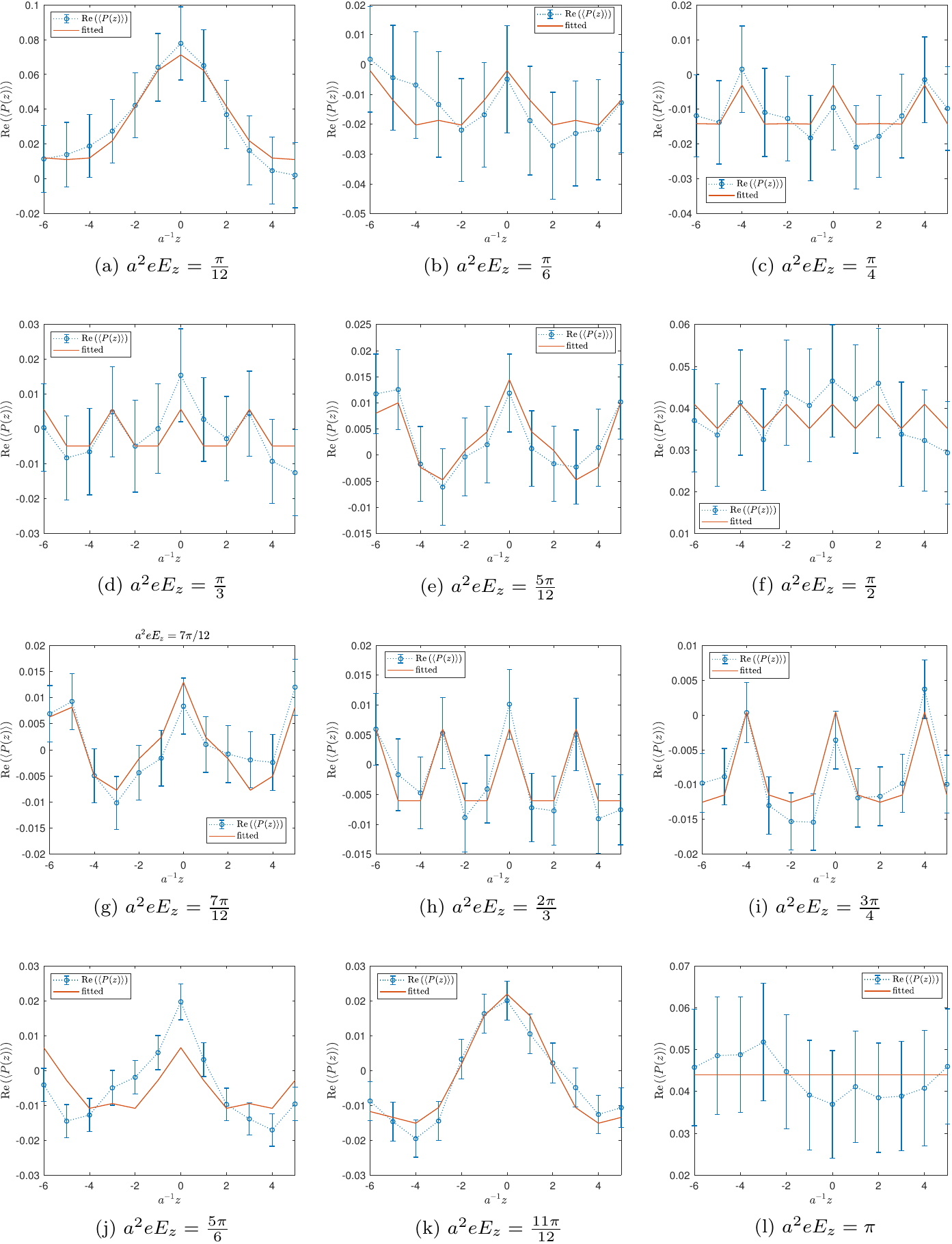}
\caption{\label{fig:polya544re}Same as Fig.~\ref{fig:cu564fit} but for ${\rm Re}\left(\langle P(z)\rangle \right)$ at $\beta=5.42$.}
\end{center}
\end{figure}

\begin{figure}
\begin{center}
\includegraphics[width=0.99\textwidth]{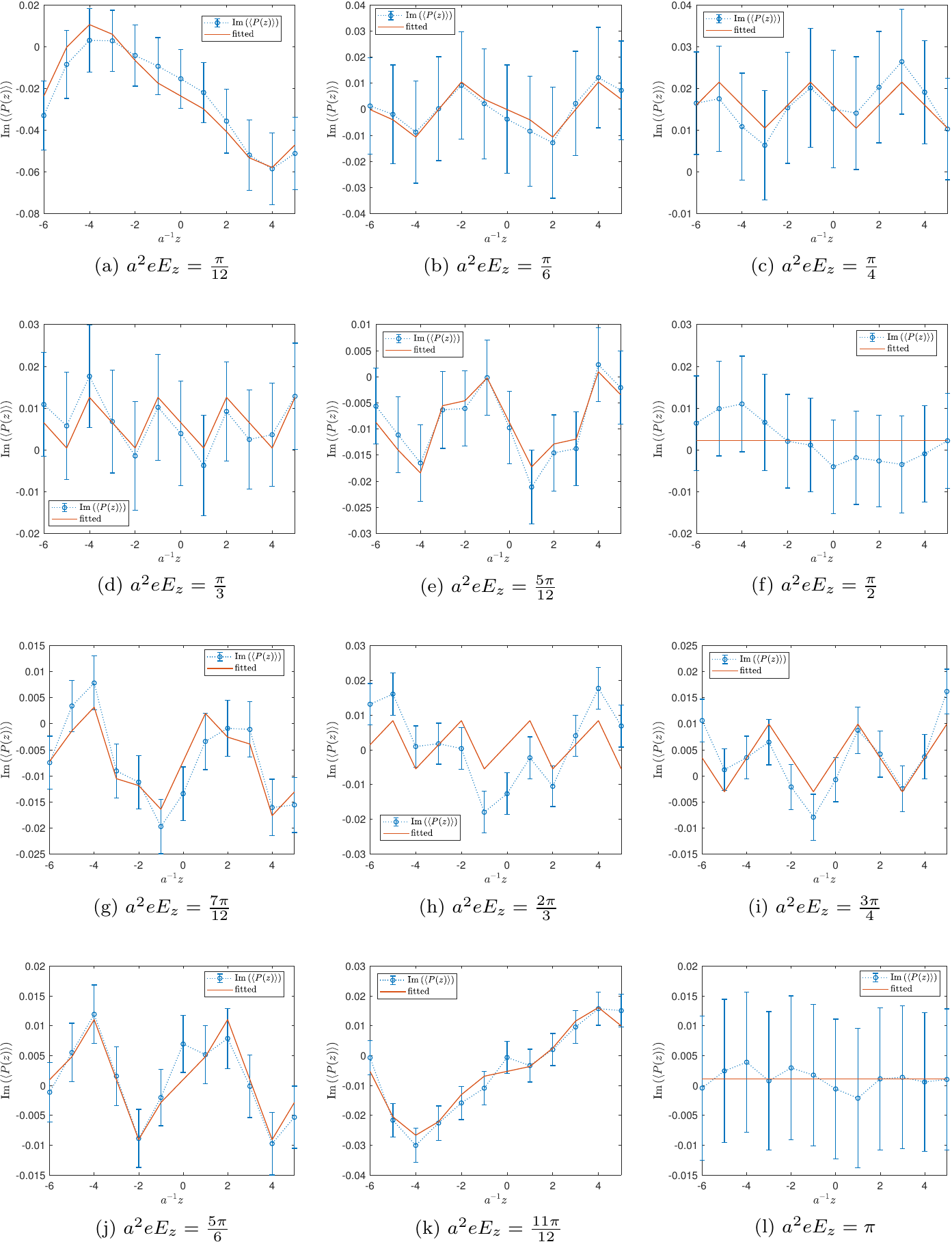}
\caption{\label{fig:polya544im}Same as Fig.~\ref{fig:cu564fit} but for ${\rm Im}\left(\langle P(z)\rangle \right)$ at $\beta=5.42$.}
\end{center}
\end{figure}

For $\beta = 5.3 \sim 5.64$, the $\langle P(z) \rangle$ are fitted, and we find $\chi ^2 /d.o.f.=0.10\sim 1.88$.
To compare with the ansatz in Eqs.~(\ref{eq.phaseansatz}) and (\ref{eq.absansatz}), using Eq.~(\ref{eq.polyaansatz}), $\chi ^2 /d.o.f.=0.67\sim 1.88$ for $\beta=5.3 \sim 5.38$, and $\chi ^2 /d.o.f.=0.24$ for $\beta=5.64$.
Taking the case of $\beta=5.42$ which lies in the middle of the region $5.4\leq \beta \leq 5.44$ as an example~($\chi ^2 /d.o.f.=0.47$), the results are shown in Figs.~\ref{fig:polya544re} and \ref{fig:polya544im}.
Eq.~(\ref{eq.polyaansatz}) is able to describe the pattern of $\langle P(z)\rangle$.

\subsubsection{\label{sec3.3.3}A criterion to distinguish the different behaviors of the Polyakov loop}

\begin{figure}
\begin{center}
\includegraphics[width=0.6\textwidth]{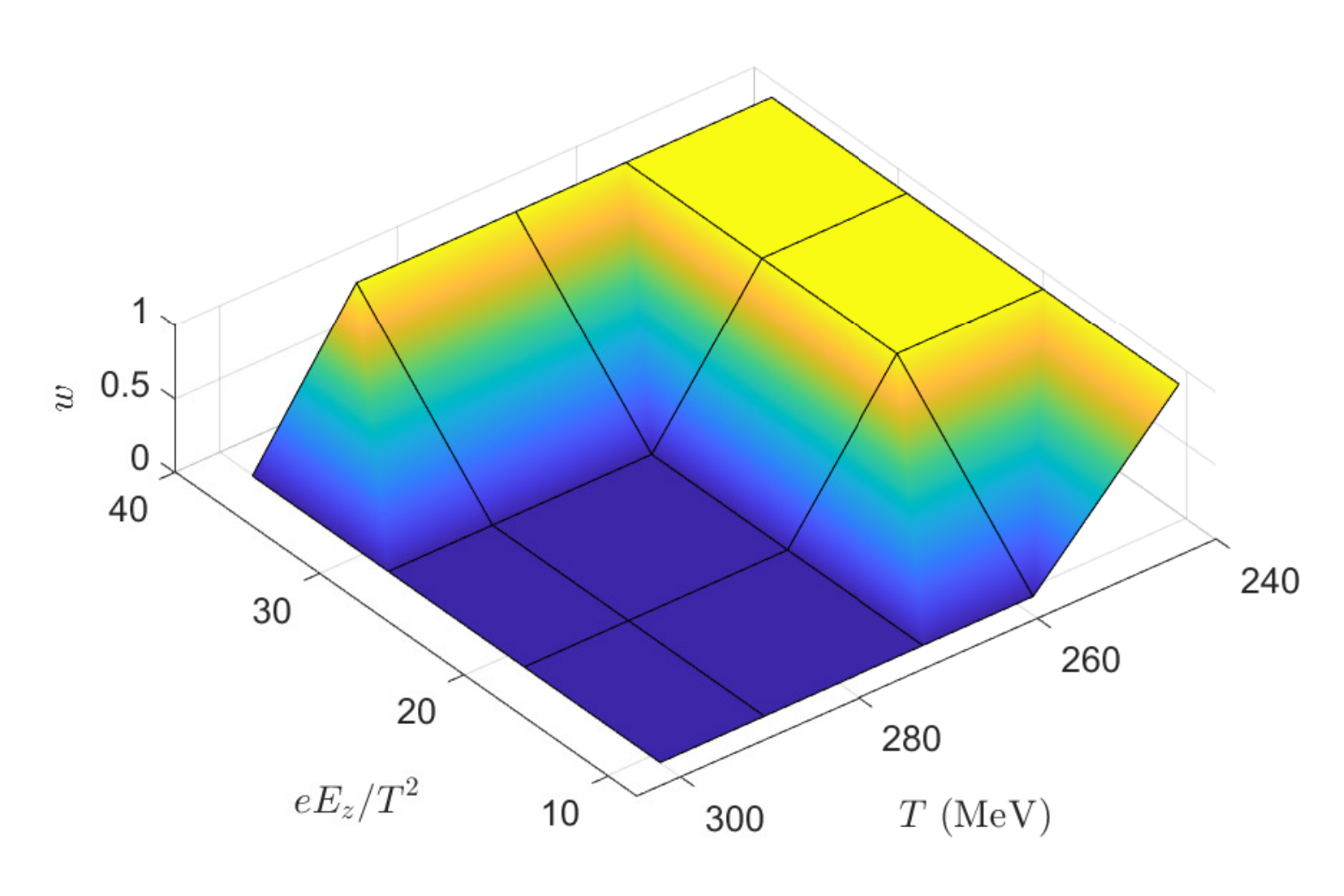}
\caption{\label{fig:criterion}$w$ as a function of $T$ and $eE_z/T^2$.}
\end{center}
\end{figure}

\begin{figure}
\begin{center}
\includegraphics[width=0.6\textwidth]{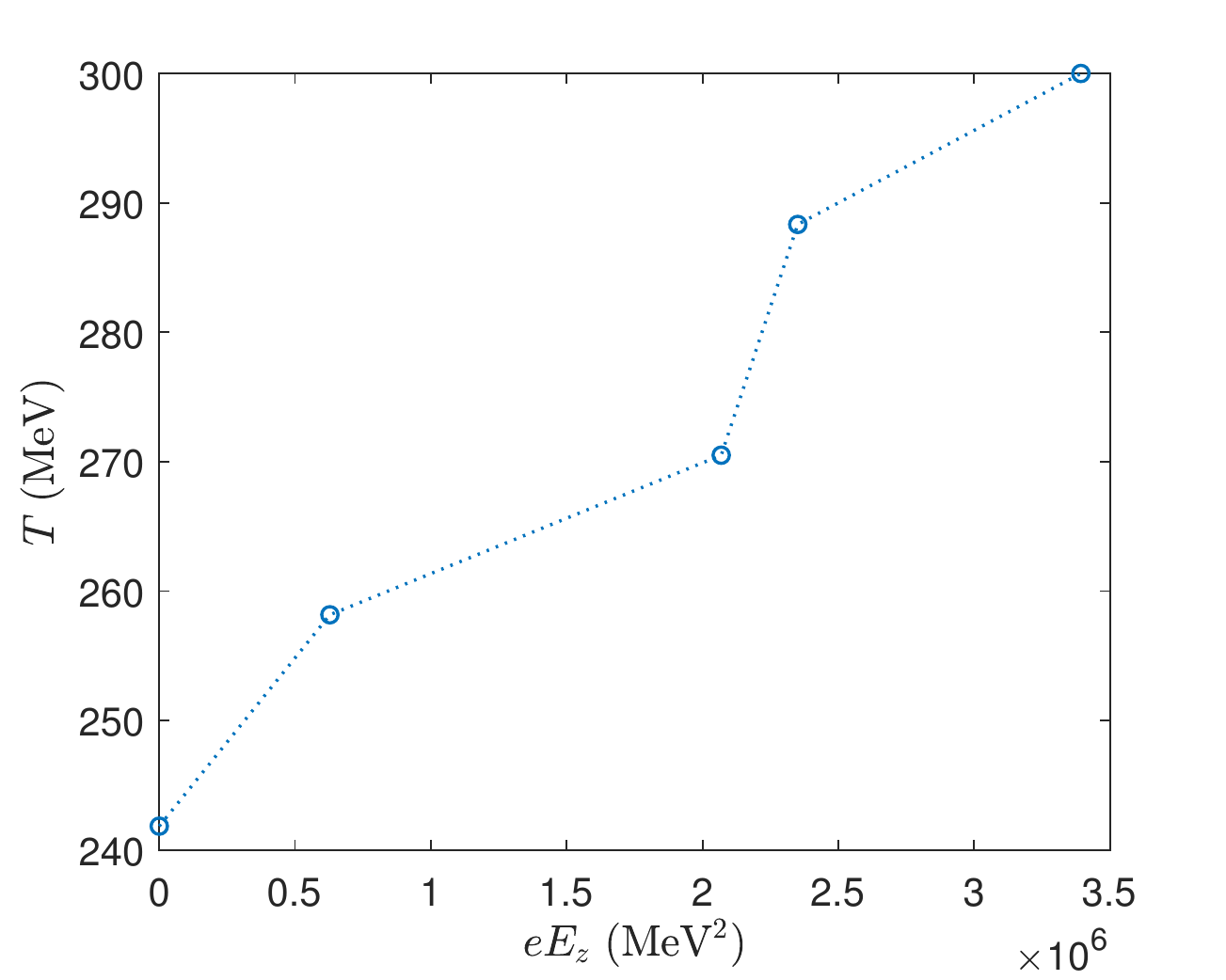}
\caption{\label{fig:phasediagram}A boundary to distinguish the different behaviors of the Polyakov loop.}
\end{center}
\end{figure}

In previous works, the susceptibility of the imaginary of the Polyakov loop is used to find out the phase diagram of the R-W transition.
However, in our study, we did not see a clear signal of phase transition in this approach.
This can be understood because from the point of view of the R-W phase transition, the phase transition point is about $\mu/T=\pi/3$ which is $a\mu=\pi/(3L_t)=\pi/18$.
For our study, $a^2\Delta eE_z=\pi/12$, so the case of the smallest electric field is already in the R-W phase.
In terms of the chiral phase transition due to the strong electric field, the phase transition point is about $\left(500\;{\rm MeV}\right)^2$ magnitude in the previous study~\cite{electricAndMagnetic2}.
For the case of the smallest lattice spacing, which is $\beta=5.3$, $\Delta eE_z\approx \left(621.5 \;{\rm MeV}\right)^2$, and the case of the smallest electric field is already in chiral symmetry restored phase.

On the other hand, comparing the chiral condensation at high and low temperatures, or the behavior of Polyakov loop, we can find clear differences.
At high temperatures, the chiral condensation and charge density oscillate with $z$ coordinate.
Another very clear difference is that if the ansatz in Eq.~(\ref{eq.polyaansatz}) is correct, then the size relationship between $|A_p|$ and $C_q$ causes a significant difference in Polyakov loop phase behavior as shown in Fig.~\ref{fig:complex}.

We use the winding number to distinguish different behaviors of the phases of Polyakov loops, which is defined as
\begin{equation}
\begin{split}
&w=-\frac{1}{2\pi i}\int _0^{2\pi}\frac{2iC_u\exp(2ix)-iC_d\exp(-ix)}{A_p+C_d\exp(-ix)+C_u\exp(2ix)}dx.\\
\end{split}
\label{eq.winding}
\end{equation}
$w$ is shown in Fig.~\ref{fig:criterion}.
The case of $\beta=5.38$ and $a^2eE_z=\pi/12$ is special.
The absolute value of the phase of $\langle P\rangle$ can exceed $\pi/2$ as shown in right panel of Fig.~\ref{fig:arglowez1}, not because in the complex plane, $\langle P\rangle$ encloses the origin, but because the two angles on the left side of the pointed triangle poke into the half plane where ${\rm Re}[\langle P\rangle] < 0$.

Using $w \neq 0$ as a criterion, the boundary of possible transition can be obtained, which is shown in Fig.~\ref{fig:phasediagram}.
Whether there is a phase transition needs more exploration.
Note that, this boundary also coincides with the boundary that the charge density start to oscillate as shown in Fig.~\ref{fig:c4oscillation}.

\section{\label{sec4}Summary}

The strong electric fields in heavy ion collisions provide a unique opportunity to study the effect of an external electromagnetic field on the quark matter.
The case of classical electric field is considered which is free from the `sign problem'.
In the case of ${\bf E}$ along ${\bf z}$ direction and in axial gauge, and neglecting the boundary condition, the electric field is equivalent as inhomogeneous imaginary chemical potential varies along the $z$ coordinate.

In this paper, we investigate the properties of the R-W phase caused by an external uniform classical electric field using lattice QCD with $N_f=1+1$ staggered fermions.
In the simulation, $am_q=0.1$ is a constant, and $\beta$ ranges from $5.3$ to $5.64$.
The simulation is carried out on a $12^3\times 6$ lattice, $a^2eE_z$ is chosen as $a^2eE_z=k\pi/12$ where $k$ is an integer and $0\leq k \leq 12$.

It is found that, at high temperatures, chiral condensation oscillates over $z$ coordinates.
Note that the action is actually translational invariant accompanied by a gauge transformation.
The oscillation over $z$ coordinates partially breaks the translational invariance.
$c_q$ at high temperatures can be well-fitted by the ansatz $A_c+B_c\cos \left(iL_{\tau}aQ_qzeE_z\right)$.
The analytical extension supports the conclusion that the chiral symmetry is restored by the external electric field.
The charge density also oscillate over $z$ coordinates with a same frequency.
Apart from that, a weak signal of rho meson condensation is observed, but it is also possible that this is a fake phenomenon from discretization errors.

The imaginary part of the Polyakov loop shows up as expected, which indicates the presence of the R-W transition.
At low temperatures and small electric field strength, the phase of Polyakov loop has plateaus at $2n\pi/3$.
When the widths of plateaus are neglected, $\arg\left(\langle P(z)\rangle\right) \approx -2iazeE_z$.
At high temperatures, the phase of the Polyakov loop is restricted to $(-\pi/2 , \pi/2)$, and the absolute value of the phase decreases with the growth of temperature.
Meanwhile, the absolute value of the Polyakov loop starts to oscillate over $z$ coordinate.

It is verified that, the Polyakov loop can be described by ansatz $A_p+\sum _{q=u,d}B_q\exp \left(L_{\tau}iaQ_qzeE_z\right)$.
From low temperature to high temperature, the size relation between $|A_p|$ and $C_q$ changes, which results in different behavior of the Polyakov loop.
The boundary to distinguish whether the Polyakov loop enclose the origin is obtained and is found to be close to the boundary that charge density starts to oscillate.
Since the behavior of the phase of Polyakov loop is very different based on whether the origin is enclosed, there is a possible phase transition, at $E_z$ much larger than the expected R-W transition or chiral transition.

\section*{ACKNOWLEDGMENT}

\noindent
We are grateful to Gergely Endr\H{o}di for useful discussions.
This work was supported in part by the National Natural Science Foundation of China under Grants No. 12147214, the Natural Science Foundation of the Liaoning Scientific Committee No.~LJKZ0978 and the Outstanding Research Cultivation Program of Liaoning Normal University (No.21GDL004).

\bibliography{em}
\bibliographystyle{JHEP}

\end{document}